\documentclass[fleqn,usenatbib]{mnras}

\usepackage{newtxtext,newtxmath}

\usepackage[T1]{fontenc}
\usepackage{ae,aecompl}

\usepackage{graphicx}	
\usepackage{amsmath}	
\usepackage{amssymb}	

\title[Stellar Parameters from INNs]{Stellar Parameter Determination from Photometry using Invertible Neural Networks}

\author[Ksoll et al.]{
Victor F. Ksoll,$^{1,2}$\thanks{E-mail: v.ksoll@stud.uni-heidelberg.de}
Lynton Ardizzone,$^{3}$
Ralf Klessen,$^{1,2}$
Ullrich Koethe,$^{3}$
\newauthor
Elena Sabbi,$^{5}$
Massimo Robberto,$^{5}$
Dimitrios Gouliermis,$^{1,4}$
Carsten Rother, $^{3}$
\newauthor
Peter Zeidler$^{5,6}$
and Mario Gennaro$^{5}$
\\
$^{1}$Universit\"{a}t Heidelberg, Zentrum f\"{u}r Astronomie, Institut f\"{u}r Theoretische Astrophysik,  Albert-Ueberle-Str. 2, 69120 Heidelberg, Germany\\
$^{2}$Universit\"{a}t Heidelberg, Interdisziplin\"{a}res Zentrum f\"{u}r Wissenschaftliches Rechnen, Im Neuenheimer Feld 205, 69120 Heidelberg, Germany\\
$^{3}$Universit\"{a}t Heidelberg, Heidelberg Collaboratory for Image Processing, Visual Learning Lab,  Berliner Str. 43, 69120 Heidelberg, Germany\\
$^{4}$Max Planck Institute for Astronomy, K\"{o}nigstuhl\,17, 69117 Heidelberg, Germany\\
$^{5}$Space Telescope Science Institute, 3700 San Martin Drive, Baltimore, MD 21218, USA\\
$^{6}$Department of Physics and Astronomy, Johns Hopkins University, Baltimore, MD 21218, USA\\
}

\date{Accepted XXX. Received YYY; in original form ZZZ}

\pubyear{2020}

\begin{document}
\label{firstpage}
\pagerange{\pageref{firstpage}--\pageref{lastpage}}
\maketitle

\begin{abstract}
Photometric surveys with the Hubble Space Telescope (HST) allow us to study stellar populations with high resolution and deep coverage, with estimates of the physical parameters of the constituent stars being typically obtained by comparing the survey data with adequate stellar evolutionary models. This is a highly non-trivial task due to effects such as differential extinction, photometric errors, low filter coverage, or uncertainties in the stellar evolution calculations. These introduce degeneracies that are difficult to detect and break. To improve this situation, we introduce a novel deep learning approach, called conditional invertible neural network (cINN), to solve the inverse problem of predicting physical parameters from photometry on an individual star basis and to obtain the full posterior distributions. We build a carefully curated synthetic training data set derived from the PARSEC stellar evolution models to predict stellar age, initial/current mass, luminosity, effective temperature and surface gravity. We perform tests on synthetic data from the MIST and Dartmouth models, and benchmark our approach on HST data of two well-studied stellar clusters, Westerlund 2 and NGC\,6397. For the synthetic data we find overall excellent performance, and note that age is the most difficult parameter to constrain. For the benchmark clusters we retrieve reasonable results and confirm previous findings for Westerlund 2 on cluster age ($1.04_{-0.90}^{+8.48}\,\mathrm{Myr} $), mass segregation, and the stellar initial mass function. For NGC\,6397 we recover plausible estimates for masses, luminosities and temperatures, however, discrepancies between stellar evolution models and observations prevent an acceptable recovery of age for old stars.
\end{abstract}




\section{Introduction}

Machine learning (ML) employs statistical models to predict the characteristics of a dataset using samples of previously collected data without relying on physical models of the system. The introduction of ML for solving regression, classification and clustering problems has revolutionised scientific research, and in particular has provided effective methods for analyzing big astronomical data \citep{FeigelsonBabu2012, Ivezic2014}. In order to construct a model from observed data, machine learning methods rely on human-defined classifiers or 'feature extractors' \citep{Hastie2009}. However, complex problems require algorithms that automate the creation of feature extractors using large amounts of data. These algorithms represent a family of ML techniques, named deep learning, and they are based on the construction of artificial neural networks \citep[NNs;][]{GoodBengCour16}. While training NNs requires significant computational power, they achieve far higher levels of accuracy than classic ML for many non-linear problems. In this pilot study we employ invertible NNs to infer stellar ages and masses from Hubble Space Telescope (HST) imaging of two well-studied stellar clusters. Our aim is to explore the efficiency of NNs in extracting stellar physical parameters from photometry alone. We train our networks using modeled-observable properties relations provided by theoretical evolutionary models.

Star clusters, the building blocks of galaxies, are the signposts guiding our understanding of the formation and evolution of stars. This understanding stems from the physical properties of stars in clusters, being deduced from detailed comparisons of photometric observations to theoretical evolutionary models. The interface where observations meet theory is often provided by the observational colour–magnitude diagram (CMD) and its theoretical counterpart, the Hertzsprung-Russell diagram (HRD). In the HRD two physical properties of stars, the effective temperature and the luminosity, are compared to stellar evolutionary models to determine fundamental stellar parameters, the initial mass and the age of the star, which are not directly accessible by observations alone. This comparison can be directly performed through fitting of isochronal evolutionary models to the observed CMDs. This method, however, lacks proper statistical basis because the relations between observables and physical properties may present degeneracies that need to be accounted for. More advanced methods, based on Bayes statistics, derive probabilistically the cumulative properties of stellar populations, such as the mean age, in terms of posterior probability distribution functions of the properties of individual stars, e.g. the age \citep[see][and references therein]{VallsGabaud2014}. These methods provide a significant improvement by tackling the intrinsic model degeneracies through priors on the stellar initial mass function, binary fraction, or extinction distribution \citep[e.g.][]{JorgensenLindegren2005, DaRio2010}.

Bayesian inference encompasses a specific class of machine learning models, i.e. those based on strong prior intuitions. However, these priors do not add significant value in the case of big data, and are computationally expensive and slow. As a consequence, other ML methods are employed to infer stellar physical parameters from photometry. The most successful techniques developed so far are generally based on time-domain observations, such as light curves using photometric-brightness variations \citep[e.g.][]{Miller2015} or time-series asteroseismic observations \citep[e.g.][]{Bellinger2016}. These methods make use of various instances of each specific target star in time, a dataset which cannot be easily obtained for rich stellar samples in compact clusters. Investigations of stars in clusters normally rely only on 'static', rather than time-dependent imaging, which cannot be addressed by classic ML methods. Moreover, it is now well understood that parameter degeneracies encoded in the evolutionary models make the problem of inferring stellar masses and ages from photometric measurements a non-linear problem. The solution of such problems calls for the employment of artificial NNs.

There have been several recent studies that employ neural network approaches to solve prediction tasks in astronomy similar to the problem that we analyze in this paper. \cite{2020MNRAS.491.2280S} train a convolutional neural network on a suite of spectral libraries in order to classify stellar spectra according to the Harvard scheme and successfully apply their approach to data from the Sloan Digital Sky Survey (SDSS) database. \cite{2020arXiv200407261K} leverage Gaia DR2 photometry and parallaxes to construct a neural network that predicts age, extinction and distance of stellar clusters in the Milky Way, allowing them to study the star formation activity in the spiral arms. \cite{2020A&A...640A...1C} use a similar neural network approach, also predicting physical parameters of stellar clusters from Gaia data, but use 2D histograms of the observed CMDs as inputs. \cite{2020AJ....159..182O} use a deep convolutional neural network to predict surface temperature, metallicity and surface gravity of young stellar objects (YSOs) based on spectra from APOGEE. Within their training set construction they employ another convolutional neural network to infer physical parameters of YSOs, i.e. ages, masses, extinction, surface temperature/gravity, from photometry in 9 bands of the Gaia system, as well as distance, stellar radius and luminosity. This auxiliary network is trained on synthetic isochrone data and successfully recovers surface temperatures for YSOs on real Gaia observations.

For many applications in natural sciences, the forward process of determining measurements from a set of underlying physical parameters is well-defined, whereas the inverse problem is ambiguous because multiple parameter sets can result in the same observation (i.e., degeneracies). Classical neural networks attempt to address this ambiguity by solving the inverse problem directly. However, to fully characterise degeneracies, the full posterior parameter distribution, conditioned on an observed measurement, must be determined. A particular class of neural networks, so-called invertible neural networks (INNs), is well suited for this task \citep[e.g.][]{Ardizzone2019a}. Unlike classical neural networks, INNs learn the forward process, using additional latent output variables to capture the information otherwise lost. This invertibility allows a model of the corresponding inverse process to be learned implicitly, providing the full parameter posterior distribution for a given observation and corresponding distribution of the latent variables. INNs are therefore a powerful tool in identifying multi-modalities, parameter correlations, and unrecoverable parameters.

In this paper we present the application of invertible neural networks to the regression problem of predicting physical parameters of individual stars based on observed photometry. Note that we do not perform an exhaustive analysis of the approach, but rather aim to provide an introduction to the method, highlighting our first successes. This paper is the first in a series, in which we adapt and develop the approach, as well as explore its limitations. 

As mentioned above, in general this regression task is prone to errors due to the many sources of degeneracy in the mapping from physical to observable space, such as metallicity, extinction, variability, binarity and the intrinsic overlap of certain phases in stellar evolution in the observable space, e.g. the red giant branch and the pre-main-sequence. Since our primary goal is to test the viability of the method, in this paper we neglect some of these factors, adopting the following \textit{simplifying assumptions}: 1) We only deal with single metallicity populations, 2) we obtain an estimate of the individual stellar extinction of the query stars, 3) we assume perfect observations, so we do not include photometric errors, and 4) we exclude effects from variability or binarity.

We train and test our method on synthetic data from the PARSEC stellar evolutionary models \citep{Bressan2012}. Furthermore, we conduct additional synthetic tests on data from the MIST \citep{Dotter2016_MIST} and Dartmouth \citep{Dotter2008_Dartmouth} models. Lastly, we perform a benchmark study on real observational data from the Hubble Space Telescope (HST) of the young star forming cluster Westerlund 2 and the old globular cluster NGC\,6397. These clusters are chosen for our pilot study due to their well-defined single ages \citep{Zeidler2016, Brown2018}, allowing for an accurate evaluation of our results. 

In Section \ref{sec:Data} we summarise the physical properties of our benchmark targets and the reduction of the observational data from their respective surveys. Furthermore, we outline the construction of our training sets from the synthetic data provided by the PARSEC models. In the following Section \ref{sec:NN_setup} we elaborate the background of the invertible neural network approach and provide details of the final architecture of our models as well as the performance measures used to evaluate their success. Section \ref{sec:Results} summarises the performance of the cINN on the PARSEC synthetic test data for each of our four training sets and details the results of the application to the MIST and Dartmouth data. In Section \ref{sec:prediction} we present the prediction outcome on the real observational data for both Westerlund 2 and NGC\,6397. We discuss possible future extensions of our approach beyond the simplifications assumed for this work in Section \ref{sec:extensions}. The final Section \ref{sec:conclusion} summarises our key findings.

\section{Data selection and Preparation}
    \label{sec:Data}
    \subsection{Observational Data}
        \label{sec:obs_data}
        
        \begin{figure}
            \centering
            \includegraphics[width = 0.49\linewidth]{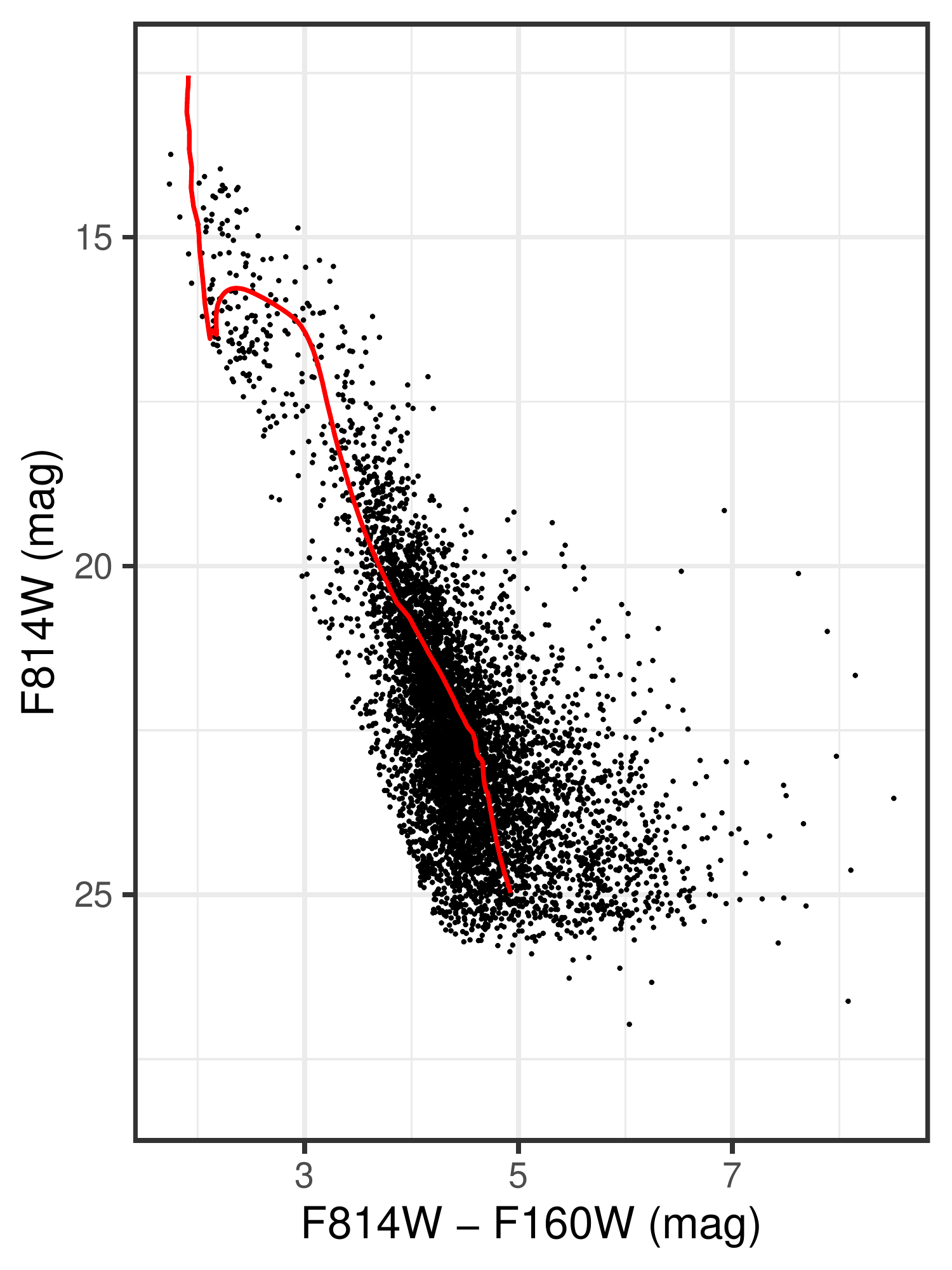}
            \includegraphics[width = 0.49\linewidth]{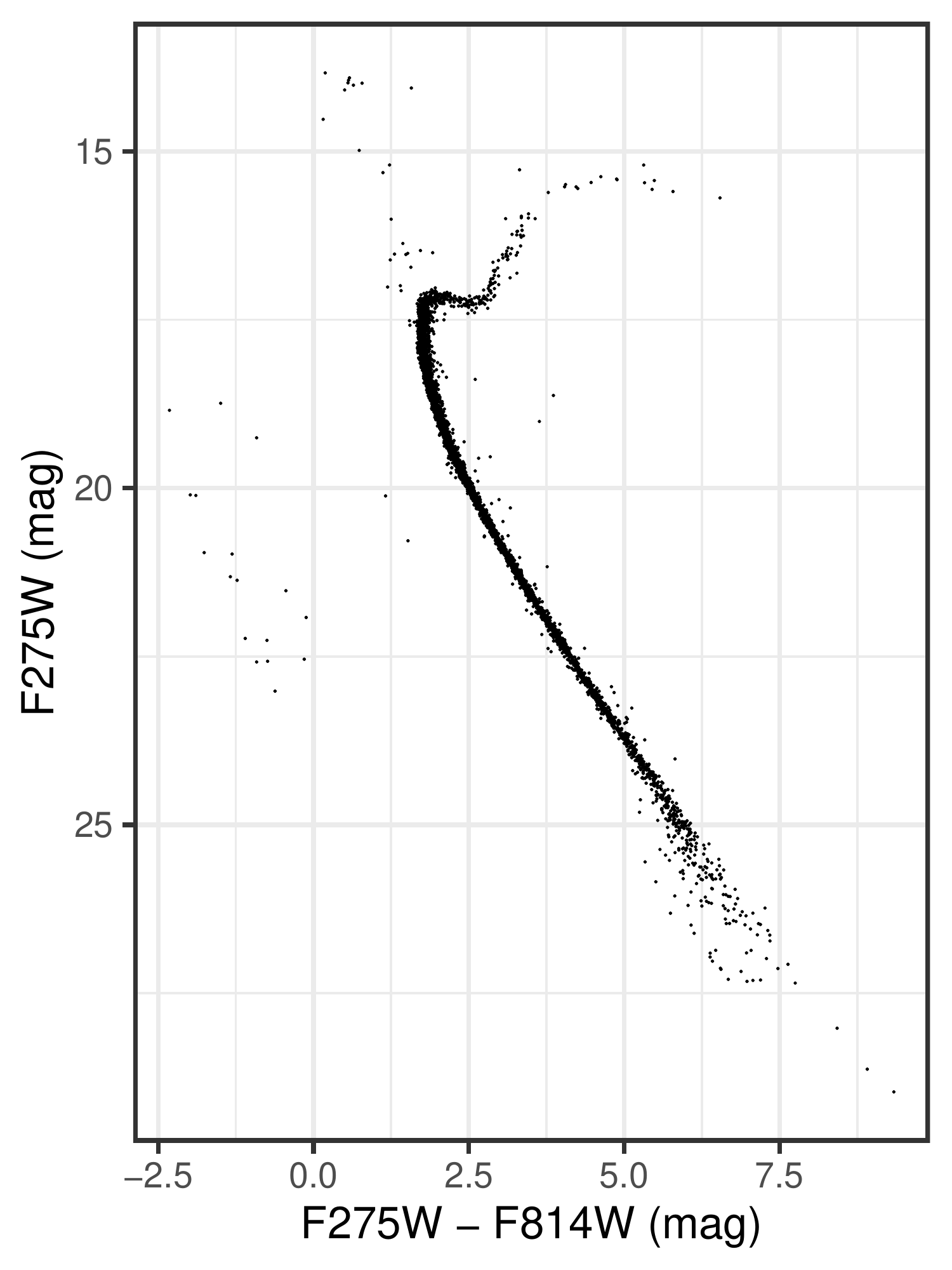}
            \caption{\textbf{Left:} Colour magnitude diagram of our Westerlund 2 data set. The red line represents a $1\,\mathrm{Myr}$ PARSEC \citep{Bressan2012} isochrone corrected for the median stellar extinction and distance modulus of Wd2. \textbf{Right:} UV-I colour magnitude diagram of the NGC\,6397 photometric catalogue from the HUGS project.}
            \label{fig:Obs_CMDs}
        \end{figure}
        To test our neural-network-based approach to predicting physical parameters of stars on real observational data we use two 'well behaved', supposedly single age (or close to) stellar clusters for which very-high spatial resolution HST observations are available, namely the young massive star forming region Westerlund 2 (hereafter referred to as Wd2) located within the Milky Way and the old globular cluster NGC\,6397 belonging to the galactic halo. Since this paper serves only as an introduction to the INN approach to gain initial insights into the systematics of the method we do not conduct an exhaustive study of the full range of the cluster mass, age, and metallicity distribution, but we consider only the two extremes in age (i.e. very young and very old).
        \subsubsection{Westerlund\,2}
            \label{sec:obs_Wd2}
            Wd2 is one of the most massive star forming clusters in the Milky Way, harboring a total stellar mass larger than $10^4\,\mathrm{M_\odot}$ \citep{Ascenso2007}. It is located in the Carina-Sagittarius arm at a distance of $4.16 \pm 0.33\,\mathrm{kpc}$ \citep{Zeidler2015} from the Sun. At an age of $1.04 \pm 0.72\,\mathrm{Myr}$ \citep{Zeidler2016} Wd2 makes for an excellent example of a young massive cluster at solar metallicity still in its early star formation stages within close proximity to the Sun. While Wd2 exhibits an average total-to-selective extinction of $R_\mathrm{V} = 3.95 \pm 0.135$ that is larger than the galactic average $R_\mathrm{V} = 3.1$, the cluster is only affected by relatively low differential reddening with $E(B-V)_\mathrm{g} = 1.87\,\mathrm{mag}$ \citep[median color excess of the gas;][]{Zeidler2015}. For our following considerations we adopt $R_\mathrm{V} = 3.8$ to be both in agreement with the findings of \cite{Zeidler2015} as well as the spectroscopic observations of \cite{Carraro2013} and \cite{VargasAlvarez2013} who suggest $R_\mathrm{V} = 3.85 \pm 0.07$ and $R_\mathrm{V} = 3.77 \pm 0.09$, respectively. Thus, the corresponding median gas extinction of Wd2 lies at $A_\mathrm{V,g} = 7.1$\,mag. \\
            
            Combining multi epoch HST images taken with the Wide Field Camera 3 (WFC3) in F814W with previously obtained UVIS-IR data in F160W (PI Nota, GO-13038) \cite{Sabbi2020} compile the photometric catalogue that we employ for this study. Due to the long 350s exposure times in F814W, this photometric catalogue does unfortunately not contain the brightest objects of Westerlund 2, i.e. the most massive upper main sequence (UMS) constituents, as they were saturated. Disregarding these missing UMS sources the \cite{Sabbi2020} photometric catalogue consists of 9267 stars, of which 6268 are thought to belong to Wd2. The remaining stars in the sample can be tentatively classified as lower main-sequence (LMS) fore- or background contaminants that fall into the line of sight. 
            The left panel in Figure \ref{fig:Obs_CMDs} shows the CMD of the 6268 cluster stars. 
            
            Adopting the \cite{Zeidler2015} gas extinction map of Wd2, we can derive individual stellar colour excesses $E(B-V)_*$ for 8939 stars that fall within the border of the map following their prescription: 
            \begin{equation}
                E(B-V)_* = 0.4314 \cdot E(B-V)_g + 0.7400.
            \end{equation}
            The individual stellar extinctions then follow as 
            \begin{equation}
            A_\mathrm{V} = R_\mathrm{V} \cdot E(B-V)_*.
            \end{equation}
        
        \subsubsection{NGC\,6397}
            NGC\,6397 is the nearest metal poor globular cluster, with a distance of $d=2.39 \pm 0.17\,\mathrm{kpc}$ (distance modulus $DM=11.89 \pm 0.16\,\mathrm{mag}$) derived from parallax measurements with high precision HST astrometry \citep{Brown2018}. Spectroscopic measurements indicate a metallicity of $\mathrm{[Fe/H]} = -2.02$ \citep{Kraft2003, Vulic2018}, making it a prime example of an ancient metal-poor stellar population. Fitting of the main-sequence turnoff suggests a cluster age of $13.4 \pm 1.9\,\mathrm{Gyr}$ \citep{Brown2018}. Several extinction studies indicate a moderate reddening, constraining $E(B-V)$ to a value between $0.183\,\mathrm{mag}$ \citep{Gratton2003}, $0.186\,\mathrm{mag}$ \citep{Schlegel1998} and $0.187\mathrm{mag}$ \citep{AnthonyTwarog1992}. In this work we adopt $E(B-V) = 0.185 \pm 0.002\,\mathrm{mag}$ from \citep{Brown2018}, corresponding to an average extinction of $A_\mathrm{V} = E(B-V) \cdot R_\mathrm{V} = 0.5735 \pm 0.0062 \,\mathrm{mag}$ with $R_\mathrm{V} = 3.1$. To derive individual stellar extinctions here we simply sample from a Gaussian distribution with this mean and standard deviation. 
            
            We use the photometric catalogue of NGC\,6397 from the HST legacy survey 'HST UV Globular Cluster Survey (HUGS)' \citep{Piotto2015_HUGS, Nardiello2018_HUGS}, which provides coverage in the F275W, F336W and F438W filters, observed with the WFC3/UVIS channel, as well as in F606W and F814W, imaged with ACS/WFC \citep{Nardiello2018_HUGS}. To pre-process this data we follow the prescription in Section 3 of \cite{Nardiello2018_HUGS}. We divide the photometric error and quality of fit distributions of each filter into 12 magnitude bins and find the $3.5\sigma$ clipped average of the magnitude and parameter in each bin. Here $\sigma$ refers to the standard deviation of the distribution in the given bin. In each bin $3.5\sigma$ is then added to the mean value and a linear interpolation is performed between these points. For the photometric errors we then reject all observations that lie above this interpolated line while for the quality of fit parameter we reject all instances below the line.
            Finally we limit the catalogue to observations with a sharpness value between $-0.15$ and $0.15$ in all five filters. Following these selection criteria we obtain a photometric catalogue containing 4831 stars. The right panel of Figure \ref{fig:Obs_CMDs} shows the corresponding UV-I CMD. 
        
    \subsection{Synthetic Training Data}
        \label{sec:training_data} 
        \begin{figure}
            \centering
            \includegraphics[width = \linewidth]{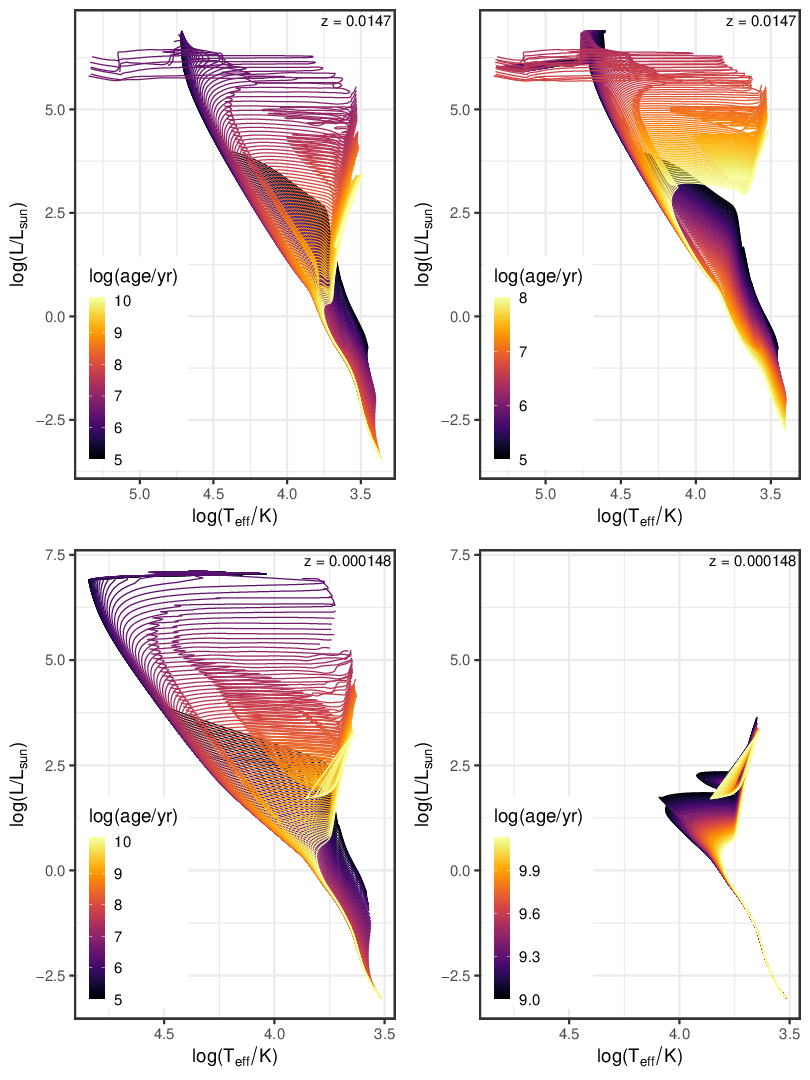}
            \caption{Hertzsprung-Russel Diagrams for the PARSEC 1.2s isochrone tables used as basis for our training sets. The \textbf{top row} shows set 'Wd2\_I' with isochrones from 5 to 10.1 in $\log(\mathrm{age/yr})$ in steps of 0.05 dex (left) and set 'Wd2\_II' with isochrones in the range of 5 to 8 in 0.025 dex (right). In the \textbf{bottom row} are the corresponding HRDs of the sets 'NGC6397\_I'(left), containing isochrones from 5 to 10.13 in 0.05 dex, and 'NGC6397\_II' with isochrones from 9 to 10.13 in steps of 0.01 dex. All isochrones are colour coded according to their $\log(\mathrm{age})$.}
            \label{fig:IsochroneModels}
        \end{figure}
        
        \begin{figure*}
            \centering
            \includegraphics[width = \linewidth]{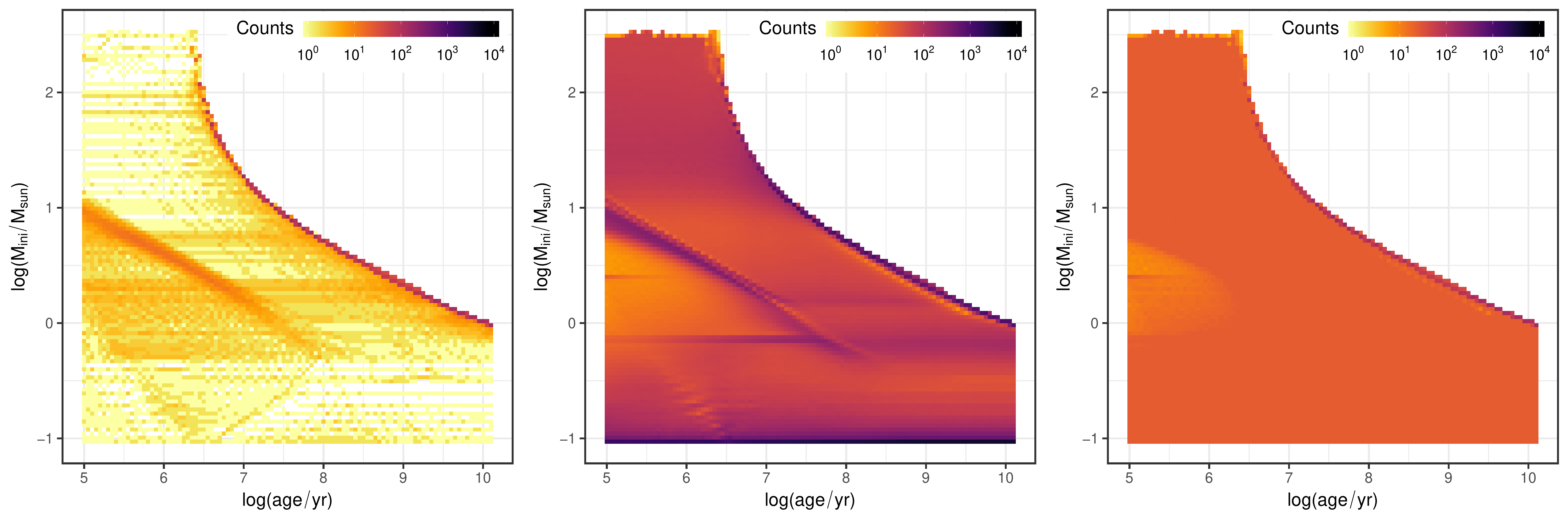}
            \caption{2D Histograms of the age vs. initial mass distributions for the original PARSEC isochrone tables (\textbf{left}), the case where each isochrone is oversampled 10,000 times in regular intervals in terms of the logarithm of its arclength in $M_\mathrm{ini}$-$L$-$T_\mathrm{eff}$ space (\textbf{middle}) and our final training set base (\textbf{right}). In the latter each age-mass-bin that contains less than 30 model points in the original table is reinforced with additional samples from the oversampled isochrones to reach 30 examples. This example is from the 'Wd2\_I' training set.}
            \label{fig:IsochronesAgeVsMass}
        \end{figure*}
        
        In order to train the neural network for the purpose of predicting physical parameters given photometric observations of individual stars, a large training set is required that contains both the physical parameters and the corresponding photometric observations of each star. Since at present such a training dataset is not readily available, we build it from theoretical stellar evolutionary models. In particular, we use version 1.2s of the PARSEC stellar evolutionary tracks \citep{Bressan2012, Tang2014, Chen2015, Chen2014} and more specifically the isochrone tables derived for the HST photometric systems 'WFC3 wide' and 'ACS WFC'. Since our observational test cases Wd2 and NGC\,6397 differ both in metallicity and HST filter coverage, we have to construct individual training sets for each cluster. This is consistent with the fact that our neural network structure can only deal with single metallicity cases. An artificial training set is also appropriate since our neural network cannot deal with missing observational features. 
        
        At this point it is important to note that using synthetic training data comes with caveats. In particular it is known that the photometry interpolated from the stellar evolution models may show minor discrepancies in colors as the approximation to the real bandpasses may be imprecise. Consequently, the synthetic photometry can never perfectly match real observations. Additionally, the models themselves may be discrepant, e.g. for very low mass stars \citep{2018MNRAS.476.3245J} or YSOs \citep{2020AJ....159..182O}, or may exhibit physically questionable properties such as the large gap in surface temperature for pre-main-sequence stars at 4000\,K in the PARSEC models. Nevertheless, e.g., \cite{2020AJ....159..182O} find that a neural network approach, trained on synthetic data, can recover realistic physical properties for YSOs on real data, where traditional isochrone fitting approaches fail due to the model discrepancies. In any case, given the task we aim to solve here, the use of synthetic training data is simply unavoidable. Therefore, we proceed keeping these caveats in mind for our real data benchmarks.
        
        For both clusters we construct two training sets, one agnostic to prior knowledge of the stellar ages and one where we constrain the stellar ages to a range close to the supposed cluster ages derived in previous studies. 
        
        The first two training sets 'Wd2\_I' and 'NGC6397\_I' thus consider isochrones with $\log(\mathrm{age/yr})$ in the range $5$ to $10.1$ in steps of $0.05\,\mathrm{dex}$. The 'NGC6397\_I' set also specifically entails the $\log(\mathrm{age/yr}) = 10.13$ isochrone to include the supposed age of NGC\,6397 of $13.4\,\mathrm{Gyr}$ \citep{Brown2018}. For the other two training sets 'Wd2\_II' and 'NGC6397\_II' we restrict the isochrones to $\log(\mathrm{age/yr})$ ranges of $5$ to $8$ in $0.025\,\mathrm{dex}$, and $9.0$ to $10.13$ in $0.01\,\mathrm{dex}$, respectively. Figure \ref{fig:IsochroneModels} shows the HRD corresponding to these training sets. We do not impose a restriction on the range of initial stellar mass $M_\mathrm{ini}$ so that the full mass range of the PARSEC models ($0.09$ to $350\,\mathrm{M}_\odot$) is available in all but the 'NGC6397\_II' training set, where the range has been reduced to $0.09$ to $1.837\,\mathrm{M}_\odot$ due to the fact that the more massive stars have already died at these ages. The other physical parameters that we consider for prediction are current mass $M_\mathrm{curr}$, luminosity $L$, effective temperature $T_\mathrm{eff}$ and surface gravity $g$. Again, we do not limit these parameters so that the respective ranges depend on the isochrones included in each training set. 
        
        For these training sets we do not perform population synthesis based on the isochrone tables, but instead we consider each point of the isochrones as an individual example star, aiming at performing parameter prediction on a star by star basis. To this purpose we need to populate the physical parameter space in the training set as evenly as possible, since overpopulated regions could introduce biases in the training process, so that our trained model might in the end generalise poorly when predicting parameters for a star that falls in a less populated area in parameter space. We face this problem with the isochrone models. While the PARSEC models provide perfectly evenly spaced isochrones in $\log(\mathrm{age})$, \cite{Bressan2012} perform an interpolation when generating the isochrones from the stellar evolutionary tracks that aims to produce smooth isochrone curves, resulting in a severe oversampling of certain masses due to the fact that very small mass variations can cause a significant change of position in the HRD on the post main sequence part of the isochrones. Figure \ref{fig:IsochronesAgeVsMass} shows an example of this mass oversampling for the 'Wd2\_I' case. The left diagram highlights how the interpolation strategy of the PARSEC isochrone tables results in a severe oversampling of masses along the ridge where the models stop, because stars of a given mass die away. Consequently, there are several regions, e.g. the old low mass stars and young super massive stars, where the age-mass space is strongly underpopulated.
        
        To remedy this problem we have devised a procedure to augment the isochrone tables so that the density differences between the over- and underpopulated regions in age-mass space are reduced. We begin by oversampling each isochrone in $M_\mathrm{ini}$ space, first performing a linear spline interpolation in the $M_\mathrm{ini}$ - $L$ - $T_\mathrm{eff}$ space to determine its arc-length, i.e. the length of the path along the isochrone from the lowest mass model point to the most massive one. Then we find 10,000 equidistant (in terms of the logarithm of the arc-length) $M_\mathrm{ini}$ points along each isochrone. For these points we determine the remaining parameters ($L$, $T_\mathrm{eff}$, $g$ and magnitudes) by performing a linear interpolation between the nearest lower and nearest higher initial mass neighbour on the original isochrone. 
        
        The resulting age-mass distribution of these oversampled isochrones is shown in the middle diagram of Figure \ref{fig:IsochronesAgeVsMass}. The plot indicates that this procedure does not solve the issue of oversampled mass bins directly, in fact, it further highlights those regions. But at the same time it manages to populate previously sparsely sampled regions. To finally produce an evenly sampled training set we then augment the original isochrone tables by adding random samples from our oversampled isochrones until every age-mass bin contains at least 30 example stars (this value is chosen to roughly represent the number of the least populated bins in the oversampled data set). If the oversampled table does not contain enough additional examples to augmented the original isochrones to 30 examples in a given bin, we simply include all available additional examples. We also ensure to only augment with examples that do not appear in the original tables. The resulting distribution in age-mass-space is depicted in the right panel of Figure \ref{fig:IsochronesAgeVsMass}, showing that this approach achieves a mostly even sampling across the whole parameter range.
        
        There are two reasons why we do not achieve a perfectly even sampling. First, subsampling the overpopulated bins would result in a significant information loss in the HRD and CMD as several post-main-sequence evolutionary tracks fall into these bins. Second, oversampling the isochrones and then augmenting the original tables to a degree that all bins reach the level of the originally most populated bin would result in a data set so large that it becomes not manageable for our remaining processing. 
        
        The last step in our training set construction procedure is to augment the data taking extinction into account. We do so for each star in the training set by including additional copies of it at different amounts of extinction $A_\mathrm{V}$ and altering their observable features, i.e. magnitudes in HST filters, accordingly. For Wd2 we consider an extinction range from $0$ to $12$ in steps of $0.2\,\mathrm{mag}$ and for NGC\,6397 from $0$ to $3$ in steps of $0.05\,\mathrm{mag}$ in accordance with the Wd2 gas extinction map from \cite{Zeidler2015} and the suggested average extinction of NGC\,6397 by \cite{Brown2018}. For the extinction law we use the diffuse Milky Way extinction curve by \cite{Cardelli1989}, deriving the $A_\lambda/A_\mathrm{V}$ values in dependence of $R_\mathrm{V}$ for the HST filters according to
        \begin{equation}
            \label{eq:alambda}
            \frac{A_\lambda}{A_\mathrm{V}} = a_\lambda + \frac{b_\lambda}{R_\mathrm{V}},    
        \end{equation}
        where $a_\lambda$ and $b_\lambda$ denote wavelength dependent coefficients defined by \cite{Cardelli1989}. Table \ref{tab:ex_law} in the Appendix provides the derived $A_\lambda/A_\mathrm{V}$ values for all filters. \\
        
        In conclusion, each training set contains the six physical parameters: age, initial mass $M_\mathrm{ini}$, current mass $M_\mathrm{curr}$, luminosity $L$, effective temperature $T_\mathrm{eff}$, surface gravity $g$, extinction $A_\mathrm{V}$ and magnitudes in filter combinations corresponding to our real observations. These are $\mathrm{F814W}_\mathrm{WFC3}$ and $\mathrm{F160W}_\mathrm{WFC3}$ for Wd2, and $\mathrm{F275W}_\mathrm{WFC3}$, $\mathrm{F336W}_\mathrm{WFC3}$, $\mathrm{F438W}_\mathrm{WFC3}$, $\mathrm{F606W}_\mathrm{ACS}$, $\mathrm{F814W}_\mathrm{ACS}$ for NGC\,6397. In total our training sets contain 12,481,881, 20,903,602, 12,356,282 and 16,817,090 example stars for 'Wd2\_I', 'Wd2\_II', 'NGC6397\_I' and 'NGC6397\_II', respectively. Figure \ref{fig:TS_Priors} in the Appendix shows the corresponding prior distributions of all physical parameters for these training sets.
        
\section{Neural Network Setup}
    \label{sec:NN_setup}
    \subsection{INN and cINN}
        \begin{figure}
            \centering
            \includegraphics[width = \linewidth]{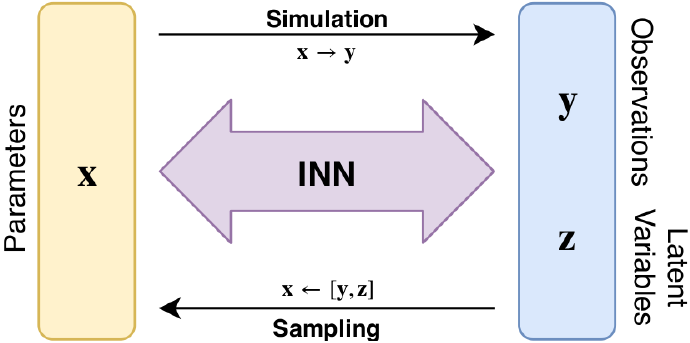}
            \caption{Schematic overview of the invertible neural network approach for solving an inverse problem. Adapted from \citep{Ardizzone2019a}.}
            \label{fig:INN_Concept}
        \end{figure}
        
        \begin{figure*}
            \centering
            \includegraphics[width = 0.9\linewidth]{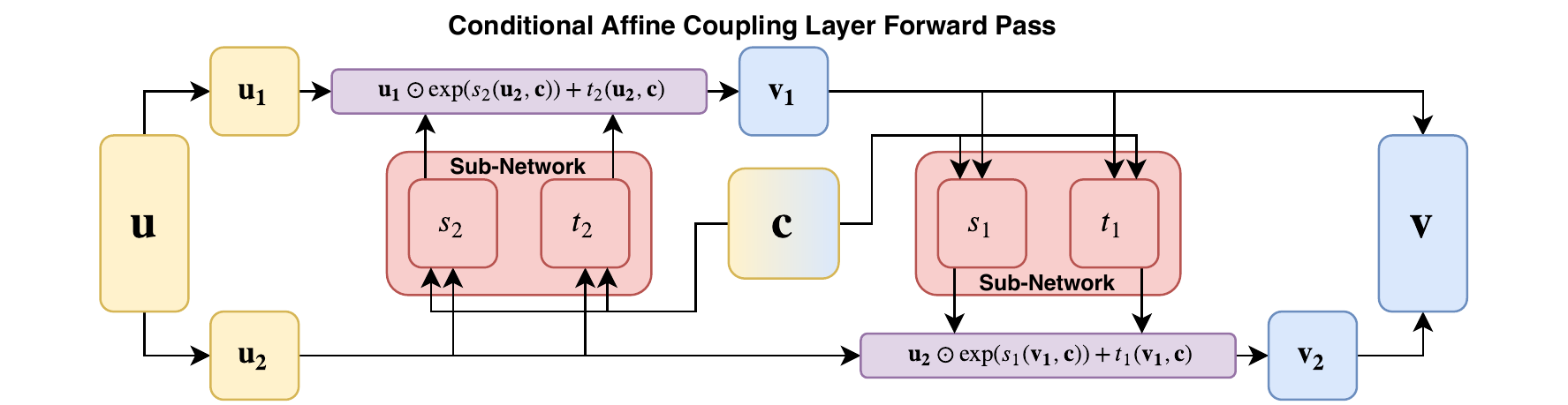}
            \includegraphics[width = 0.9\linewidth]{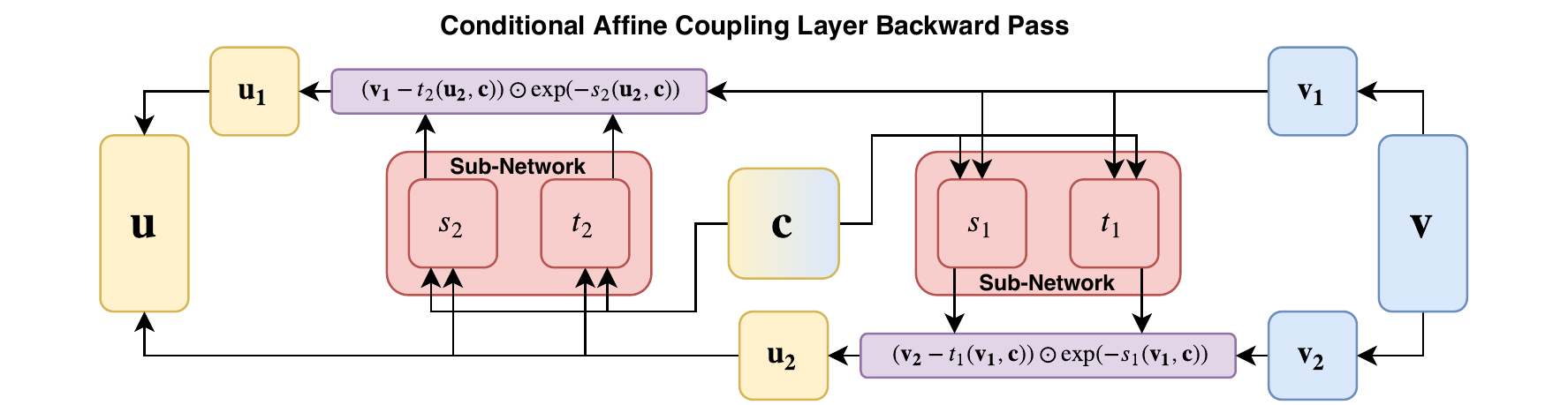}
            \caption{Schematic overview of the architecture of the conditional affine coupling blocks used in the cINN. In particular we show the GLOW \citep[Generative Flow;][]{Kingma2018} configuration, where the outputs $s_\mathrm{i}()$ and $t_\mathrm{i}()$ are computed by a single subnetwork (for each $i$).  The \textbf{top} panel shows how data is passed through the block in the forward direction (from $\mathbf{x}$ to $\mathbf{z}$), while the \textbf{bottom} panel displays the inverted case following the affine transformations in Equations\eqref{eq:INN_forward} and \eqref{eq:INN_backward}.}
            \label{fig:cINN_ABC}
        \end{figure*}
    
        In this paper we solve the inverse problem of predicting physical parameters of stars from HST photometry employing an invertible neural network (INN) as described in \cite{Ardizzone2019a, Ardizzone2019b}. This INN approach provides an inverse solver that estimates the \textit{complete posterior} distribution of physical parameters \textit{conditioned} on an observation. Figure \ref{fig:INN_Concept} outlines the concept of the INN methodology. Given a well-understood simulation that maps physical parameters $\mathbf{x}$ to observations $\mathbf{y},$ we assume that this forward process entails an inherent information loss, such that $\mathbf{y}$ does \textit{not} explain all variance of $\mathbf{x}$ and \textit{degeneracies} occur in the mapping. To retain this information that would be otherwise lost additional latent variables $\mathbf{z}$ are introduced to encode all the variance of $\mathbf{x}$ that is not captured in $\mathbf{y}$. 
        
        A benefit of a network with an invertible architecture with regard to our current regression problem is that once it has been trained to approximate the known forward process $f$, it provides a solution for the inverse process $f^{-1}$ for free. In the application outlined here the INN will thus learn how to associate physical parameter values $\mathbf{x}$ to unique pairs $[\mathbf{y}, \mathbf{z}]$ of observations and latent variables, as it trains to optimise the forward mapping $f(\mathbf{x}) = [\mathbf{y}, \mathbf{z}]$ and then implicitly finds the inverse $\mathbf{x} = f^{-1}(\mathbf{y}, \mathbf{z}) = g(\mathbf{y}, \mathbf{z})$ \citep{Ardizzone2019a}. For simplicity the prior distribution of the latent variables $p(\mathbf{z})$ is assumed (and enforced during training) to be Gaussian. The desired posterior distribution $p(\mathbf{x}|\mathbf{y})$ is represented by the function $g(\mathbf{y}, \mathbf{z}) = \mathbf{x}$, which, given the condition $\mathbf{y}$, transforms the known distribution $p(\mathbf{z})$ to $\mathbf{x}$-space \citep{Ardizzone2019a}. In practise this means that for a given observation $\mathbf{y}$ the posterior distribution $p(\mathbf{x}|\mathbf{y})$ is determined by sampling the latent variables. 
        
        In \cite{Ardizzone2019a} the invertibility of the network is achieved by a series of reversible blocks based on the architecture proposed by \cite{Dinh2016}. These blocks split their input vector $\mathbf{u}$ into two halves $\mathbf{u_1}$ and $\mathbf{u_2}$ and then apply two complementary affine transformations with element-wise multiplication $\odot$ and addition $+$,
        \begin{equation}
            \label{eq:INN_forward}
            \begin{split}
                \mathbf{v_1} &= \mathbf{u_1} \odot \exp(s_2(\mathbf{u_2})) + t_2(\mathbf{u_2}),\\
                \mathbf{v_2} &= \mathbf{u_2} \odot \exp(s_1(\mathbf{v_1})) + t_1(\mathbf{v_1}),
            \end{split}      
        \end{equation}
        where $s_\mathrm{i}$ and $t_\mathrm{i}$ are mappings that can be arbitrarily complex functions of $\mathbf{u_2}$ and $\mathbf{v_1}$ that do not need to be invertible themselves and can even be represented by neural networks. These affine transformations are easily inverted given the output $\mathbf{v} = [\mathbf{v_1}, \mathbf{v_2}]$,
        \begin{equation}
            \label{eq:INN_backward}
            \begin{split}
                \mathbf{u_2} &= (\mathbf{v_2} - t_1(\mathbf{v_1})) \odot \exp(-s_1(\mathbf{v_1})),\\
                \mathbf{u_1} &= (\mathbf{v_1} - t_2(\mathbf{u_2})) \odot \exp(-s_2(\mathbf{u_2})).
            \end{split}    
        \end{equation}
        Based on the \cite{Ardizzone2019a} method, \cite{Ardizzone2019b} present an extension to their original INN approach, the conditional invertible neural network (cINN). Here they adapt the affine coupling block architecture to accept additional conditioning inputs $\mathbf{c}$. Since the mappings $s_\mathrm{i}$ and $t_\mathrm{i}$, also when represented by neural networks, are only evaluated in the forward direction, even when inverting the network, it is possible to concatenate these conditioning inputs with the regular inputs of the sub-networks without compromising the INNs invertibility, e.g. by replacing $s_2(\mathbf{u_2})$ with $s_2(\mathbf{u_2}, \mathbf{c})$ etc. in Equations \eqref{eq:INN_forward} and \eqref{eq:INN_backward}.
        Figure \ref{fig:cINN_ABC} shows an illustration for the forward (top) and backward (bottom) pass of this conditional affine coupling layer design in the GLOW \citep[Generative Flow; proposed by][]{Kingma2018} configuration (see Section \ref{sec:ArchitectureDetails} for details). In this setting the forward mapping is modified to $f(\mathbf{x};\mathbf{c}) = \mathbf{z}$ and the inverse to $\mathbf{x} = g(\mathbf{z}; \mathbf{c})$. The invertibility is given for \textit{fixed} condition $\mathbf{c}$ as 
        \begin{equation}
            f(\,\cdot\,;\mathbf{c})^{-1} = g(\,\cdot\,;\mathbf{c}).
        \end{equation}
        In our regression problem the conditioning is given by the observations. Therefore, as for the standard INN, \textit{during training given an observation the network will learn to encode all information about the physical parameters in the latent variables that was not contained in the observation.} Also analogous to the standard INN, we retrieve the desired posterior distribution $p(\mathbf{x}|\mathbf{y})$ for a given observation $\mathbf{y}$ by sampling the latent variables according to their Gaussian priors and using the inverted network $g$:
        \begin{equation}
            \mathbf{x_{posterior}} = g(\mathbf{z}; \mathbf{c} = \mathbf{y}),\,\,\mathrm{with}\, z \sim p_Z(\mathbf{z}) = \mathcal{N}(\mathbf{z}, 0, \mathbf{I}),
        \end{equation}
        where $\mathbf{I}$ is the $K\times K$ unity matrix with $K = \mathrm{dim}(\mathbf{z})$.
        
        One of the cINN benefits over the standard INN architecture is that no zero padding \citep[as described in][]{Ardizzone2019a} is necessary if the dimension of $[\mathbf{y}, \mathbf{z}]$ were to exceed that of $\mathbf{x}$, as the conditioning input $\mathbf{c}$ can be arbitrarily large in this approach and the dimension of $\mathbf{z}$ simply matches that of $\mathbf{x}$.
        
    \subsection{Architecture Details}
    
        \begin{figure*}
            \centering
            \includegraphics[width = 0.8\linewidth]{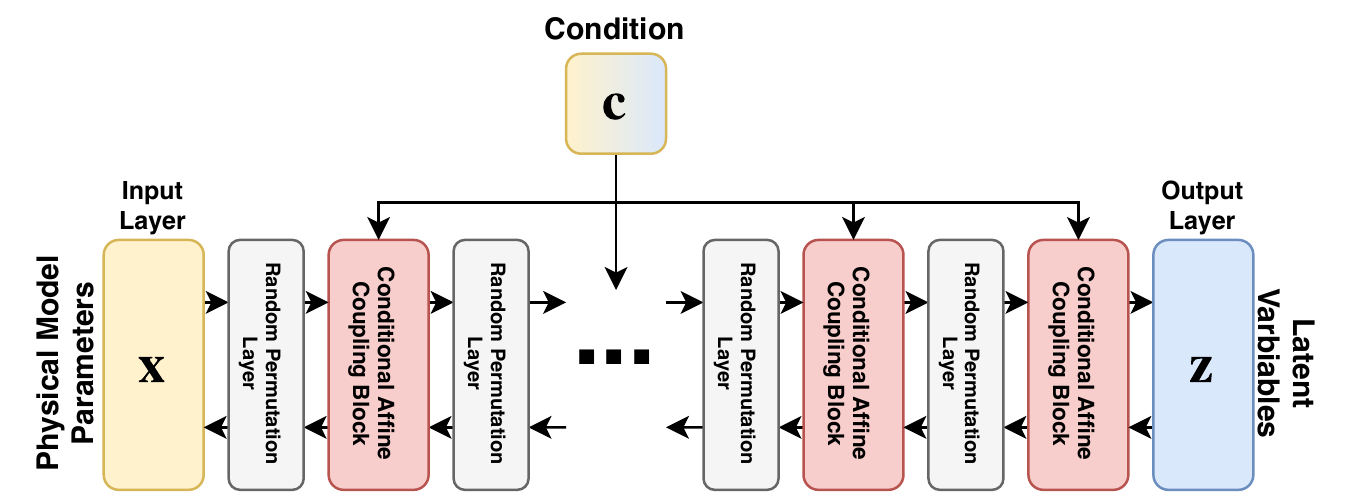}
            \caption{Schematic representation of the cINN architecture used for the physical parameter prediction based on photometry. In total we use 16 conditional affine coupling blocks interchanged with random permutation layers.}
            \label{fig:cINN_TopLevel}
        \end{figure*}
    
        \label{sec:ArchitectureDetails}
        To implement the cINN for our purposes we use the 'Framework for Easily Invertible Architectures' (FrEIA) for python \citep{Ardizzone2019a, Ardizzone2019b} based on the 'pytorch' library \citep{Paszke2017_pytorch}.  
        
        In our problem the input $\mathbf{x}$ is given by the six physical parameters of the isochrone tables, so that, following the cINN architecture, we also have six latent variables $\mathbf{z}$. Our cINN is conditioned on the observables, 2 and 5 magnitudes for Wd2 and NGC\,6397, respectively, and the individual stellar extinctions, so that the condition $\mathbf{c}$ has the dimension 3 in the Wd2 cases and 6 for NGC\,6397. \cite{Ardizzone2019b} also introduce a 'conditioning' network which transforms the input condition into some intermediate representation and is trained jointly with the cINN. We do not use this additional network in our setup, as we find that given the few observables in our problem the cINN tends to overfit to the synthetic training data when employing a feature extraction network, resulting in poor performance on the real benchmark data. 
        
        Our cINN consists of 16 conditional affine coupling blocks, each in the GLOW configuration \citep{Kingma2018}, which reduces computational cost and speeds up learning by jointly predicting the subnetwork outputs $s_\mathrm{i}()$ and $t_\mathrm{i}()$ using a single subnetwork. As in \cite{Ardizzone2019b} we introduce an additional nonlinear transformation of the scale coefficients $s$, 
        \begin{equation}
            s_\mathrm{clamp} = \frac{2 \alpha}{\pi} \arctan\left(\frac{s}{\alpha}\right),
        \end{equation}
        where $\alpha = 1.9$, so that $s_\mathrm{clamp} \approx s$ for $|s| \ll \alpha$ and $s_\mathrm{clamp} \approx \pm \alpha$ for $|s| \gg \alpha$, in order to avoid instabilities induced by large magnitudes of the  exponential $\exp\left(s_\mathrm{clamp}\right)$.
        
        We alternate the conditional affine coupling blocks with random permutation layers. The latter consist of random orthogonal matrices which mix the information between the two streams $\mathbf{u_1}$ and $\mathbf{u_2}$ in the coupling blocks. Following \cite{Ardizzone2019b}, these matrices are fixed during training and cheaply invertible. The combination of these permutation layers with the interlocked affine transformations of the affine coupling blocks ensures that the network cannot ignore the conditioning input when learning the forward mapping. 
        The subnetworks in the conditional affine coupling layers are simple fully connected feed-forward networks with three hidden layers of width 512 with rectified linear units (ReLU) as activation functions. Figure \ref{fig:cINN_TopLevel} provides a schematic overview of our setup for the cINN.
        
        We train the cINN models as described in \cite{Ardizzone2019b} by minimization of the \textit{maximum likelihood loss}
        \begin{equation}
            \mathcal{L} = \mathbb{E}_\mathrm{i}\left[ \frac{\left\vert \left\vert f(\mathbf{x_i}; \mathbf{c_i}, \theta)\right\vert \right\vert_2^2}{2} - \log \left\vert J_\mathrm{i}\right\vert  \right],
        \end{equation}
        where $\mathbf{x_i}$ is a training example with its corresponding condition $\mathbf{c_i}$ and $J_\mathrm{i}$ denotes the determinant of the Jacobi matrix $J_\mathrm{i} = \det \left( \left.\frac{\partial f}{\partial \mathbf{x}}\right\vert_\mathbf{x_i}\right)$ evaluated at $\mathbf{x_i}$. 
        
        For each training set the cINN is trained until the loss curve converges, but at least long enough that the model has seen each training example multiple times. 
        
    \subsection{Data Pre-Processing}
        In preparation for training the cINN we split our training data into physical parameters $\mathbf{x}$ (age, $M_\mathrm{ini}$, $M_\mathrm{curr}$, $L$, $T_\mathrm{eff}$, $g$) and observables $\mathbf{y}$ (magnitudes + $A_\mathrm{V}$). To avoid issues in the training process that can occur due to their broad range of values, the physical parameters are transformed to logarithmic space. This serves not only to even out magnitude differences, but it has the general benefit of implicitly enforcing that these quantities can only be positive. Since all our observables are photometric magnitudes and thus already a logarithmic quantity, this step is not necessary there. On top of that we add a small amount of Gaussian noise (standard deviation of $1\times10^{-5}$) to the strongly discretised $\mathrm{log(\mathrm{age})}$ parameter. This form of data augmentation through a small amount of noise serves to smooth out discretization artifacts of the input \citep{Ardizzone2019b}. The remaining parameters are sampled unevenly enough that augmentation with noise is unnecessary. 
        
        After that we re-scale each parameter so that their resulting distribution has zero mean and unit standard deviation, following the \textit{linear} transformation
        \begin{equation}
            \hat{x_\mathrm{i}} = (x_\mathrm{i} - \mu_{x_i}) \cdot \frac{1}{\sigma_{x_i}},
        \end{equation}
        where $\mu_{x_i}$ and $\sigma_{x_i}$ are the mean and standard deviation of the distribution of the physical parameter $x_i$.
        At prediction time these linear re-scaling operations are easily inverted in order to retrieve the correct predicted physical parameters $x_\mathrm{i,pred}$ from the predicted $\hat{x}_\mathrm{i,pred}$ as
        \begin{equation}
            x_\mathrm{i, pred} = \hat{x}_\mathrm{i,pred} \cdot \sigma_{x_i} + \mu_{x_i}.
        \end{equation}
        For the observables, after first centring the data ($\tilde{y_\mathrm{i}} = y_\mathrm{i} - \mu_{y_i}$), we perform a matrix whitening procedure \citep{Hyvarinen2000} on the $N \times M$ matrix $\mathbf{\tilde{Y}}$, where $N$ is the total number of examples in the training set and $M$ the number of observables. The resulting \textit{linearly} transformed matrix $\mathbf{\hat{Y}}$ has the properties that all its columns $\mathbf{\hat{y_i}}$ have unit variance and that its covariance matrix $\mathbf{\Sigma}_\mathbf{\hat{Y}}$ is equal to the unity matrix. $\mathbf{\hat{Y}}$ is calculated as follows:
        \begin{equation}
            \mathbf{\hat{Y}} = \mathbf{W}_\mathbf{\tilde{Y}} \mathbf{\tilde{Y}} = \mathbf{E} \mathbf{D}^{-\frac{1}{2}} \mathbf{E}^T \mathbf{\tilde{Y}},
        \end{equation}
        where $\mathbf{E}$ is the orthogonal matrix of eigenvectors of the covariance matrix $\mathbf{\Sigma}_\mathbf{\tilde{Y}}$ of $\mathbf{\tilde{Y}}$ and $\mathbf{D}^{-\frac{1}{2}} = \mathrm{diag}(d_1^{-\frac{1}{2}}, ..., d_m^{-\frac{1}{2}})$ with $d_i$ being the $i$th eigenvalue of $\mathbf{\Sigma}_\mathbf{\tilde{Y}}$.
        In practise we add a fudge factor $\epsilon = 1\times10^{-7}$ in the calculation of $\mathbf{D^{-\frac{1}{2}}}$ to avoid over-amplification of eigenvectors associated with small eigenvalues
        \begin{equation}
            \mathbf{D}^{-\frac{1}{2}} = \mathrm{diag}\left(\frac{1}{\sqrt{d_1 + \epsilon}}, ..., \frac{1}{\sqrt{d_m + \epsilon}}\right).
        \end{equation}
        The scaling parameters $\mu_{x_i}$, $\sigma_{x_i}$, $\mu_{y_i}$ and $\mathbf{W}_\mathbf{\tilde{Y}}$ are calculated from our entire synthetic data set, before we perform the split in training and test set. At prediction time of the real data from Wd2 and NGC\,6397 the observational data is scaled using the \textit{same} scaling parameters derived from the synthetic data the respective models were trained on (e.g. if we train the cINN on the synthetic data set 'Wd2\_I', the real observations are scaled using the scaling parameters derived from that data set).  
    
    \subsection{Evaluating Training Success}
        \label{sec:training_evaluation}
        After training our models until the maximum likelihood loss converges, we evaluate the performance of these trained models on a held-out subset of the training data. In all our cases these randomly chosen test subsets contain 20,000 observations. On a given test set we begin verifying if the cINN has converged to a good solution by confirming that the predicted distribution $\mathbf{Z_\mathbf{test}}$ of the latent variables actually follows the multivariate normal distribution we prescribed as the target. This is easily checked by calculating the covariance matrix $\mathbf{\Sigma_\mathbf{Z_\mathrm{test}}}$ of $\mathbf{Z_\mathbf{test}}$ and determining if its close enough to the unity matrix, as well as checking that all columns follow a normal distribution with zero mean. 
        
        To ascertain the quality of the predicted posterior distributions for each of the physical parameters we compute the median \textit{calibration error} $e_\mathrm{cal}^\mathrm{median}$. For a given confidence interval $q$ the calibration error over a set of $N$ observations is defined as 
        \begin{equation}
            e_\mathrm{cal} = q_\mathrm{inliers} - q,
        \end{equation}
        where $q_\mathrm{inliers} = \frac{N_\mathrm{inliers}}{N}$ indicates the fraction of observations for which the true value falls within the $q$-confidence interval of the corresponding predicted posterior distribution. Negative values of $e_\mathrm{cal}$ indicate that the model is overconfident, predicting too narrow posterior distributions, while a positive $e_\mathrm{cal}$ describes an under-confident model that predicts posteriors that are too broad \citep{Ardizzone2019a}. We calculate $e_\mathrm{cal}^\mathrm{median}$ as the median of the absolute values of the calibration errors over a range of confidence intervals from 0.01 to 0.99 in steps of 0.01. 
        
        Apart from the calibration error we also measure the cINN model accuracy for point estimations $\hat{x}$, i.e. maximum a posteriori (MAP) estimates, of each physical parameter by computing the root mean square error (RMSE) with respect to the ground truth $x^*$ over the entire test set
        \begin{equation}
            \label{eq:RMSE}
            \mathrm{RMSE} = \sqrt{\frac{\sum_{i=1}^{N}(\hat{x}_i - x_i^*)^2}{N}}.
        \end{equation}
        In order to better compare the RMSEs of the four different models we train, we also compute a normalised RMSE (NRMSE). We derive this quantity for each physical parameter $x$ by dividing the RMSE by the range $\bar{x} = x_\mathrm{max}^\mathrm{ts} - x_\mathrm{min}^\mathrm{ts}$ covered in the training set, i.e.
        \begin{equation}
            \label{eq:NRMSE}
        	\mathrm{NRMSE} = \frac{\mathrm{RMSE}}{\bar{x}}.
        \end{equation}
        To derive these performance measures for all of the 20,000 observations in the test sets for each posterior we sample 4,096 times from the latent space $\mathbf{Z}$.

    \subsection{Determining MAP estimates}
        \label{sec:MAP_estimates_theory}
        In order to assess the point estimate accuracy (Section \ref{sec:training_evaluation}) on our test set, as well as on the predicted physical parameters for the real observations presented in Section \ref{sec:Results}, we compute maximum a posteriori (MAP) estimates. To this purpose, given a posterior distribution for a physical parameter, we first perform a kernel density estimation on the posterior using a Gaussian kernel function and then we find the parameter value at which this density estimate has a maximum. In practise we evaluate the density on a regularly spaced grid of 1,024 points ranging from the minimum to the maximum of the given posterior. To derive a suitable bandwidth $h$ for this kernel density estimation we use Silverman's rule of thumb,
        \begin{equation}
            h = 1.06 \cdot \min\left(\sigma, \frac{\mathrm{IQR}}{1.34}\right) \cdot n^{-\frac{1}{5}},
        \end{equation}
        where $\mathrm{IQR}$ denotes the interquartile range, $\sigma$ the standard deviation of the data and $n$ the number of data points \citep{Silverman1986}. We choose this bandwidth estimator for its computational efficiency in order to quickly derive MAP estimates for our test observations, keeping in mind that this estimator is prone to suggest sub-optimal bandwidths for density distributions that differ strongly from unimodal Gaussians.
        
    \subsection{Re-Simulation Error}
        To verify whether the predicted posterior distributions are correct and not just cINN artefacts one usually performs a re-simulation. Here either the MAP estimates of the physical parameters or individual samples of the predicted posteriors are put into the simulation, that maps the physical to the observable space, to derive the associated re-simulated observables $\mathbf{y}_\mathrm{i,resim}$. They are then compared with the cINN input condition $\mathbf{y}_\mathrm{i,input}$ of the given star. Using the MAP estimates one can compute a MAP re-simulation error over the test set following
        \begin{equation}
            \mathrm{RMSE_{resim}^{MAP}} = \sqrt{\frac{\sum_{i=1}^{N}(\mathbf{y}_\mathrm{i,resim} - \mathbf{y}_\mathrm{i,input})^2}{N}}.
        \end{equation}
        Unfortunately we do not have direct access to the stellar evolution code that our training data is based on, just the publicly available isochrone tables. Therefore, we cannot perform a full re-simulation for our predictions. 
        
        To still get an idea of the re-simulation error of our approach we adopt a simple approximation instead. For a given MAP estimate or sample prediction of the physical parameters we do a nearest neighbour search in the $\mathbf{x}$ + $A_\mathrm{V}$ space on the training data (\textit{after} the test split). Even though we do not predict the extinction, we have to include it in this nearest neighbour search to select the correct copy of the data point closest to our query. We note that also in a full re-simulation we would have to input extinction to correctly retrieve the magnitudes. This approach allows us to report the approximate MAP re-simulation error $\mathrm{RMSE_{resim}^{MAP}}$ on our synthetic training data (see Table \ref{tab:performance_overview}). 
        
        It is important to keep in mind though that this is only an \textit{approximation}, so that in cases where the distance to the nearest training data point is large in this 7 dimensional parameter space, the associated magnitudes might not necessarily represent the true re-simulated observables of a given prediction. It is therefore likely that this approximation tends to overestimate the re-simulation error.

\section{Training Results}
    \label{sec:Results}
    
    \begin{table*}
        \centering
        \caption{Overview of the performance on a test set of 20,000 cases for the four cINN models we have trained. Reported are the calibration error, median uncertainty at 68\% confidence (width of the 68\% confidence interval), the standard (RMSE) and normalised root mean squared error (NRMSE) of the MAP estimates, as well as the total MAP re-simulation error $\mathrm{RMSE_{resim}^{MAP}}$ from our nearest neighbour approximation.}
        \begin{tabular}{lcccc}
             \hline
             \hline
              & \multicolumn{4}{c}{Training Set} \\
             Performance Measure & Wd2\_I & Wd2\_II & NGC6397\_I & NGC6397\_II \\
             \hline
             Calibration Error & & & & \\
             \hspace{10pt} $\mathrm{\log age}$      & 0.005 & 0.001 & 0.005 & 0.011  \\
             \hspace{10pt} $\mathrm{\log M_{ini}}$  & 0.009 & 0.006 & 0.006 & 0.004  \\
             \hspace{10pt} $\mathrm{\log M_{curr}}$ & 0.009 & 0.004 & 0.007 & 0.007  \\
             \hspace{10pt} $\mathrm{\log L}$        & 0.068 & 0.048 & 0.003 & 0.007  \\
             \hspace{10pt} $\mathrm{\log T_{eff}}$  & 0.028 & 0.020 & 0.007 & 0.003  \\
             \hspace{10pt} $\mathrm{\log g}$        & 0.013 & 0.003 & 0.006 & 0.007  \\
             \hline
             Median Uncertainty at 68\% confidence & & & & \\
             \hspace{10pt} $\mathrm{\log age}$      & 0.199 & 0.049 & 0.065 & 0.120  \\
             \hspace{10pt} $\mathrm{\log M_{ini}}$  & 0.004 & 0.002 & 0.004 & 0.001  \\
             \hspace{10pt} $\mathrm{\log M_{curr}}$ & 0.004 & 0.003 & 0.004 & 0.002  \\
             \hspace{10pt} $\mathrm{\log L}$        & 0.002 & 0.002 & 0.005 & 0.001  \\
             \hspace{10pt} $\mathrm{\log T_{eff}}$  & 0.001 & 0.001 & 0.001 & 0.001  \\
             \hspace{10pt} $\mathrm{\log g}$        & 0.006 & 0.004 & 0.004 & 0.002  \\
             \hline
             RMSE & & & & \\
             \hspace{10pt} $\mathrm{\log age}$      & 0.572 & 0.379 & 0.481 & 0.1659 \\
             \hspace{10pt} $\mathrm{\log M_{ini}}$  & 0.065 & 0.120 & 0.018 & 0.0036 \\
             \hspace{10pt} $\mathrm{\log M_{curr}}$ & 0.064 & 0.074 & 0.019 & 0.0036 \\
             \hspace{10pt} $\mathrm{\log L}$        & 0.093 & 0.154 & 0.008 & 0.0011 \\
             \hspace{10pt} $\mathrm{\log T_{eff}}$  & 0.041 & 0.071 & 0.003 & 0.0002 \\
             \hspace{10pt} $\mathrm{\log g}$        & 0.131 & 0.200 & 0.021 & 0.0034 \\
             \hline
             NRMSE & & & & \\
             \hspace{10pt} $\mathrm{\log age}$      & 0.1122 & 0.1263 & 0.0938 & 0.1468 \\
             \hspace{10pt} $\mathrm{\log M_{ini}}$  & 0.0180 & 0.0334 & 0.0050 & 0.0028 \\
             \hspace{10pt} $\mathrm{\log M_{curr}}$ & 0.0179 & 0.0207 & 0.0053 & 0.0028 \\
             \hspace{10pt} $\mathrm{\log L}$        & 0.0091 & 0.0160 & 0.0008 & 0.0002 \\
             \hspace{10pt} $\mathrm{\log T_{eff}}$  & 0.0207 & 0.0366 & 0.0023 & 0.0003 \\
             \hspace{10pt} $\mathrm{\log g}$        & 0.0191 & 0.0291 & 0.0038 & 0.0007 \\
             \hline
             $\mathrm{RMSE_{resim}^{MAP}}$ & 0.071 & 0.123 & 0.078 & 0.043 \\
             \hline
             \hline
        \end{tabular}
        \label{tab:performance_overview}
    \end{table*}

    For all four of our models the cINN training process converges quickly, the training time being usually within 1 to 2 hours when making use of GPU acceleration with a NVIDIA GTX 1080 graphics card. Once trained the prediction of posterior distributions is very rapid. For the 20,000 observations in our test sets generating the posterior distributions, sampling each 4,096 times, takes in total about 10 minutes, averaging around 35 predicted posterior distributions per second. This makes the cINN approach a very time efficient predictor.

    \subsection{Performance Overview}
        Across all four cINN models we were able to achieve well converged model solutions. Both the covariance of the latent variables, as well as their distributions, evaluated on the respective test sets, reach their targets of unity and standard normal distribution, respectively. Figure \ref{fig:Wd2_I_Z_covariance} in the Appendix shows an example of the achieved covariance matrix and latent variable distributions for the 'Wd2\_I' cINN model.
        Table \ref{tab:performance_overview} gives an overview of our remaining performance measures, namely the median calibration error, the median uncertainty at 68\% confidence, the RMSE and NRMSE of the MAP point estimate (see Equations \ref{eq:RMSE} and \ref{eq:NRMSE}), as well as our approximation of the total re-simulation error across all four trained models. 
        
        In terms of the median calibration error we find that all four models reach calibrated solutions for their predicted posterior distributions, as the largest error across all parameters and models is only about 6.8\%. Given the similar magnitude of the errors for all four models, there is no clear influence of the training set size or feature abundance on the cINN's ability to converge to a well calibrated solution. In particular, there is no significant difference between the models trained on the full training sets 'Wd2\_I' and 'NGC6397\_I' vs. their counterparts 'Wd2\_II' and 'NGC6397\_II'. As the latter include prior knowledge about the age of the clusters, they should theoretically allow for more accurate solutions of the regression problems (i.e. less degenerate mappings). The only notable difference between the Wd2 and NGC\,6397 models in terms of the median calibration error is that we find slightly better calibrated solutions for the luminosity and effective temperature prediction for the NGC\,6397 models.
        
        Concerning the median uncertainty at the 68\% confidence level, an indicator of the average width of the predicted posterior distributions, we find that all four trained cINN models can constrain all physical parameters, except for the age, remarkably well with uncertainties on the order of only a few 0.001 dex on average. Again, the availability of more features or the prospect of less degenerate mappings by including prior knowledge does not significantly improve the result. 
        
        Judging by the uncertainty values, the stellar age appears to be the most difficult parameter to constrain. Of the six parameters, age is also the only one where the prediction is influenced by the amount of available features. The 'NGC6397\_I' cINN model constrains the age to distributions that are about 0.1\,dex narrower than the similar one trained on 'Wd2\_I', despite the fact that both training sets cover basically the same physical parameter space (albeit at different metallicity). For the age prediction we also observe a difference between the 'Wd2\_I' and 'Wd2\_II' model, as the cINN trained on the data set including prior knowledge returns narrower age posterior distributions. The lower uncertainty is likely influenced by the overall smaller range in possible predicted ages, but could also be a result of the missing degeneracies in 'Wd2\_II'. Interestingly, we do not observe the same effect between 'NGC6397\_I' and 'NGC6397\_II', where in fact the median uncertainty increases for the model trained on the much narrower age range. This could indicate that constraining the age distribution for these old stars (above 1\,Gyr) may not facilitate the regression problem, while the reverse may be true for the young stars. 
        
       The point estimate accuracy, as measured by the RMSE between the MAP prediction and the true values, confirms that age is the most difficult parameter to predict for all our models. With RMSEs of a few 0.01 to 0.1 dex, the cINN predicts the remaining five physical parameters very well, while the RMSE for the age prediction, on the order of 0.5\,dex, is about a magnitude larger.  For comparison, a predictor that returns a random value drawn from a uniform distribution within the age range of 'Wd2\_I' achieves a RMSE of about 2.1 (NRMSE of 0.41). The $\sim0.1$\,dex differences in the RMSEs between the models trained on 'Wd2\_I' and 'NGC6397\_I' suggest that an increased feature abundance (i.e. number of observables) improves the point estimation accuracy of the model. Interestingly, while the 'Wd2\_II' model decreases the age RMSE by about 0.2 dex, the error of the point estimate for all remaining physical parameters increases. Comparing the NRMSEs between 'Wd2\_I' and 'Wd2\_II', however, we find that both models perform evenly well and all flat RMSE differences are likely effects of the different parameter ranges. We find a similar behaviour between 'NGC6397\_I' and 'NGC6397\_II' for all parameters except the age again, where the 'NGC6397\_II' model actually performs the worst across all models. As previously indicated by the uncertainty, this supports the finding that the age prediction within the range from 1\, to 13\, Gyr is the most difficult task on the synthetic data.
       
       Finally, for our approximation of the total MAP re-simulation error we find excellent results for all of our models, with values on the order of only 0.1\,mag and below. Considering that our approximation likely overestimates this error because we have to rely on the observables of a nearest neighbour proxy, errors this small are more than satisfactory. The corresponding comparisons of the 're-simulated' and observed magnitudes show almost perfect 1-to-1 correlations with very few outliers in both the MAP and entire posterior re-simulations. Therefore, we are very confident that, even though we could not perform a true re-simulation, our predicted posterior distributions are true and not just numerical artefacts. Importantly this also indicates that the overall broader age posteriors are generally not caused by an underperforming cINN but rather due to actual intrinsic degeneracies in the age prediction, correctly captured by the cINN.
       
        \begin{figure*}
            \centering
            \includegraphics[width = 0.2625\linewidth]{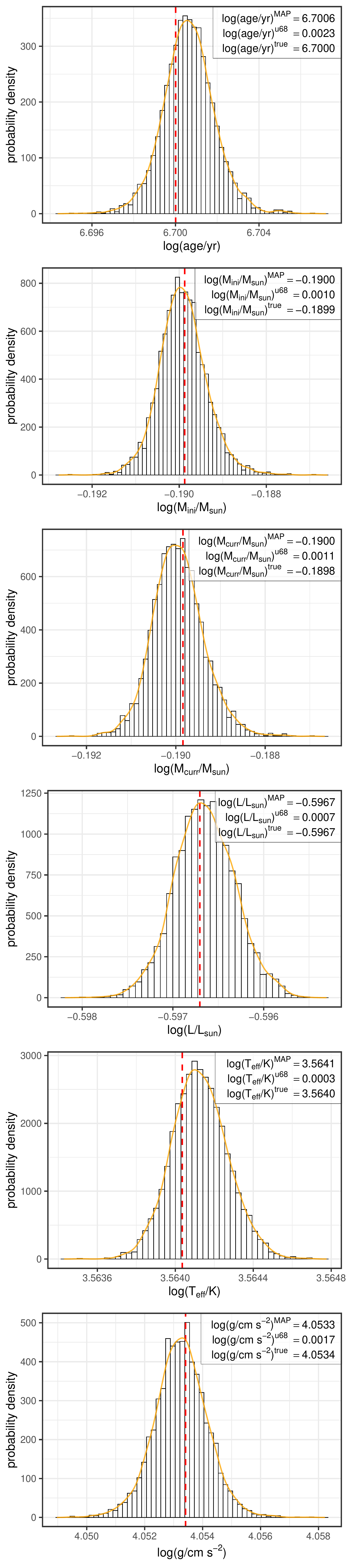}
            \includegraphics[width = 0.2625\linewidth]{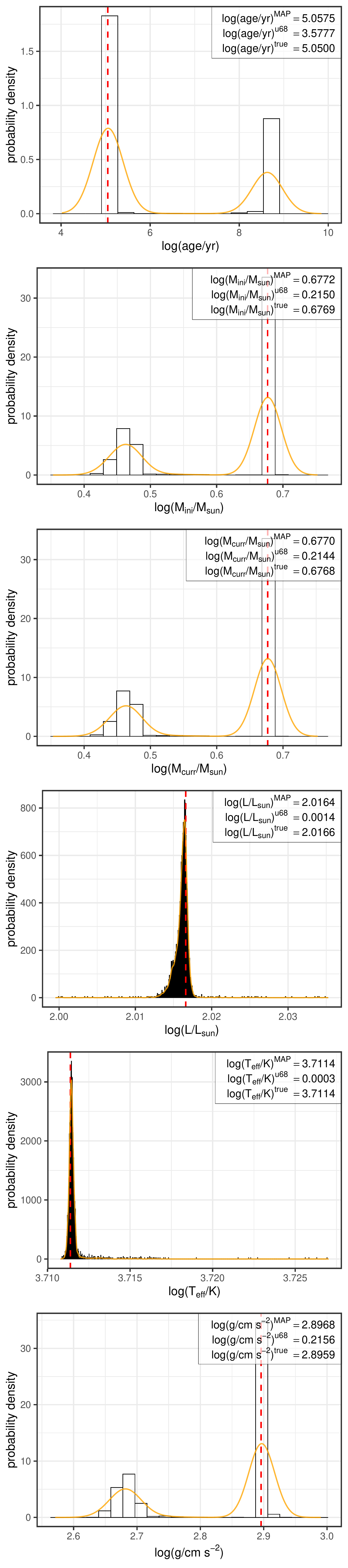}
            \includegraphics[width = 0.2625\linewidth]{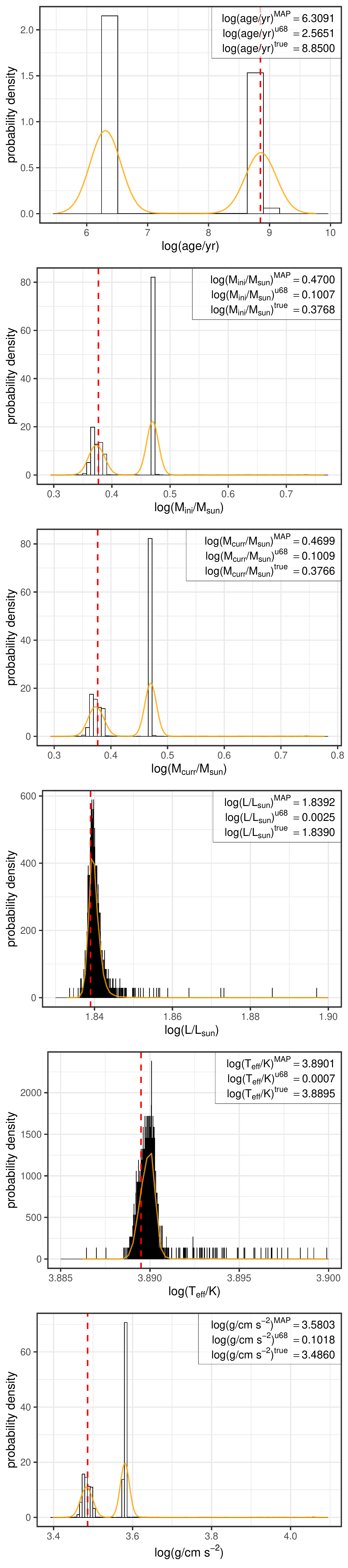}
            \caption{Predicted posterior distributions for three test examples (column-wise) as predicted by the cINN trained on 'Wd2\_I'. The red dotted line in each histogram indicates the known true value for the given test observation. The orange line represents the kernel density estimate of the predicted distribution used to locate the MAP solution. The \textbf{left column} shows an example case where the cINN is able to constrain the physical parameters of this observation extremely well. The remaining two columns show degenerate examples where the predicted posterior distributions of some parameters (e.g. age and mass) show multi-modalities as a consequence. The \textbf{middle column} test observation shows an example case where the MAP of the predicted bi-modal distribution coincides with the true value, while in the \textbf{right column} case the true value falls onto the second peak of the distribution. Note the different scaling in each column.}
            \label{fig:Wd2_I_TestPosteriorExamples}
        \end{figure*}
        
        \begin{figure*}
            \centering
            \includegraphics[width = \linewidth]{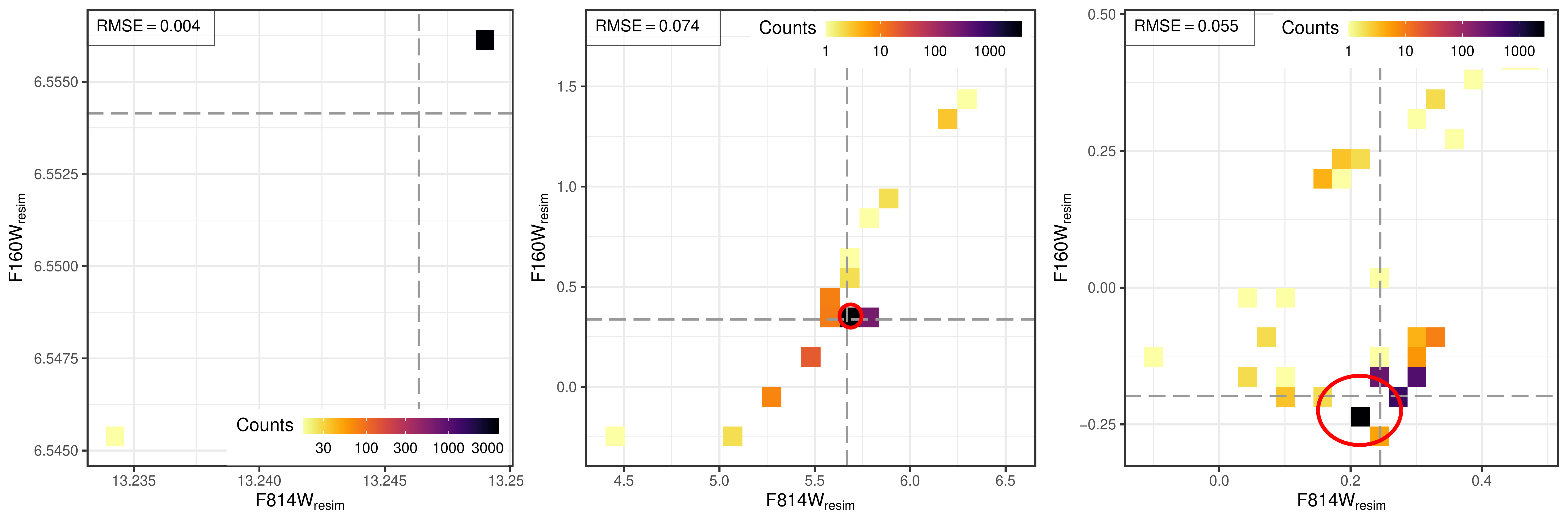}
            \caption{2D histograms of the 're-simulated' magnitudes for the example posteriors in Figure \ref{fig:Wd2_I_TestPosteriorExamples} (columns match accordingly). The grey dashed lines indicate the observed magnitudes, while the red circles in the middle and right panel indicate the area in which 60\% of the samples are located that have the lowest distances to the nearest neighbour chosen as the re-simulation proxy. Again, it should be mentioned that the axis scaling is very different in each plot.}
            \label{fig:Wd2_I_TestPosteriorExamplesResim}
        \end{figure*}
        
    \subsection{Wd2\_I and Wd2\_II}
        \begin{figure*}
            \centering
            \includegraphics[width = 0.85\linewidth]{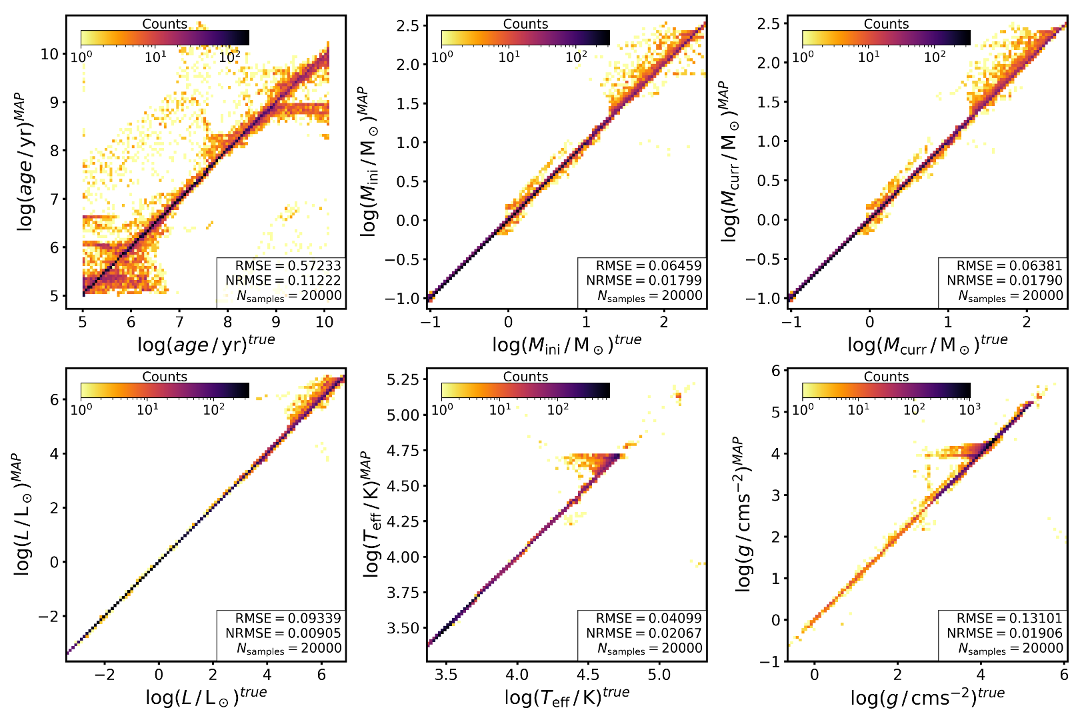}
            \caption{2D histograms of the MAP estimates plotted against the true values for the six physical parameters we predict with the cINN trained on 'Wd2\_I' for 20,000 cases from our test set. From top left to bottom right we show age, $M_\mathrm{ini}$, $M_\mathrm{curr}$, $L$, $T_\mathrm{eff}$ and $g$. }
            \label{fig:Wd2_I_MAP_vs_true}
        \end{figure*}
        
        \begin{figure*}
            \centering
            \includegraphics[width = 0.85\linewidth]{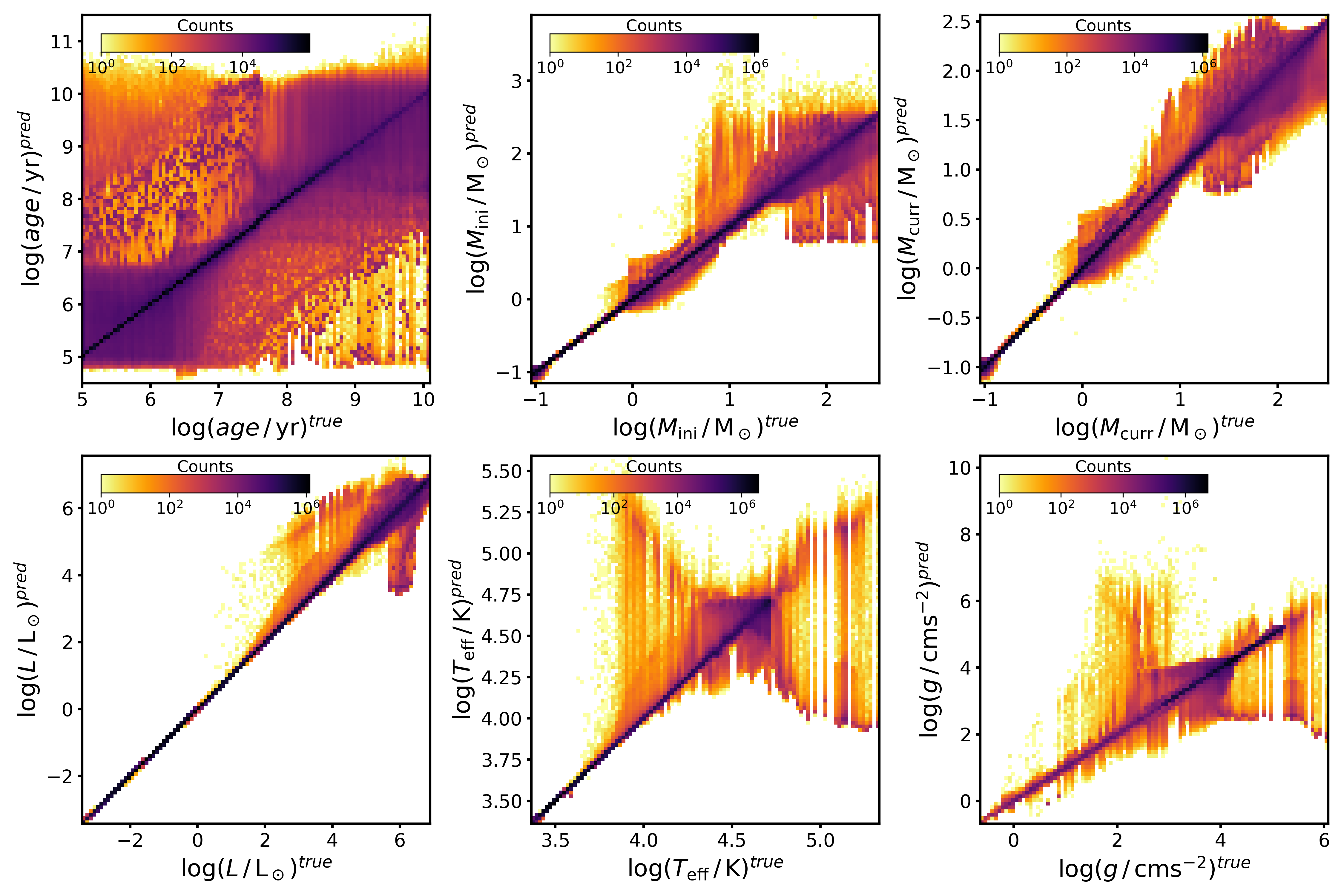}
            \caption{2D histograms of the entire predicted posteriors plotted against the true value of the six physical parameters of the cINN trained on 'Wd2\_I' for 20,000 cases from the test set. From top left to bottom right age, $M_\mathrm{ini}$, $M_\mathrm{curr}$, $L$, $T_\mathrm{eff}$ and $g$ are shown.}
            \label{fig:Wd2_I_Posteriors_vs_true}
        \end{figure*}
        
        \begin{figure*}
            \centering
            \includegraphics[width = \linewidth]{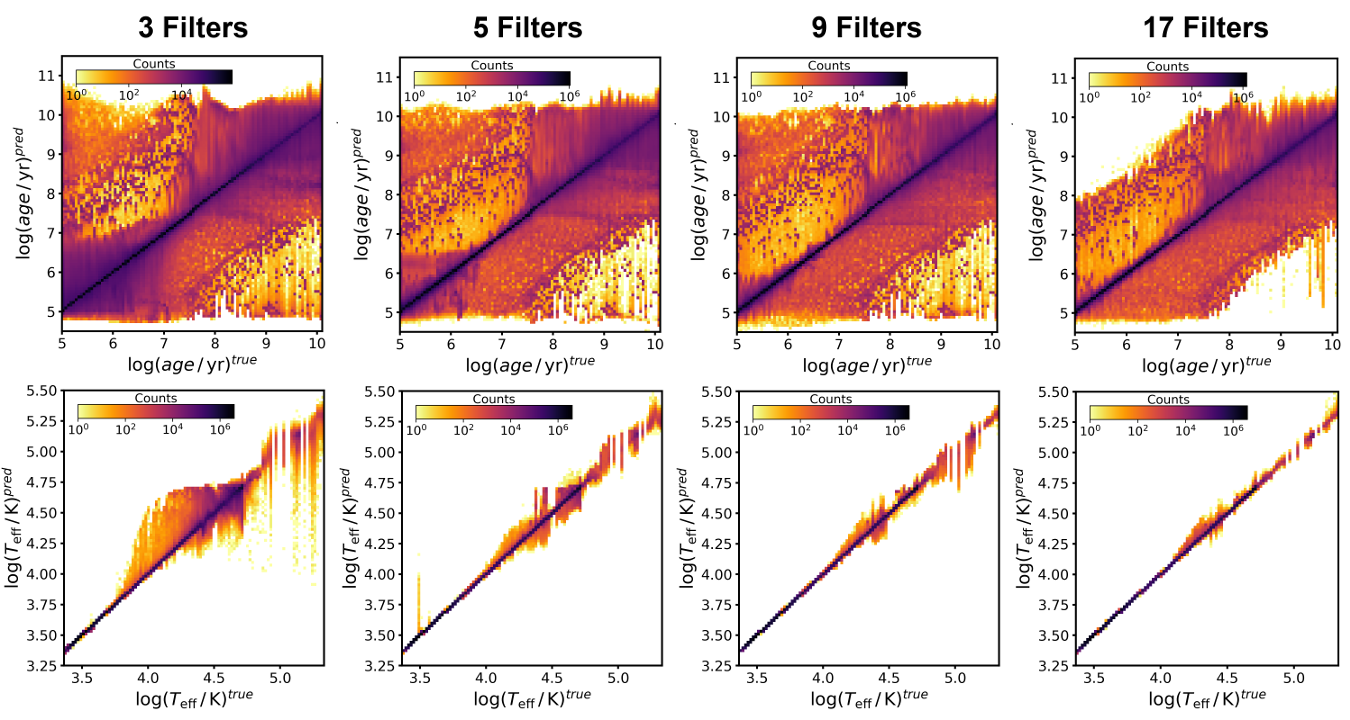}
            \caption{2D histograms of the predicted posterior distributions plotted against the true value for age (\textbf{top}) and $T_\mathrm{eff}$ (\textbf{bottom}) for 20,000 test observations as predicted by cINN models trained on 'Wd2\_I' with increased numbers of photometric filters compared to our standard 'Wd2\_I' setup. The 3 filter case (\textbf{first column}) entails the HST filters F555W, F814W, F160W, the 5 filter one (\textbf{second column}) adds F275W and F336W to that, and the 9 filter setup (\textbf{third column}) further includes F438W, F606W, F775W and F110W on top of the previous. The final 17 filter case (\textbf{fourth column}) entails all previous filters in addition to F218W, F225W, F390W, F475W, F625W, F105W, F125W and F140W. Compared to our standard 'Wd2\_I' with only 2 filters in Figure \ref{fig:Wd2_I_MAP_vs_true} this sequence highlights how the increased feature abundance improves the predictive capability of the cINN as the 'arrow'-like artifacts in $T_\mathrm{eff}$ disappear and the age posterior distributions decrease in width, especially so for the young test observations.}
            \label{fig:Wd2_I_PostVsTrue_multifilter}
        \end{figure*}
        
        As indicated by the summary statistics in Table \ref{tab:performance_overview} the 'Wd2\_I' and 'Wd2\_II' cINN models perform very well. We look at this in more detail. Figure \ref{fig:Wd2_I_TestPosteriorExamples} shows example posterior distributions for all six physical parameters for three held-out test observations predicted by the 'Wd2\_I' model. This plot exhibits some of the typical posterior distributions that the cINN returns on the synthetic data in this regression problem. 
        
        The first case, shown in the left column, is an example where the cINN constrains all physical parameters of the star extremely well with very narrow posterior distributions centred around the known true value. As the low median uncertainties at 68\% confidence for all parameters except age already suggest, this kind of prediction is among the most common results for the synthetic test set. The left panel in Figure \ref{fig:Wd2_I_TestPosteriorExamplesResim} presents the approximate re-simulation for the full posterior of this example. Evidently we match the input observation almost exactly (note the small axis range and error), confirming the validity of the predicted posterior. The observed deviation is a direct result of the nearest neighbour approximation and the discreteness of the training set. The latter is also the reason why the two re-simulation solutions appear so 'far' apart, as there simply are no models in between. The samples with a greater discrepancy to the true observation (bottom left corner) have a larger distance to the nearest neighbour than the others. The re-simulation approximation is therefore less precise for these samples as the distance is a direct measure of similarity between the nearest neighbour proxy and the given query samples.
        
        In contrast, the middle column exhibits an example of the kind of degeneracy that we frequently find in this regression problem, with bi-modal solutions within the predicted posterior distributions. The age and mass distributions indicate that this observation could be explained by a $\sim2.9$ $\mathrm{M_\odot}$ star that is $\sim425$ Myr old, so likely well within its post-main-sequence phase. Or it could be a very young ($\sim0.1$\,Myr) more massive $\sim4.75\,\mathrm{M_\odot}$ pre-main-sequence star. Due to the overlap of the post-main-sequence and pre-main-sequence evolution in observable space, especially so in the presence of extinction, this is one of the major degeneracies that make the prediction of stellar physical parameters from photometry such a difficult regression problem. In this example the cINN prediction reveals that this degeneracy is not broken with only two passbands, but also finds that the young $4.75\,\mathrm{M_\odot}$ star is the most likely solution as indicated by the MAP estimates. Therefore, the cINN successfully recovers the true solution for this synthetic star. 
        
        The middle panel of Figure \ref{fig:Wd2_I_TestPosteriorExamplesResim} shows the approximate re-simulated magnitudes for this example posterior in comparison with the true input observations. Overall we find very good agreement, except for a few outlier cases. The red circle indicates the area populated by the 60\% of the samples (containing instances from both peaks) with the lowest distance to the nearest neighbour used as a re-simulation proxy. This set matches the true observations almost perfectly. All of the outliers exhibit larger nearest neighbour distances (especially the far outliers). Consequently our re-simulation approximation is less precise for these objects, which likely explains the offset from the true observation. Therefore this diagram confirms the validity of the predicted posterior and the identified degeneracy.  
        
        The final example in the right column shows another degenerate case that could be explained as either a younger $\sim3\,\mathrm{M_\odot}$ or a much older $\sim2.4\,\mathrm{M_\odot}$ star. Here the most likely explanation of the observation as given by our MAP estimate is in fact not the true one, which falls into the secondary peak. 
        This result may seem unsatisfying at first glance, but a true posterior distribution describes all \textit{possible} physical parameters that can explain the given observation. That means that the most likely combination does not necessarily have to be the one that generated the observation. In fact these two degenerate examples show the great strength of the cINN approach for this type of degenerate regression problem, as even in the second case the true solution is part of the posterior distribution as the second most likely result. The interpretation of cases like these, as always, benefits from additional astrophysical constraints. 
        
        The right panel of Figure \ref{fig:Wd2_I_TestPosteriorExamplesResim} provides the re-simulation approximation for this example. Again we find a good match with the observation for the objects for which our approximation is the most precise. Only objects with large distance to the nearest neighbour, so less precise re-simulation approximation, deviate more significantly. 
        
        To assess the limitations of the method tested on the synthetic data we compare the predicted point estimates with the true values (as previously summarised by the RMSEs in Table \ref{tab:performance_overview}) for the 20,000 test observations in Figure \ref{fig:Wd2_I_MAP_vs_true}. The figure highlights how well the cINN predicts $M_\mathrm{ini}$, $M_\mathrm{curr}$, $L$, $T_\mathrm{eff}$ and $g$ as only very few predictions (note the logarithmic colour scale) fall off a perfect 1-to-1 correlation between the predicted and true values. However, for $T_\mathrm{eff}$ and $g$ we observe some structure (around $\log(T_\mathrm{eff}/\mathrm{K}) \approx 4.75$ and $\log(g/\mathrm{cm\,s^{-2}}) \approx 4$, respectively) that seems systematic in nature. For the effective temperature there is also a deviation from the 1-to-1 correlation for $\log(T_\mathrm{eff}/\mathrm{K}) > 4.75$. 
        
        We find the largest scatter in the age prediction, confirming that this parameter is the most difficult to predict for the cINN. 
        It has the most trouble with predicting ages for the very young ($\log(\mathrm{age/yr}) < 6.5$) and the oldest ($\log(\mathrm{age/yr} > 8.5)$ objects in the test set, as we find the most deviations from the perfect correlation here. Still, even in this regime there is a majority of good predictions (note again the logarithmic colour scale). 
        
        The difficulty in predicting the correct age becomes further apparent when visualising the posterior distributions in relation to the true values, as in Figure \ref{fig:Wd2_I_Posteriors_vs_true}. Here we plot the spread of the posterior distribution of every physical parameter against the true value for all 20,000 test observations. Again we observe that for all physical parameters except age the cINN provides well constrained posterior distributions that are in many cases quite narrow and symmetric around the true value. Similar to the systematic structures in the MAP estimates we find 'arrow'-like 'artefacts' for $T_\mathrm{eff}$ and $g$ here. For $\log(T_\mathrm{eff}/\mathrm{K}) > 4.75$ we also discover two 'branches', indicating a strong bi-modal degeneracy in this range that explains the deviation from the 1-to-1 correlation in Figure \ref{fig:Wd2_I_MAP_vs_true}, as the MAP estimates seemingly tend to fall into this lower branch.
        
        The age posterior distributions appear to be much wider, although the visual effect is amplified in Figure \ref{fig:Wd2_I_Posteriors_vs_true} by the logarithmic colour scaling, chosen to better visualise outliers. Most of the predicted posterior distributions are also well centred on the true value, but nevertheless we find many more wide outliers here, indicating ample degeneracy. Analogous to the MAP estimates, the posterior distributions narrow down within the intermediate age range and widen for the youngest and the oldest stars, also exhibiting the multi-modalities previously highlighted in Figure \ref{fig:Wd2_I_TestPosteriorExamples}. Despite the slightly discouraging look of the age posteriors it is important to note that in 99.8\% of the cases the true value \textit{is part} of the predicted posterior distribution. 
        
        To evaluate whether the ``arrow'' artefacts observed in the MAP estimates and posteriors of $T_\mathrm{eff}$ and $g$ are a cINN model specific issue, we re-trained the cINN model on modified versions of training set 'Wd2\_I', where we increase the number of observables with additional photometric filters. Within the synthetic datasets these additional filters are readily available.
        Figure \ref{fig:Wd2_I_PostVsTrue_multifilter} shows the results of this experiment. It provides the posterior against true value diagrams for age and surface temperature for different numbers of additional photometric filters. This sequence shows that the 'arrow'-like structures in $T_\mathrm{eff}$ and $g$, as well as the second branch in $T_\mathrm{eff}$, are in fact a result of the limited number of photometric filters in our study, as the effect already decreases when the F555W filter is added and basically disappears when we use 9 photometric filters. Not surprisingly, the predictions also improve as more observational information is gained, the posterior distributions narrowing down noticeably. Especially interesting for the age prediction is that we already observe a considerate improvement with 5 filters (F275W, F336W and F555W on top of F814W and F160W). Specifically the spread for very young objects ($\log(\mathrm{age/yr}) < 6.5)$ decreases significantly. We observe the same improvement in the point estimates (see Appendix Figure \ref{fig:Wd2_I_MAPvsTrue_Age_multifilter}). Still, even with the 'ideal' information of the full complement of 17 photometric filters of the 'HST WFC3 wide' photometric system used by the PARSEC isochrones, the prediction for old stars is not perfect. The age prediction of old stars thus remains the most challenging task within this regression problem (see also the discussion in Sections \ref{sec:Test_NGC6397} and \ref{sec:NGC6397_prediction} ). In any case, based on this performance analysis, we \textit{recommend} to use at least five photometric filters in addition to extinction if they are available. 
        
        The model trained on 'Wd2\_II' does not show significant difference with respect to the 'Wd2\_I' cINN model within their range of overlap. The corresponding diagrams of the point estimate and posteriors against the true values for 'Wd2\_II', as well as a more detailed discussion can be found in Appendix \ref{app:Wd2_II}, Figures \ref{fig:Wd2_II_MAP_vs_true} and \ref{fig:Wd2_II_Posteriors_vs_true} respectively. 
    	
    \subsection{NGC6397\_I and NGC6397\_II}
        \label{sec:Test_NGC6397}
        Overall the training results of model 'NGC6397\_I' match those of 'Wd2\_I', except for the previously described slight improvements in accuracy. Judging by our performance experiments in dependence of filter coverage carried out on 'Wd2\_I', these improvements are likely caused by the larger number of photometric filters, five instead of the two used for 'Wd2\_I' (see Appendices \ref{app:Wd2_I} and \ref{app:NGC6397_I}).
        
        In general, of all trained models 'NGC6397\_II' provides the smallest RMSEs across all predicted physical parameters and lowest median uncertainty for all parameters but age. Given how well the 'Wd2\_I' and 'NGC6397\_I' models already constrain the posterior distributions for all parameters (except age), this extra performance gain can be attributed to the more limited physical parameter space. The NRMSE of this cINN model confirms again that the age prediction for very old stars (1 Gyr and above) is the most difficult part of this regression problem. We find that the age posterior distributions tend to be quite broad and that the cINN has a tendency to extrapolate with predicted posterior distributions ranging from $\log(\mathrm{age/yr}) = 8$ to above $\log(\mathrm{age/yr}) = 11$ (see Figure \ref{fig:NGC6397_II_Posteriors_vs_true} in the Appendix), outside the boundaries of the training set range of 9 to 10.13. This extrapolatory behaviour within the 1 to 10 Gyr range appears in the 'Wd2\_I' and 'NGC6397\_I' models as well, but to a lesser degree. From the age MAP estimate against true plot of the 'NGC6397\_II' model, we also find that, while most predictions fall on the ideal 1-to-1 correlation, there is a faint trace of an almost flat 'branch' at $\log(\mathrm{Age/yr}) \approx 9.6$ (see Figure \ref{fig:NGC6397_II_MAP_vs_true} in the Appendix). This might suggest that the cINN has a slight tendency to predict something akin to a mean age value (9.6 is exactly the average) over the trained range when it encounters a star with uncertain age. 
    
    \subsection{On the age prediction of main-sequence stars}
        \begin{figure*}
            \centering
            \includegraphics[width = 0.85\linewidth]{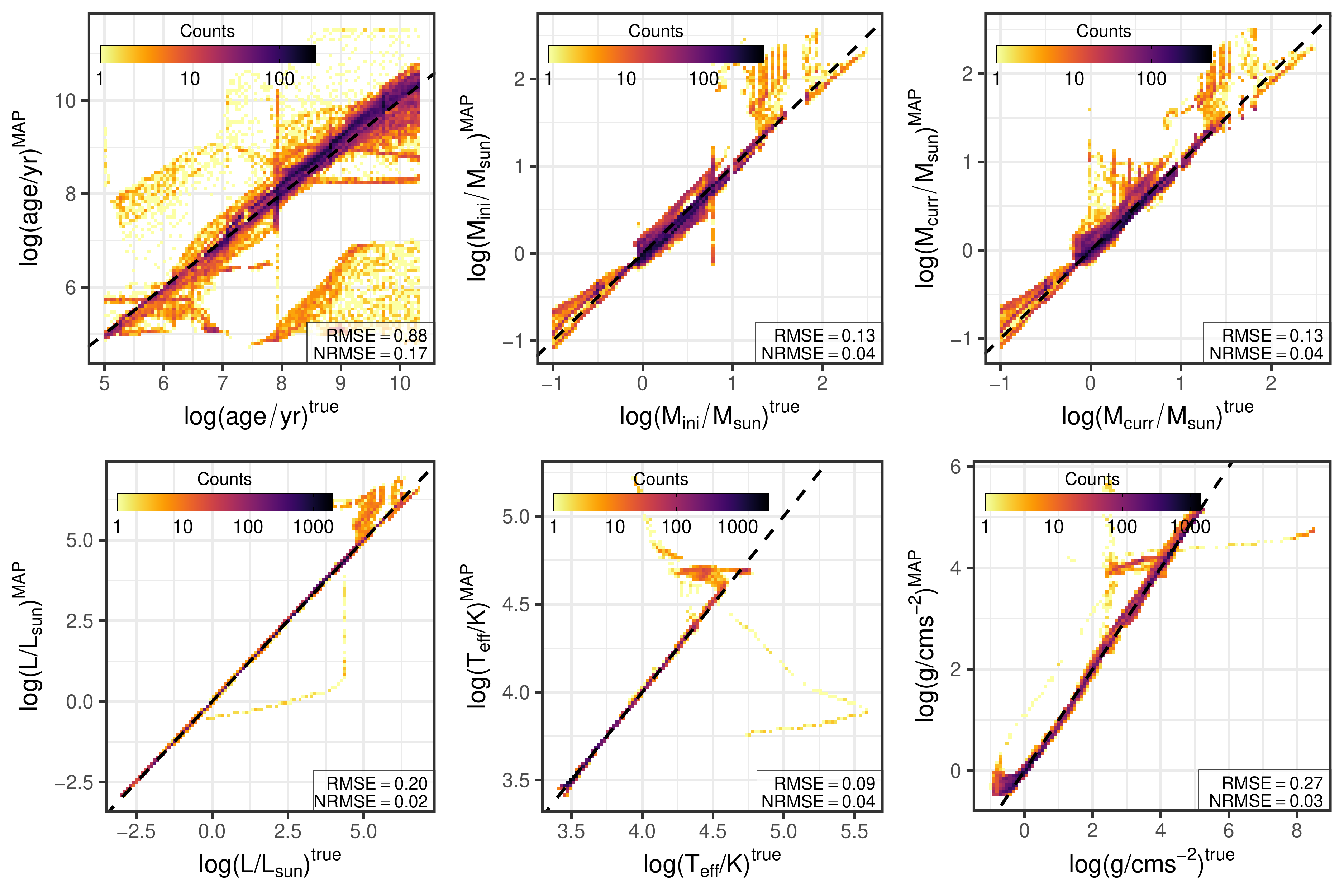}
            \caption{2D histograms of the MAP predictions for the six physical parameters on 40,000 samples from the MIST isochrone tables as predicted by the 'Wd2\_I' cINN model. Note that the NRMSEs are normalised to the parameter ranges of the MIST ground truth here instead of the ranges in the 'Wd2\_I' training set.}
            \label{fig:Wd2_I_MIST_MAP_vs_True}
        \end{figure*}
        
        \begin{figure*}
            \centering
            \includegraphics[width = 0.8\linewidth]{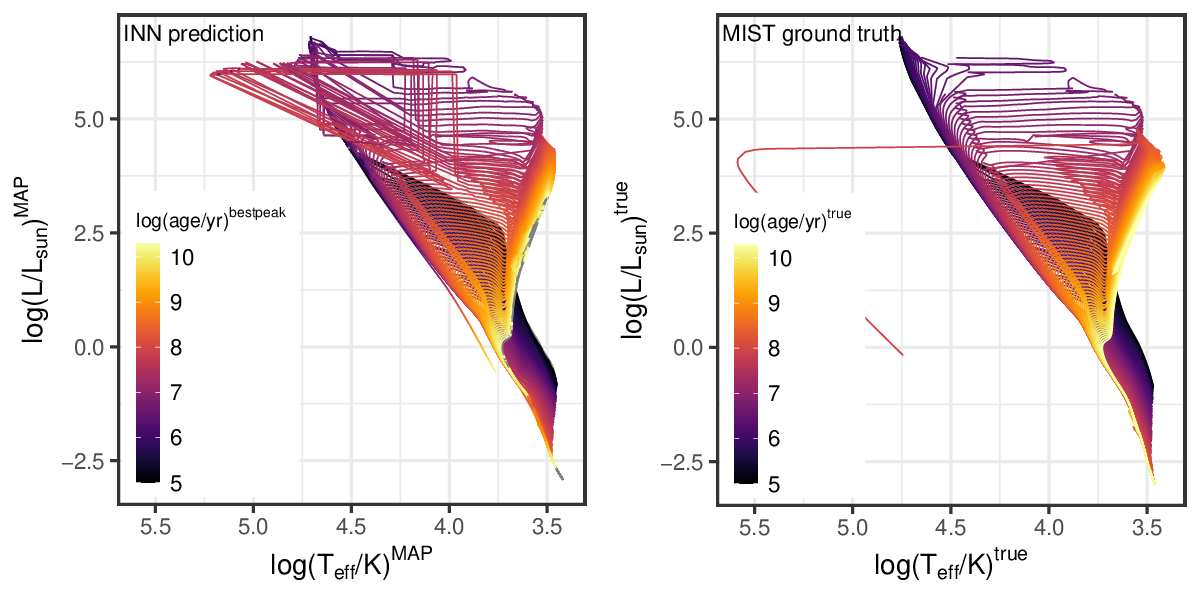}
            \caption{\textbf{Left:} cINN prediction for the HRD of the MIST isochrones. Note that MAP estimates are used for luminosity and effective temperature here, but the colour code that indicates the predicted age does not correspond to the MAP age prediction but rather the best fitting peak of the predicted age posterior. The latter is done to take multi-modal age posteriors into account. \textbf{Right:} Ground truth HRD of the MIST isochrones, colour coded according to their age.}
            \label{fig:Wd2_I_MIST_HRD}
        \end{figure*}
    
        \label{sec:synth_ms_age_pred}
        One matter we have not discussed in detail so far is the age prediction for main-sequence objects. With traditional isochrone fitting methods recovering the age of a main-sequence star from photometry alone is a notoriously difficult, if not impossible task. Our approach, on the other hand, successfully predicts ages across the entire spectrum of objects, including synthetic main-sequence stars. Given the difficulties traditional approaches have, this could be an indication that our cINN models achieve this task only by overfitting the synthetic training data. To ascertain whether this is the case we perform a test prediction with the 'Wd2\_I' and 'NGC6397\_I' models on synthetic data generated from different stellar evolution models, namely the MIST \citep{Dotter2016_MIST, Choi2016_MIST, Paxton2011_MESA, Paxton2013_MESA, Paxton2015_MESA} and Dartmouth \citep{Dotter2008_Dartmouth, Dotter2007_Dartmouth} isochrone tables. These models also provide synthetic photometry, but treat the underlying physics slightly different than PARSEC. Note that the Dartmouth isochrones only cover an age range of 1 to 15 Gyr, while the MIST tables are available over a similar $\log(\mathrm{age/yr})$ span of 5 to 10.3 as the PARSEC models. For the test we choose data sets matching the corresponding metallicities for 'Wd2\_I' and 'NGC6397\_I', and for simplicity only treat the zero extinction case. Additionally, for the MIST data we remove the post-AGB phase as our selection of PARSEC models (version 1.2s) does not include it.
        
        In the solar metallicity 'Wd2\_I' case we retrieve overall excellent results (see Figure \ref{fig:Wd2_I_MIST_MAP_vs_True}). In particular, for the MIST data the cINN recovers $\log(L)$, $\log(T_\mathrm{eff})$ and $\log(g)$ almost perfectly, except for a few instances of massive post-main-sequence stars. $M_\mathrm{ini}$ and $M_\mathrm{curr}$ are also recovered well, but exhibit more scatter than in our PARSEC test case. Lastly, while the age prediction also exhibits more of a spread around a perfect 1-to-1 correlation, with a median absolution deviation of only 0.2 dex the cINN correctly retrieves ages for most samples, \textit{including} main-sequence objects. Figure \ref{fig:Wd2_I_MIST_HRD} shows the predicted HRD in comparison to the MIST ground truth, highlighting the excellent performance of the cINN. Note that the predicted ages are represented by the best fitting peak in the age distribution here in order to account for multi-modal distributions found for, i.e., post-main-sequence objects. We find a similar success for the Dartmouth models, recovering luminosity, temperature and gravity near flawlessly (see Figure \ref{fig:Wd2_I_DM_MAP_vs_True} in the Appendix). The initial mass predictions are overall also fairly accurate, but exhibit a slight systematic over-prediction below $0.5\,M_\odot$, likely an effect of the known model discrepancy between Dartmouth and PARSEC in the sub-solar mass regime. The age prediction on the other hand is slightly less successful here. While we can recover ages for most post-main-sequence objects, taking multi-modalities in the posteriors into account, and for some main-sequence stars down to about $0.75\,M_\odot$, below this mass limit we find larger errors (see also Appendix, Figure \ref{fig:Wd2_I_DM_HRD}). A likely explanation for this behaviour is a combination of the fact that the cINN also struggles in the range above one Gyr on the PARSEC test data and the significant model difference of Dartmouh and PARSEC in the low mass regime.
        
        With the low metallicity 'NGC6397\_I' model we are also fairly successful on the MIST synthetic test data (Appendix, Figures \ref{fig:NGC6397_I_MIST_MAP_vs_True} and \ref{fig:NGC6397_I_MIST_HRD}). Interestingly, despite using photometry in three more filters we get overall larger errors compared to the 'Wd2\_I' test. It appears that the differences between the stellar evolution models, e.g. in the model stellar atmospheres, become more significant outside of the solar metallicity case. The 'NGC6397\_I' model also recovers luminosity and temperature well, but has more difficulties with the age prediction. Still for a large fraction of test objects, both main-sequence and non-main-sequence, a correct age is inferred (median absolute deviation of 0.3\,dex). For the Dartmouth test data we find overall the worst results with the 'NGC6397\_I' model in this experiment (see Appendix, Figures \ref{fig:NGC6397_I_DM_MAP_vs_True} and \ref{fig:NGC6397_I_DM_HRD}). While the cINN recovers luminosity, temperature and gravity decently for most test samples, we find larger systematic deviations in the low brightness regime. Likewise we find a significant discrepancy for the predicted initial masses within a range from 0.25 to 0.6 $M_\odot$ and for some objects above $0.8\,M_\odot$. Lastly the age prediction fails completely for this synthetic test set with the 'NGC6397\_I' model systematically underestimating the age. Given that the prediction performance on the MIST data is acceptable, we conclude that the significant model discrepancy between Dartmouth and PARSEC at this metallicity, especially in the synthetic photometry, is the primary reason for the cINNs difficulties.
        
        In summary these experiments provide good evidence that our cINN models have not simply overfit the synthetic PARSEC training data as they are able to recover correct ages in most cases for test data from different stellar evolution codes, including ages of main-sequence objects. Furthermore, this test shows that the cINN generalises well to slightly different populations and especially excels in recovering luminosity, temperature and surface gravity. Concerning the predictions for main-sequence stars, we believe that a combination of the latent variable approach, encoding enough of the lost information, and the fact that we are using \textit{perfect} photometry allows the cINN to correctly recover ages for these objects. Consequently, as real photometry is never perfect, we acknowledge that the cINN age prediction for any real main-sequence star needs to be treated with caution. We will further discuss this matter in our application to the real NGC6397 data, as this cluster consists primarily of main-sequence sources, contrary to the young Wd2.
        
\section{Prediction}
    \label{sec:prediction}
    With the excellent performance of the cINN on the synthetic training data for Wd2 and NGC\,6397 we can now benchmark the method on real observational data. As with the synthetic test set, to retrieve the posterior distributions we sample the latent variables 4096 times for each star and determine point estimates for all physical parameters as described in Section \ref{sec:MAP_estimates_theory}. Since we have seen no significant differences between the full models and those that entail prior knowledge about the age on the synthetic data, in the following we take the full model predictions as our primary reference and perform a short comparison with the other models at the end of each section (providing further details in the Appendix). 
    
    \subsection{Westerlund 2}
        \label{sec:prediction_Westerlund2}
        \begin{figure*}
            \centering
            \includegraphics[width = 0.75\linewidth]{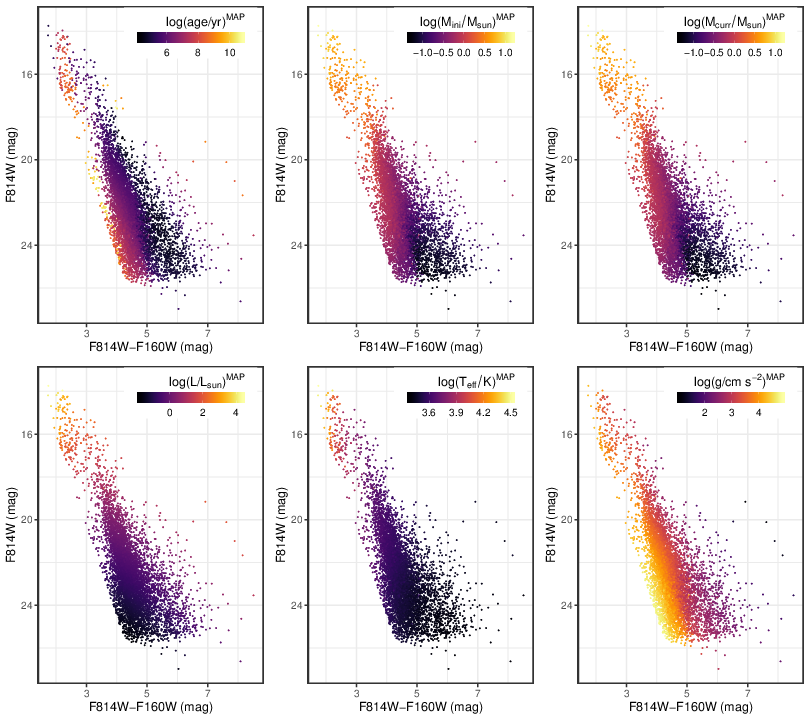}
            \caption{Optical colour magnitude diagrams of the Westerlund 2 HST data, colour coded according to the MAP estimates for the six physical parameters predicted with the cINN trained on 'Wd2\_I'.}
            \label{fig:Wd2_I_MAP_CMD}
        \end{figure*}

        \begin{figure*}
            \centering
            \includegraphics[width = 0.5333\linewidth]{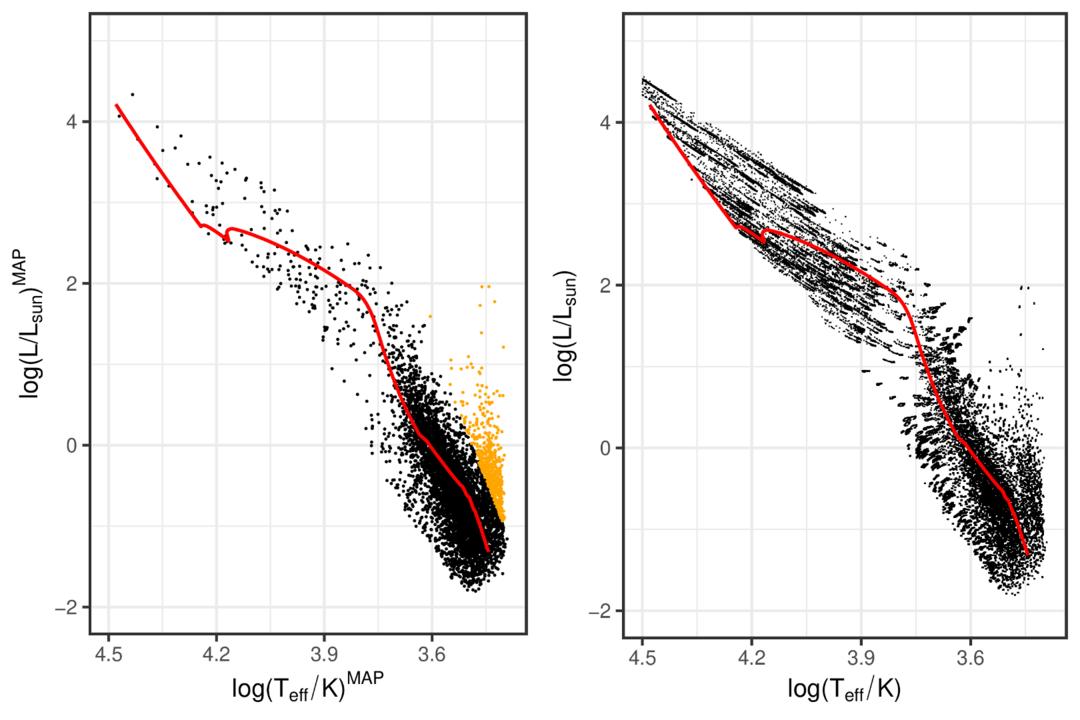}
            \includegraphics[width = 0.8\linewidth]{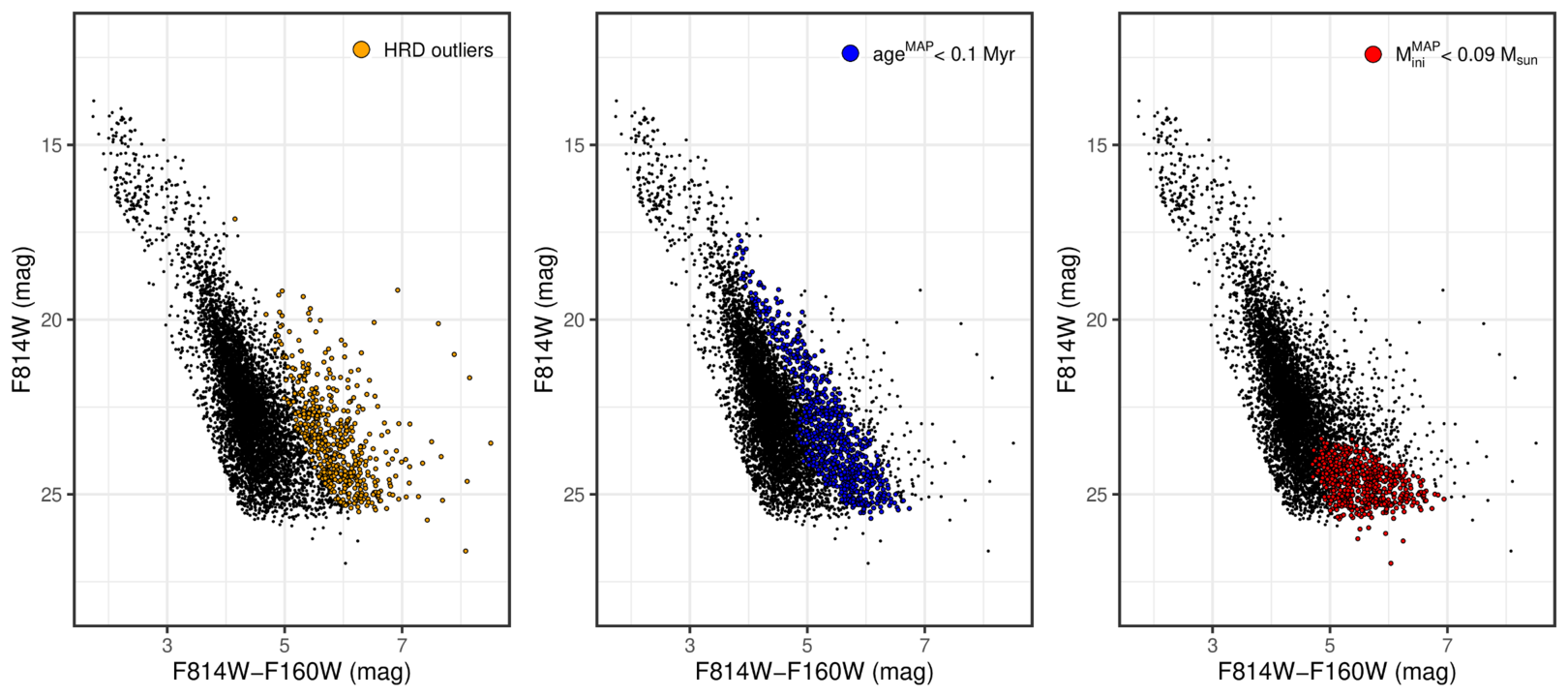}
            \caption{The top panels depict predicted Hertzsprung-Russel-Diagrams for the Westerlund 2 cluster constituents provided by the cINN model trained on 'Wd2\_I'. The \textbf{left} panel shows the HRD based on the MAP predictions for $\log(L)$ and $\log(T_\mathrm{eff})$, while in the \textbf{right} panel the entire posterior distributions of these two parameters are plotted for every star. The red line in both diagrams indicates a 1 Myr isochrone for comparison. The orange points in the first panel indicate a possible vertical artefact in the cINN prediction. The \textbf{bottom left} panel indicates the observed CMD position of these HRD outliers. In the \textbf{bottom centre} and \textbf{right} panels we indicate the observed CMD positions of a few stars for which the predicted age (blue) or initial mass (red), respectively, are below the lower limits of the training data.}
            \label{fig:Wd2_I_HRD}
        \end{figure*}
        
        \begin{figure*}
            \centering
            \includegraphics[width = \linewidth]{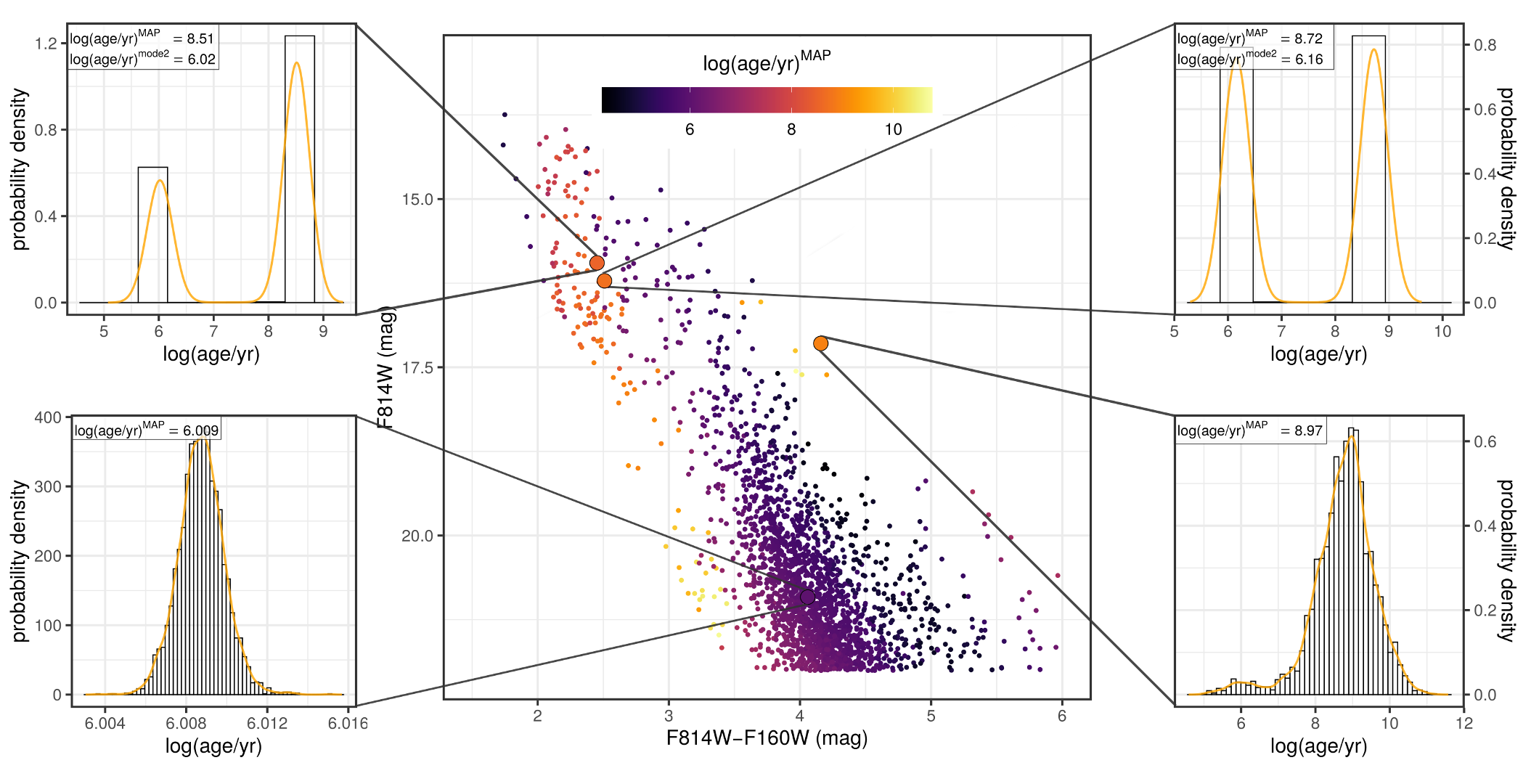}
            \caption{The \textbf{middle} panel shows a zoom-in of the optical CMD of the Westerlund 2 cluster constituents colour coded according to the MAP prediction of $\log(\mathrm{age})$. The four panels \textbf{left} and \textbf{right} show the predicted age posterior distributions of the highlighted stars in the CMD. The \textbf{bottom left} panel is an example PMS stars for which our approach provides excellent results, returning a very narrow age distribution at the proposed cluster age. The remaining three cases are taken from the 86 stars likely on the turn-on for which the MAP age estimate is significantly above the suggested age of Westerlund 2. The two posterior distributions on the top show commonly observed behaviour among these 86 stars, where we find a second peak in the age posterior distribution, either less (\textbf{right panel}) or almost equally likely (\textbf{left panel}), which is more consistent with Westerlund's suggested age. The \textbf{bottom right} panel is a rare example where the age posterior distribution shows no significant second peak and the age of the star is predicted to be too old for Wd2.}
            \label{fig:Wd2_I_TO_posteriors}
        \end{figure*}
        
        Figure \ref{fig:Wd2_I_MAP_CMD} presents our cINN prediction results for all six physical parameters, showing their MAP estimates colour coded on the optical CMD of Westerlund 2 (cluster members only).
        Overall the results are very reasonable for Westerlund 2, from sub-solar masses for low mass PMS stars to above solar masses for UMS stars, with the correct gradients of $L$, $T_\mathrm{eff}$ and $g$ vs. magnitude and colour. On top of that the median 1.27 Myr cluster age from MAP estimates is well within the previously determined age range of $1.04 \pm 0.72\,\mathrm{Myr}$. The resulting HRD, shown in Figure \ref{fig:Wd2_I_HRD} (top left panel as per MAP estimates, top right panel as per entire posteriors), also matches fairly well the 1 Myr isochrone traced in red for comparison. There is a noticeable spread around the isochrone, but most of the stars are correctly placed within the PMS regions of the diagram. Notable is only a small vertical feature at the extreme right of the predicted HRD, highlighted by the orange points in the top left panel of Figure \ref{fig:Wd2_I_HRD}, which appears to be deviating more systematically from the 1\,Myr isochrone. These 502 stars, all located at the very red edge of the CMD (bottom left panel of Figure \ref{fig:Wd2_I_HRD}) have a median photometric error of 0.15\,mag. It is quite possible that the cINN prediction entails this vertical artefact due to photometric uncertainties, which are not accounted for in our setup. 
        
        We also find other miss-predictions from the cINN. For 584 stars (179 among the HRD outliers) the initial mass MAP estimate falls below the $0.09\,\mathrm{M_\odot}$ minimum of the training set and in 292 cases even below the H-burning threshold of $0.072 \,\mathrm{M_\odot}$ \citep[Solar metallicity;][]{Chabrier2002}. With a minimum of $0.05\,\mathrm{M_\odot}$ the mass estimates for these stars (red points, Figure \ref{fig:Wd2_I_HRD} bottom centre panel) are still physically plausible for, i.e., young brown dwarfs, but this extrapolation might indicate a systematic error. Like the HRD outliers these objects are subject to a notable amount of photometric uncertainty (median of 0.2\,mag in F814W), being a likely culprit for these miss-predictions.
        
        For another 818 stars (343 also in HRD outliers) the MAP age estimate is below the 0.1\,Myr training set minimum, going down to 0.02\,Myr. Given their location at the red edge of the CMD (blue points, Figure \ref{fig:Wd2_I_HRD} bottom left panel) these results are somewhat plausible but not convincing. Aside from the photometric uncertainties, limitations of the \cite{Zeidler2015} prescription to estimate stellar extinction from gas colour excess could provide an explanation for these results, if e.g. the stellar extinction has been underestimated for these objects.
        
        Lastly, a number of stars are predicted to be unreasonably old for Wd2. These are located primarily at the very blue and red edge of the PMS population in the CMD, but we also find 86 among them on the turn-on (highlighted in Figure \ref{fig:Wd2_I_TO_posteriors}). The former could potentially be field contaminants that survived our initial rejection using Besan\c{c}on models in the direction of Wd2 \citep{Zeidler2015} and are correctly identified as old. Evidence for this hypothesis is that we identify these outliers in our age prediction primarily in the CMD region where \cite{Zeidler2015} find an overlap of the Besan\c{c}on models and the cluster constituents.
        
        For the 86 turn-on stars only the MAP age estimate is incorrect, as almost all of them show degenerate age posteriors with a prominent second peak close to the supposed cluster age. Figure \ref{fig:Wd2_I_TO_posteriors} presents three example age posteriors for these turn-on objects and one from the majority of well constrained solutions (bottom left panel) for comparison. In the top-left posterior example, a common case, an old age appears as the most likely solution, but we find a secondary maximum at the cluster age. The top-right panel represents another frequent outcome among these 86 stars, where the young and old solution are almost equally likely. The (rarely occurring) final case in the bottom right shows a  ``complete failure'', where no prominent secondary maximium exists at the cluster age. Given that (field) RGB stars can very well overlap with PMS stars within the main sequence turn-on region, these results demonstrate again the great strength of the cINN approach as it recognises and shows this possibility in the predicted age posterior distributions. At the same time these examples serve as a reminder that careful post-processing (e.g. identification of all major peaks) of the predicted posterior distributions is necessary to avoid possible false conclusions by e.g. relying only on MAP point estimates.
        
        Comparing the predictions on the Wd 2 HST data between the models 'Wd2\_I' and 'Wd2\_II' we find that they agree well with each other. See Appendix \ref{app:Wd2_II} and Figure \ref{fig:Wd2_I_vs_Wd2_II} for more details. We conclude that inclusion of prior knowledge in the form of a simple range cut of the training set does not benefit the cINN approach in the Wd2 case.
        
        \subsubsection{Cluster Age}
            \begin{figure}
                \centering
                \includegraphics[width = 0.8\linewidth]{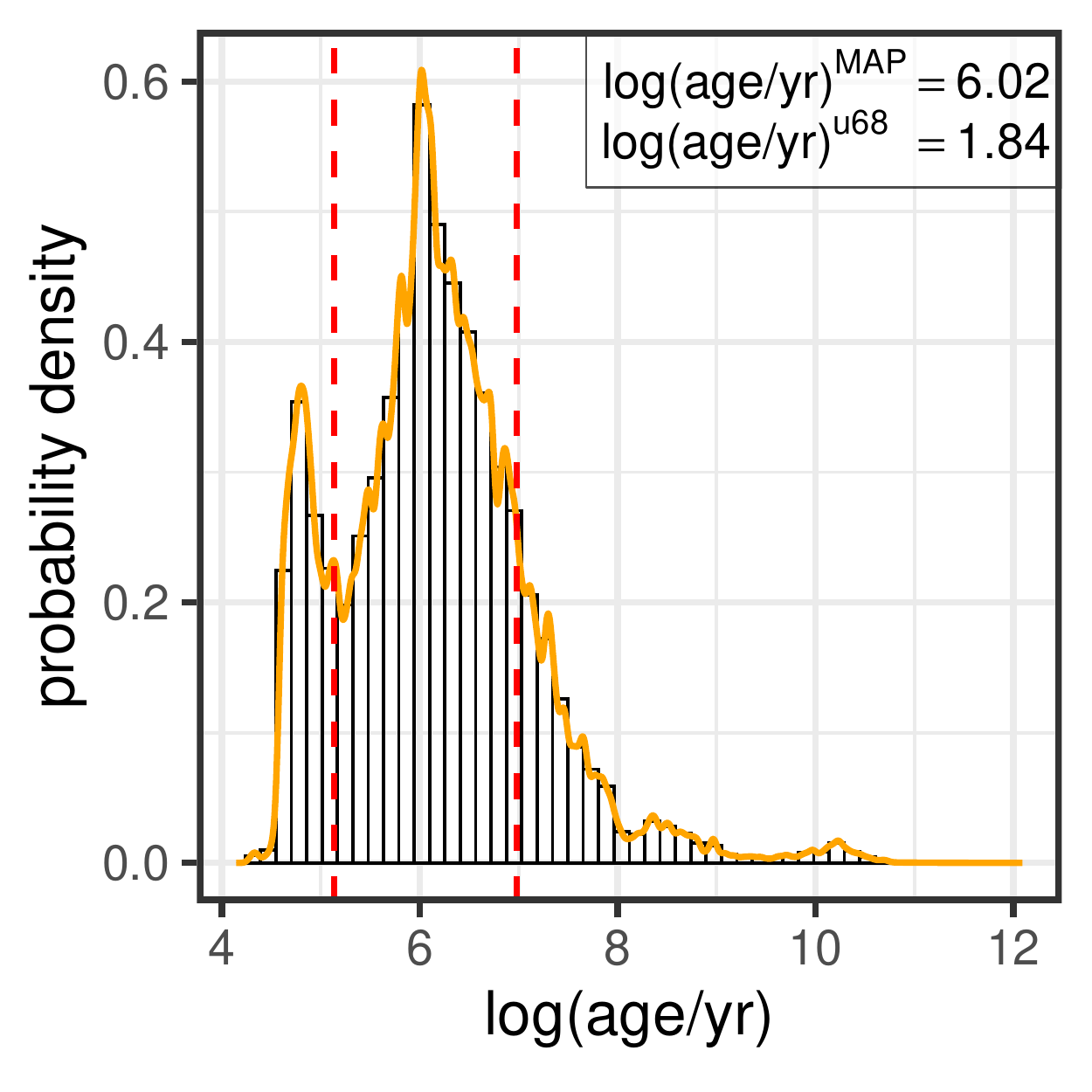}
                \caption{Histogram of the sum of the age posterior distributions of all Westerlund 2 cluster stars as predicted by the 'Wd2\_I' cINN model. The orange line indicates a kernel density fit to this 'cumulative' posterior distribution to determine the most likely cluster age. The red dashed lines mark the width of the 68\% confidence interval.}
                \label{fig:Wd2_I_logAgeKDE}
            \end{figure}
            
            \begin{figure}
                \centering
                \includegraphics[width = 0.8\linewidth]{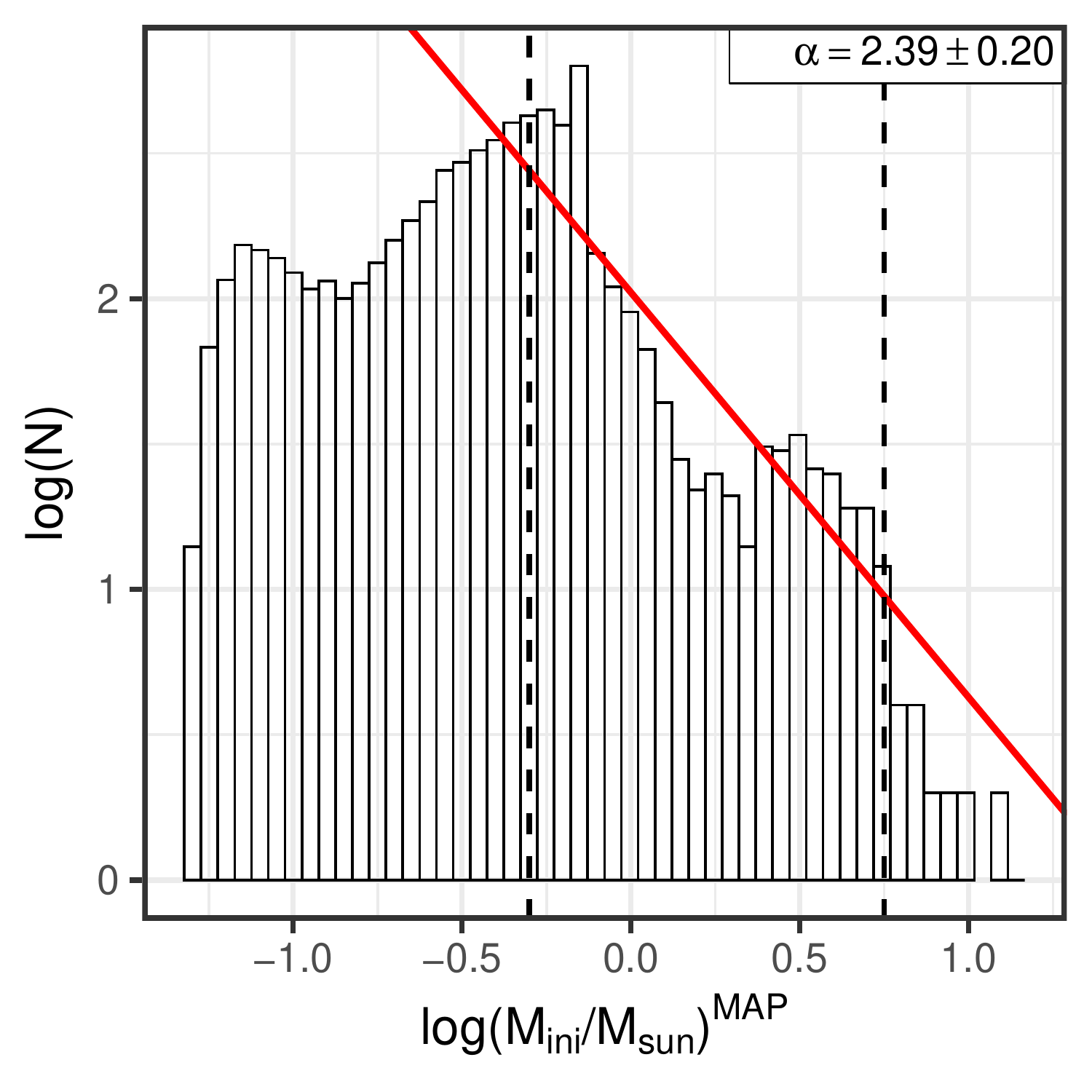}
                \caption{The predicted Initial Mass Function of Westerlund 2 based on the MAP estimates of $M_\mathrm{ini}$ of the individual stars provided by the 'Wd2\_I' cINN model. The black dashed lines indicate the range ($0.5$ to $\sim 5.6\,\mathrm{M_\odot}$) used to fit the high mass slope of the IMF. The fit with a slope of $\alpha = 2.39 \pm 0.2$ is given by the red line.}
                \label{fig:Wd2_I_IMF}
            \end{figure}
            
            \begin{figure*}
                \centering
                \includegraphics[width = 0.75\linewidth]{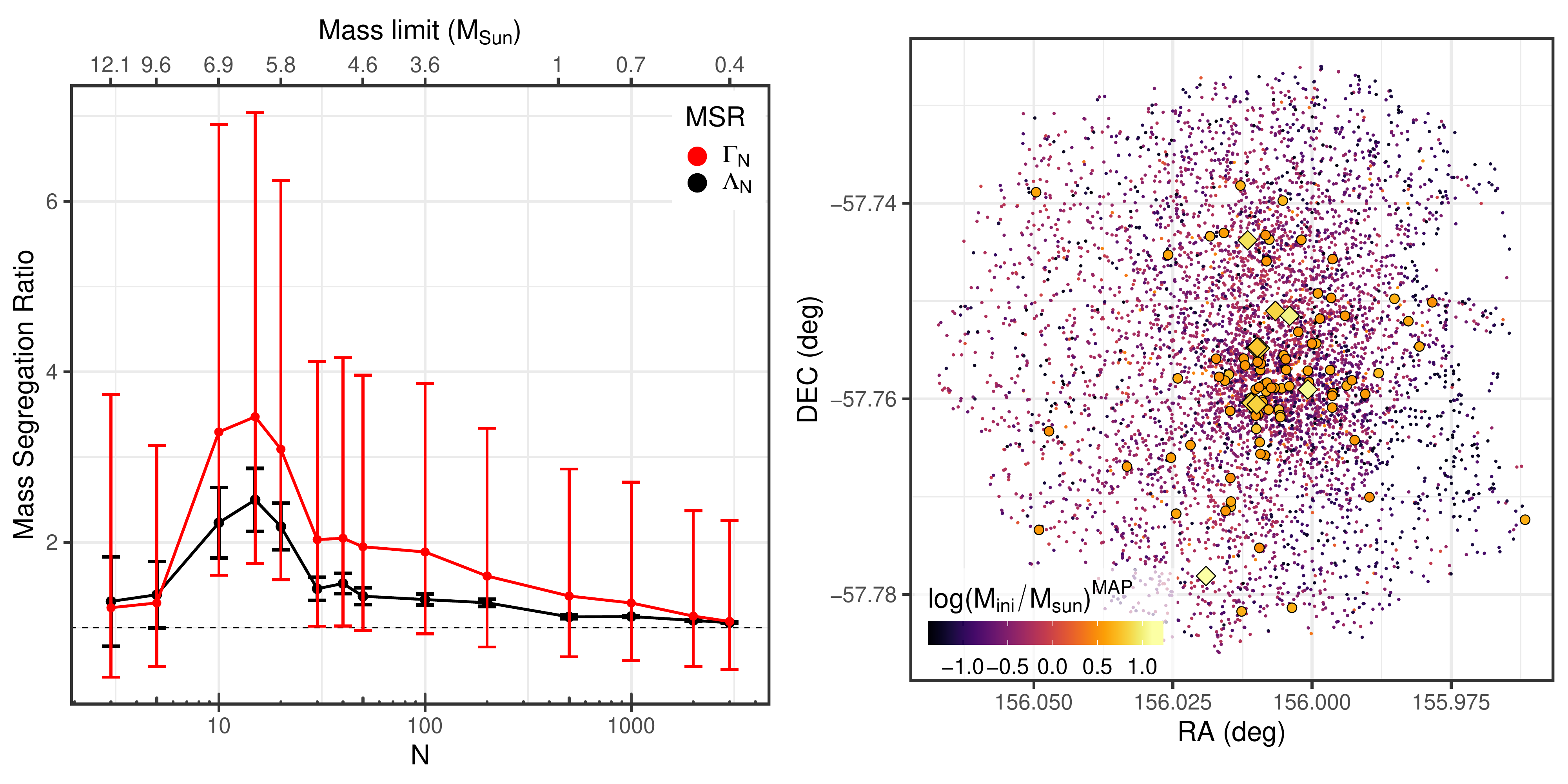}
                \caption{The \textbf{left} panel shows the mass segregation ratios $\Lambda_\mathrm{MSP}$ \citep{Allison2009} and $\Gamma_\mathrm{MSP}$ \citep{Olczak2011} for different numbers of the most massive stars of Westerlund 2 based on the MAP estimates of the initial mass of the stars. Note that the second x-axis in this diagram denotes the corresponding lower mass limits, i.e. the mass of the lowest mass star within the set of the N most massive stars. On the \textbf{right} the spatial distribution of the Westerlund 2 stars is shown colour coded according to the MAP estimate of the initial mass. The stars highlighted by the large diamond symbols are the 10 most massive stars in our prediction, while the large circles (plus the diamonds) indicate the 100 most massive stars.}
                \label{fig:Wd2_I_MassSeg}
            \end{figure*}
            Having assessed the overall satisfying prediction results of the 'Wd2\_I' cINN model we now derive some physical properties of the cluster and compare them to previous studies. 
            
            To begin we derive a cluster age from our individual stellar age predictions. As previously mentioned from the MAP stellar age estimates we find a median age of the cluster stars of $1.27_{-0.94}^{+3.62}\,\mathrm{Myr}$. Determining the cluster age as the most likely value from the sum of all the individual age posterior distributions (Figure \ref{fig:Wd2_I_logAgeKDE}) using a kernel density estimate, we find a value of $1.04_{-0.90}^{+8.48}\,\mathrm{Myr}$ (MAP and edges of 68\% confidence interval). We find an almost identical result for the same derivation with the 'Wd2\_II' model (See Appendix \ref{app:Wd2_II} and Figure \ref{fig:Wd2_II_logAgeKDE}).  While we cannot constrain the cluster age more precisely than the previous study by \cite{Zeidler2016}, both of our values match the previously derived age within their errors. This is a very satisfactory result given that our method derives the cluster age without any prior knowledge, just on the basis of the stellar magnitudes in two photometric broadband filters and an extinction estimate.
        
        \subsubsection{The stellar initial mass function}
            As our method predicts the initial mass of each star of Westerlund 2 we can also analyse the initial mass function (IMF) of the cluster, shown in Figure \ref{fig:Wd2_I_IMF}. We suffer from incompleteness at the low mass end and from saturation at the high mass end but nevertheless, using the range from $0.5\,\mathrm{M_\odot}$ to $\sim 5.6\,\mathrm{M_\odot}$ as a proxy to derive the slope of the high mass IMF, we find a value of $\alpha = 2.39 \pm 0.20$, which matches the Salpeter IMF slope of $\alpha = 2.35$ within $1\,\sigma$. \cite{Zeidler2017} determine a present day mass function (PDMF) with a slope of $\alpha = 2.53 \pm 0.05$ for the survey area of our Westerlund 2 data. Presuming that the PDMF should not deviate too much from the IMF given the young age of Westerlund 2, our slope is in good accordance with the result from \cite{Zeidler2017}.
            
        \subsubsection{Mass segregation}
            \cite{Zeidler2017} also find evidence for mass segregation in Westerlund 2 through the analysis of the PDMF within different annuli around the midpoint between the main and northern sub-cluster of Westerlund 2. Using our individual stellar mass predictions we try to confirm this finding by computing the mass segregation ratios (MSR) $\Lambda_\mathrm{MSR}$ \citep{Allison2009} and $\Gamma_\mathrm{MSR}$ \citep{Olczak2011}. These two quantities are derived by constructing a minimum spanning tree (MST) for the $N$ most massive stars within the population and comparing it with $k$ MSTs of $N$ random stars from the stellar sample. For $\Lambda_\mathrm{MSR}$ we then compute the tree length $l_\mathrm{mass}$ of the tree with the massive stars and the average tree length $\langle l_\mathrm{rand} \rangle$ of the $k$ trees of random stars, so that we find the MSR as
            \begin{equation}
                \Lambda_\mathrm{MSR} = \frac{ \langle l_\mathrm{MST}^\mathrm{rand} \rangle}{l_\mathrm{MST}^\mathrm{mass}} \pm \frac{\sigma_\mathrm{rand}}{l_\mathrm{mass}},
            \end{equation}
            where $\sigma_\mathrm{rand}$ is the standard deviation of $\langle l_\mathrm{rand} \rangle$ \citep{Allison2009}.
            
            $\Gamma_\mathrm{MSR}$ is given by the ratio between the mean edge lengths $e_\mathrm{mass}$ and $\langle e_\mathrm{rand} \rangle$: 
            \begin{equation}
                \Gamma_\mathrm{MSR} = \frac{\langle e_\mathrm{rand}\rangle}{e_\mathrm{mass}}.
            \end{equation}
            Here we proceed in a fashion similar to $\Lambda_\mathrm{MSR}$, except that we now calculate the geometric instead of the arithmetic mean. For each of the $k$ random MSTs we determine the geometric standard deviation according to
            \begin{equation}
                \Delta e_\mathrm{rand}^k = \exp \left( \sqrt{\frac{\sum_{\mathrm{i}=1}^{N} (\ln e_\mathrm{i}^k - \ln e_\mathrm{rand}^{k})^2}{N}} \right),
            \end{equation}
            where $e_\mathrm{i}^k$ are the $N$ edges of the $k$th tree \citep{Olczak2011}, and then derive the upper and lower $1\sigma$ intervals as the means of the $k$ lower and upper $1\sigma$ intervals (note that $\Delta e_\mathrm{rand}^k$ is a multiplicative standard deviation):
            \begin{equation}
                \begin{split}
                    \Gamma_\mathrm{MST}^\mathrm{upper} &= \frac{1}{k} \sum_{\mathrm{i}=1}^{k} \frac{e_\mathrm{rand}^i}{e_\mathrm{mass}} \Delta e_\mathrm{rand}^i, \\
                    \Gamma_\mathrm{MST}^\mathrm{lower} &= \frac{1}{k} \sum_{\mathrm{i}=1}^{k} \frac{e_\mathrm{rand}^i}{e_\mathrm{mass}} \frac{1}{ \Delta e_\mathrm{rand}^i}.
                \end{split}
            \end{equation}
            Values of $\Lambda_\mathrm{MST} \approx 1$ and $\Gamma_\mathrm{MST} \approx 1$ indicate that the $N$ massive and the $N$ randomly selected stars are similarly distributed, while $\Lambda_\mathrm{MST} \gg 1$ ($\Gamma_\mathrm{MST} \gg 1$) signifies mass segregation and $\Lambda_\mathrm{MST} \ll 1$ ($\Gamma_\mathrm{MST} \ll 1$) suggests inverse mass segregation where the most massive stars are more spread outwards \citep{Dib2018}. Following the suggestion in \cite{Olczak2011} we calculate the number $k$ of random population MSTs based on the number $N$ of massive stars, such that a fraction $p = 0.99$ of the total population of $M$ stars is covered according to
            \begin{equation}
                k = \mathrm{ceil}\left(\frac{\ln(1-p)}{\ln(1-\frac{N}{M})} \right),
            \end{equation}
            where ceil(x) denotes the ceiling function, i.e. the function rounding up to the next larger integer.
            
            In the left panel of Figure \ref{fig:Wd2_I_MassSeg} we present our resulting MSRs for different numbers $N$ of the most massive stars drawn from the total population. We find some evidence for mass segregation as $\Lambda_\mathrm{MST} > 1$ and $\Gamma_\mathrm{MST} > 1$ for the 10 to 100 most massive stars. With a maximum MSR of $\sim3.4$ within this range, however, our analysis suggests that the mass segregation is not strongly pronounced. The spatial distribution of the 10 (diamond markers) and the 100 (large circles + diamond markers) most massive stars shown for comparison in the right panel of Figure \ref{fig:Wd2_I_MassSeg} confirms this finding, as the most massive stars appear slightly more clustered towards the centre but not to an excessive degree. The decrease in MSR for the five and three most massive stars is likely due to the fact that the single most massive star ($M_\mathrm{ini}^\mathrm{MAP} \approx 14.7 \mathrm{M_\odot}$) in our sample is actually located away from the centre of Westerlund 2 (the southernmost diamond in the diagram), which induces large tree and edge lengths in the MST.
            
            In conclusion, our results for cluster age, slope of the IMF and observed mass segregation, derived from the cINN predictions of Westerlund 2, are in good accordance with previous studies. Therefore, the cINN method performs to a very satisfactory degree on the actual observational data of Westerlund 2. 
    
        \begin{figure*}
            \centering
            \includegraphics[width = 0.75\linewidth]{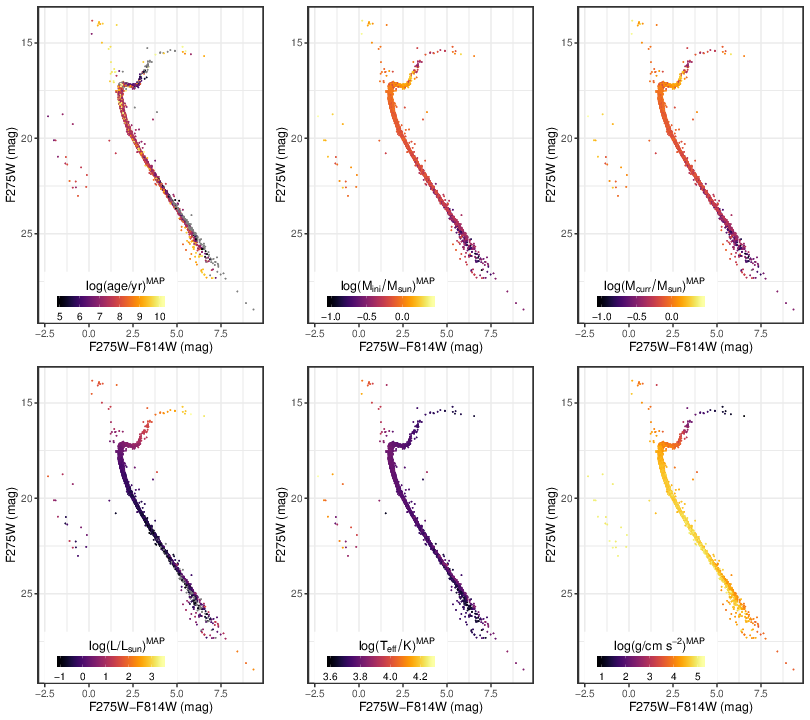}
            \caption{UV-I CMDs of NGC\,6397 colour coded according to the MAP estimates of the six physical parameters $\log(\mathrm{age})$, $\log(M_\mathrm{ini})$, $\log(M_\mathrm{curr})$, $\log(L)$, $\log(T_\mathrm{eff})$, $\log(g)$ as predicted by the cINN trained on 'NGC6397\_I'.}
            \label{fig:NGC6397_I_MAP_pred_CMD}
        \end{figure*}
    \subsection{NGC6397}
        \label{sec:NGC6397_prediction}
        
        \begin{figure*}
            \centering
            \includegraphics[width = 0.85\linewidth]{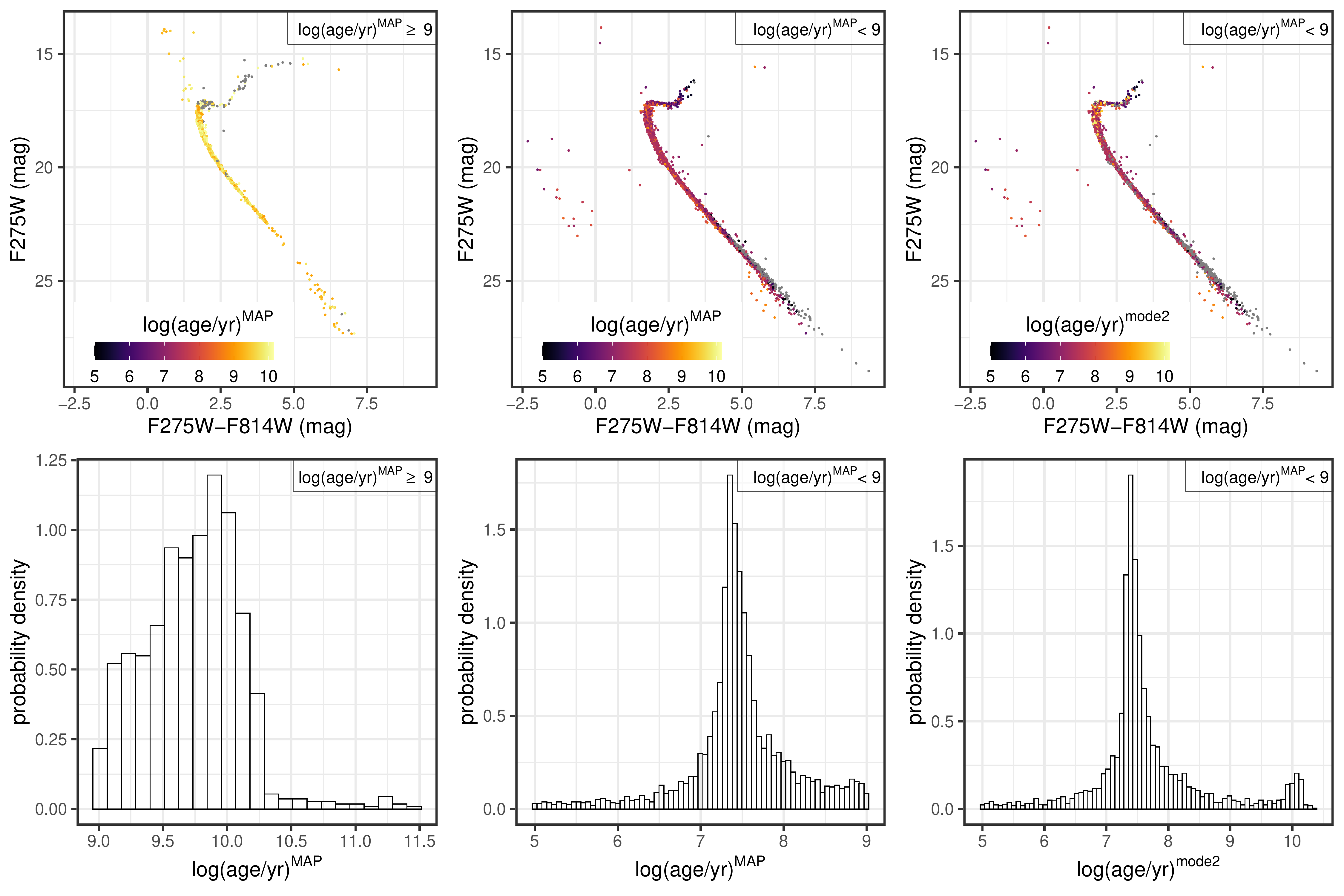}
            \caption{The \textbf{top left} panel shows the CMD of the 999 NGC\,6397 stars for which the MAP estimate of the age, derived from the 'NGC6397\_I' cINN prediction, is above 1 Gyr as indicated by the colour coding. The \textbf{top middle} panel shows the CMD position of the remaining 3832 stars, also colour coded according to $\log(\mathrm{age/yr})^\mathrm{MAP}$, for which the MAP estimate is below 1 Gyr for comparison. The \textbf{top right} panel shows the CMD of the same stars as the middle panel, now colour coded according to the second most likely mode (the second highest local maximum) of their posterior distribution. Stars that show no second mode in the predicted age posterior are indicated in grey. The \textbf{bottom row} shows the histograms of the age MAP estimates (or second mode in the age posterior in the last case) of the stars in the corresponding panels in the top row.}
            \label{fig:NGC6397_I_Age_MAP_CMD_below_above_GYR_hists}
        \end{figure*}
        
        \begin{figure}
            \centering
            \includegraphics[width = 0.8\linewidth]{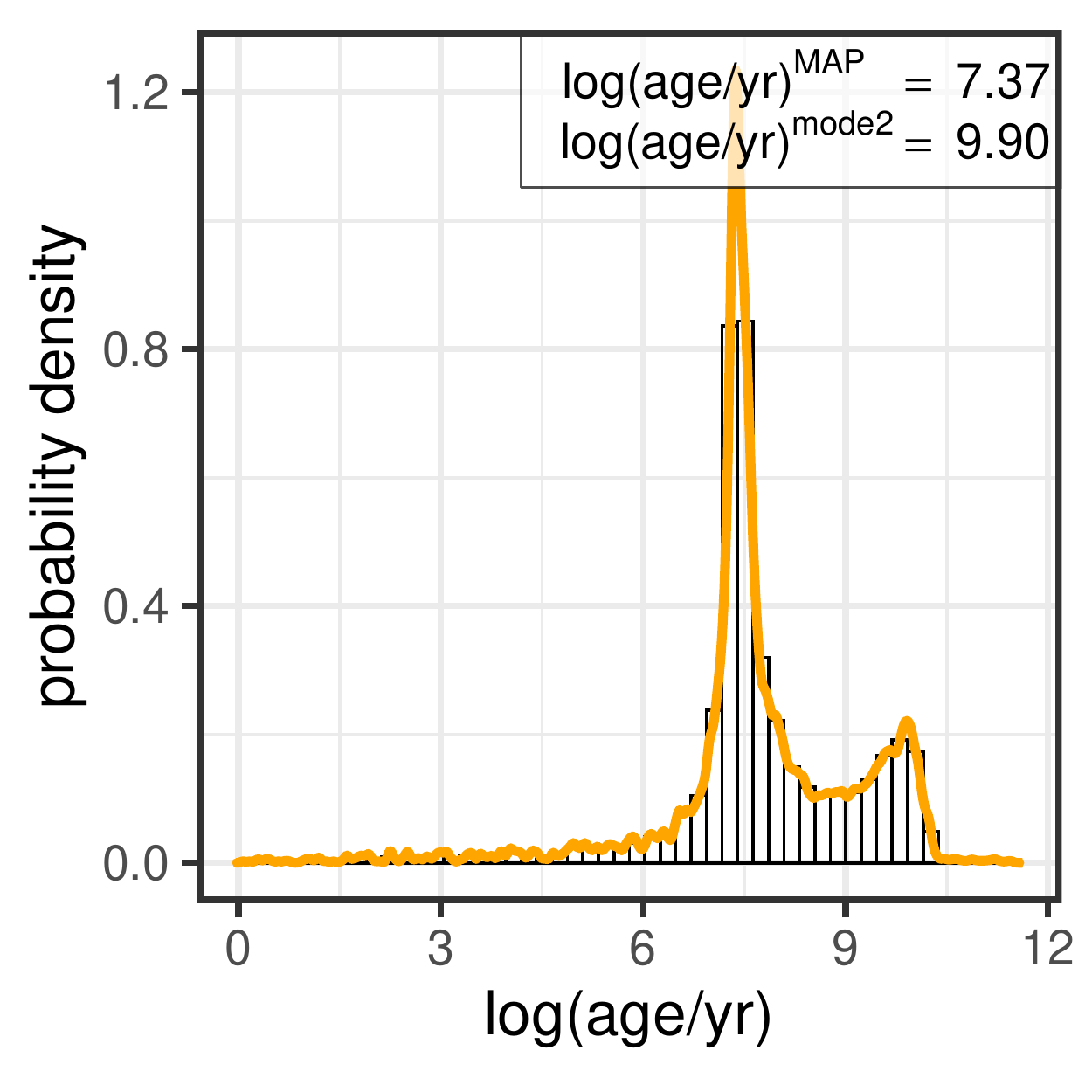}
            \caption{Histogram of the sum of the age posterior distributions of all NGC\,6397 stars, predicted by the cINN trained on 'NGC6397\_I'. The orange line indicates a kernel density fit to this cumulative posterior distribution to determine the most likely cluster age.}
            \label{fig:NGC6397_I_logAgeKDE}
        \end{figure}
        
        \begin{figure*}
            \centering
            \includegraphics[width = \linewidth]{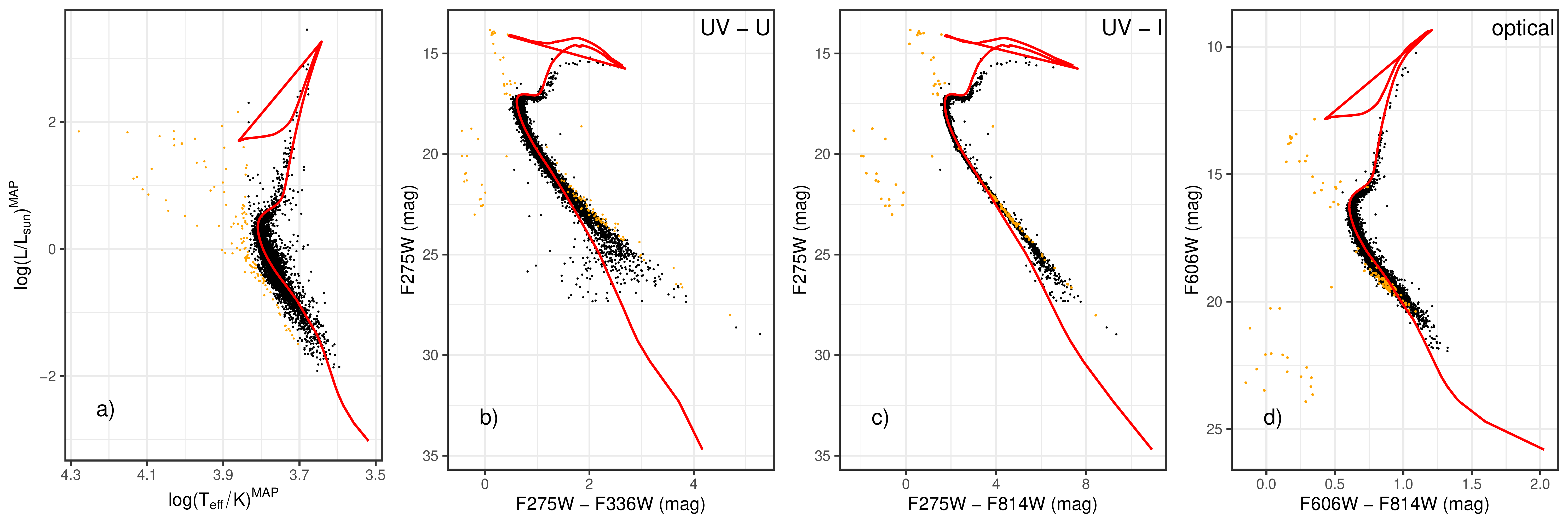}
            \caption{Panel \textbf{a)} shows the predicted HRD of NGC\,6397 based on the MAP estimates of $\log(L)$ and $\log(T_\mathrm{eff})$ by the cINN trained on 'NGC6397\_I'. The remaining panels \textbf{b)} - \textbf{d)} show in order the UV-V, UV-I and the optical CMDs of the NGC\,6397 data, respectively. The orange marked stars are those for which the cINN prediction of the HRD position deviates strongly from the supposed age of the cluster of $\sim13\,\mathrm{Gyr}$, as indicated by the red isochrone in all four diagrams. This series shows that, aside from the white dwarfs and blue stragglers, the cINN prediction fails where the PARSEC isochrone models fail to fit the data. }
            \label{fig:NGC6397_I_HRD_outliers}
        \end{figure*}
        
        \begin{figure*}
        	\centering
        	\includegraphics[width = \linewidth]{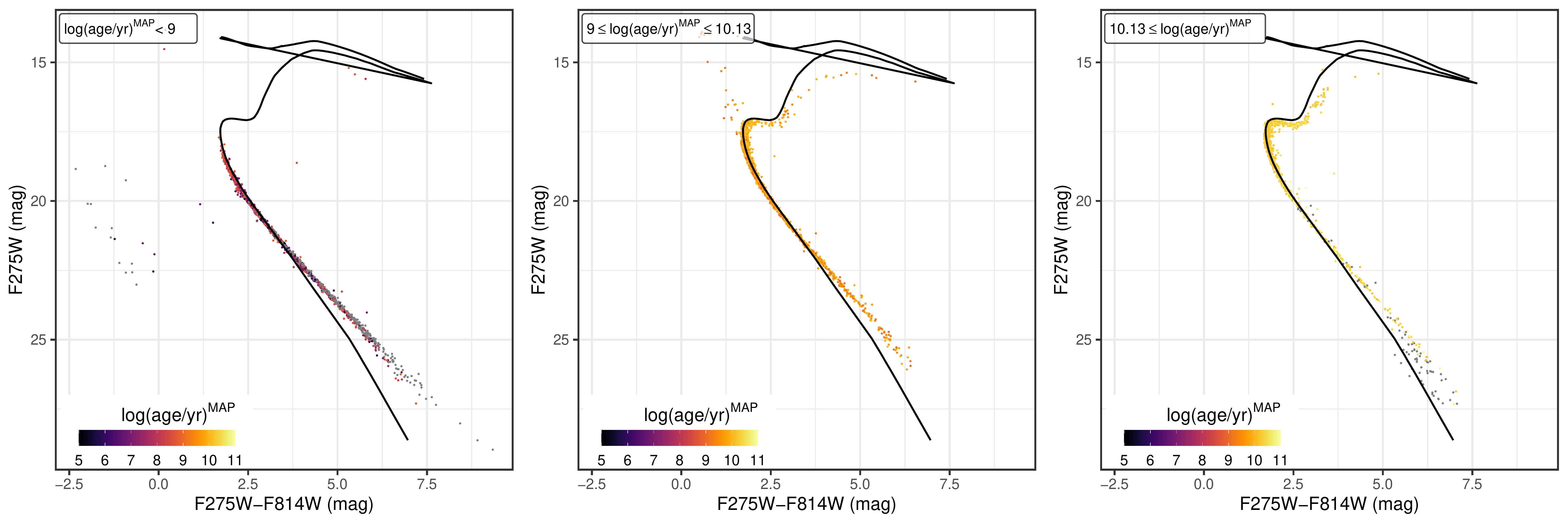}
        	\caption{UV-I CMD of the NGC\,6397 data colour coded according to the MAP age estimates predicted by the cINN trained on 'NGC6397\_II'. In order the three panels show the stars for which we find $\log(\mathrm{age/yr}) < 9$, $9 \leq \log(\mathrm{age/yr}) \leq 10.13$ and $10.13 < \log(\mathrm{age/yr})$, respectively. The red line in all diagrams indicates a 13\,Gyr isochrone for comparison. This sequence demonstrates that the age prediction of the 'NGC6397\_II' fails noticeably for those stars, where the observations deviate the most from the theoretical model.}
        	\label{fig:NGC6397_II_Age_outliers}
        \end{figure*}
        
        For NGC\,6397 our cINN predictions do not achieve the same success on the real HST data as for Wd2. Figure \ref{fig:NGC6397_I_MAP_pred_CMD} summarises our results showing the MAP estimates for the physical parameters colour coded for every star in the UV-I CMD.  Overall we find fairly plausible values for all parameters, except for age. For instance, most predicted masses are below one solar mass, which is expected for a 13 Gyr old cluster given that more massive stars should already have disappeared. With the age prediction, however, we find worse results. A large fraction of stars is predicted to be much younger than what would be reasonable for NGC\,6397, considering that some of them are located on the red giant branch (RGB) and the main sequence turn-off, the features traditionally used to date globular clusters. The top-left and top-center panels of Figure \ref{fig:NGC6397_I_Age_MAP_CMD_below_above_GYR_hists} show the age prediction more in detail, separating those stars in the CMD for which the MAP estimate is plausible (above 1 Gyr) from those where it is definitely incorrect (below 1 Gyr). Only 1/5 of the stars (999 out of 4831) have plausible MAP age estimates. Of the remaining 3832 stars, only 359 have a second or third mode in their predicted age posterior distributions that falls above 1 Gyr. The top-right panel in Figure \ref{fig:NGC6397_I_Age_MAP_CMD_below_above_GYR_hists} shows that most of these are located at the turn-off and bottom of the RGB, an indication that the cINN has learned, at least to some degree, that stars located on the turn-off may be old. But even including these 359 additional stars, where a plausible solution is part of the posterior distribution, we still find that for more than two thirds of the observational data our age prediction fails entirely. Failure may be expected for some of the stars within the NGC\,6397 sample as our training set does not include e.g. white dwarfs, so that a miss-prediction in these cases is easily explained. If we subtract the latter cases and the 359 turn-on stars with a plausible second mode, we find that the age prediction fails primarily for low-mass main (LMS) sequence stars.
        
        As we have previously discussed, predicting the age of low-mass main sequence stars is arguably an extremely difficult task as stars with a wide range of ages share very similar observational features. Even though the cINN estimates an age within a plausible range for a number of LMS stars, at least down to about 23 mag in F275W, some of these predictions are still flawed, as can be seen in the histograms in the bottom row of Figure \ref{fig:NGC6397_I_Age_MAP_CMD_below_above_GYR_hists}. There are, in particular, cases here where the MAP estimates are too large, sometimes even way above the age of the universe. With only a minority of stars with plausible MAP age estimates, deriving a cluster age as the most likely age provided by the sum of the age posteriors (Figure \ref{fig:NGC6397_I_logAgeKDE}) is not applicable. The most likely age value would be 23.4 Myr, way too low, and the barely prominent second peak, while in the vicinity of the relatively well known cluster age, still underestimates the age with a value of 7.9\,Gyr. 
        
        In contrast to the problematic age estimates, luminosity and effective temperature appear to be predicted quite well by the cINN model. The left panel in Figure \ref{fig:NGC6397_I_HRD_outliers} shows the predicted HRD for NGC\,6397 based on the MAP estimates of $\log(L)$ and $\log(T_\mathrm{eff})$. The cINN prediction traces the 13.4 Gyr isochrone (red line) overall very closely, but there are a few outliers, indicated by orange points in the diagram. Among them we find the white dwarfs and blue stragglers as revealed by the CMDs in the remaining panels of Figure \ref{fig:NGC6397_I_HRD_outliers}. Apart from these cases there are a couple of low-mass main sequence stars for which the prediction deviates noticeably from what one would expect, similar to what we found with the predicted ages. This leads us to a possible hypothesis to explain the problems our cINN approach encounters with the low-mass NGC\,6397 stars. The main sequence outliers fall primarily into a region in the observable space where the PARSEC isochrone models cannot properly fit the observed data (note the deviation between the red isochrone and the data points, most obvious in the UV-I CMD in Figure \ref{fig:NGC6397_I_HRD_outliers} c)). Given that the deviation between model and data is most severe starting e.g. at $\sim 23\,\mathrm{mag}$ in F275W in the UV-I CMD, this deficiency of the models could also explain why we find very few plausible age predictions below this magnitude. The deviation between the model and the data is not only present for the LMS, but also in the RGB, as shown by the CMDs. This could further explain the many age miss-predictions for even the RGB constituents. 
        
        If a problem with the underlying stellar evolutionary models is indeed the root of the cINN prediction shortcomings, restricting the training set to a narrower range as in our 'NGC6397\_II' model cannot be a remedy (see Figures \ref{fig:NGC6397_II_MAP_CMD}, \ref{fig:NGC6397_II_HRD_outliers} and \ref{fig:NGC6397_II_logAgeKDE} in the Appendix). While narrowing the range provides age predictions much closer to the actual age of the cluster, a large number of cases show over-prediction (1013 stars have an age MAP estimate above 13.5 Gyr) as well as several instances of extrapolation far below the minimum age of the 'NGC6397\_II' training set (1751 stars with $\mathrm{age}^\mathrm{MAP} < 1\,\mathrm{Gyr}$, going down to 0.1 Myr and below). The CMD positions of the latter outliers (Figure \ref{fig:NGC6397_II_Age_outliers}) provides further support for the hypothesis that the discrepancy between model and observations causes the prediction issues: the cINN underestimates the age predominantly for stars located where the observed LMS population deviates the most from the theoretical data. For further evidence on this matter and additional details on the 'NGC6397\_II' results we refer to Appendix \ref{app:NGC6397_II}. 
        
        Returning to our discussion in Section \ref{sec:synth_ms_age_pred} concerning the age estimation for main-sequence sources from photometry alone, it is necessary to mention here that the complexity of the task itself likely also plays a role in the NGC\,6397 prediction outcome. While the evidence for the main culprit being discrepancy between model and observations appears conclusive to us, given also its similarity to the model-model difference issues we have discovered with the prediction on the low metallicity Dartmouth isochrone data, we have to acknowledge an additional caveat here. Even where the PARSEC models do match the observations, we find that the age estimates for the LMS constituents are \textit{just} plausible but do not exactly recover the known age of NGC\,6397. Based on this consideration and with regard to the difficulty that traditional methods have with dating main-sequence objects, we have to surmise that the cINN might not necessarily outperform known approaches on this specific aspect. Consequently, our age estimates for real main-sequence stars should be treated with caution.
        
        In conclusion our prediction of physical parameters for the globular cluster NGC\,6397 does not achieve the same satisfying results as the cINN model does for Westerlund 2. While the predicted masses fall within reasonable ranges and the HRD constructed from the MAP estimates of $L$ and $T_\mathrm{eff}$ traces the theoretical position of the cluster reasonably well, we find a significant number of outliers and major problems with the prediction of the age. Here the cINN tends to overestimate the age for the stars it recognises as old while severely underestimating the age for a majority of stars of the globular cluster. Comparing the location of these outliers in the CMD with the underlying PARSEC models, we believe that these issues are primarily rooted in a miss-match, contrary to the Westerlund 2 case, bewteen the isochrones and the observations of NGC\,6397. We find large discrepancies especially for the LMS and RGB stars (see e.g. Figure \ref{fig:NGC6397_I_HRD_outliers}). This ultimately demonstrates again that even a machine learning approach as powerful as the cINN is always only as good as the underlying physical model. Therefore, it is crucial to choose models that provide the best agreement with the data.
        
\section{Possible Extensions}
    \label{sec:extensions}
    \subsection{Extinction as a physical parameter}
        \begin{figure*}
            \centering
            \includegraphics[width = \linewidth]{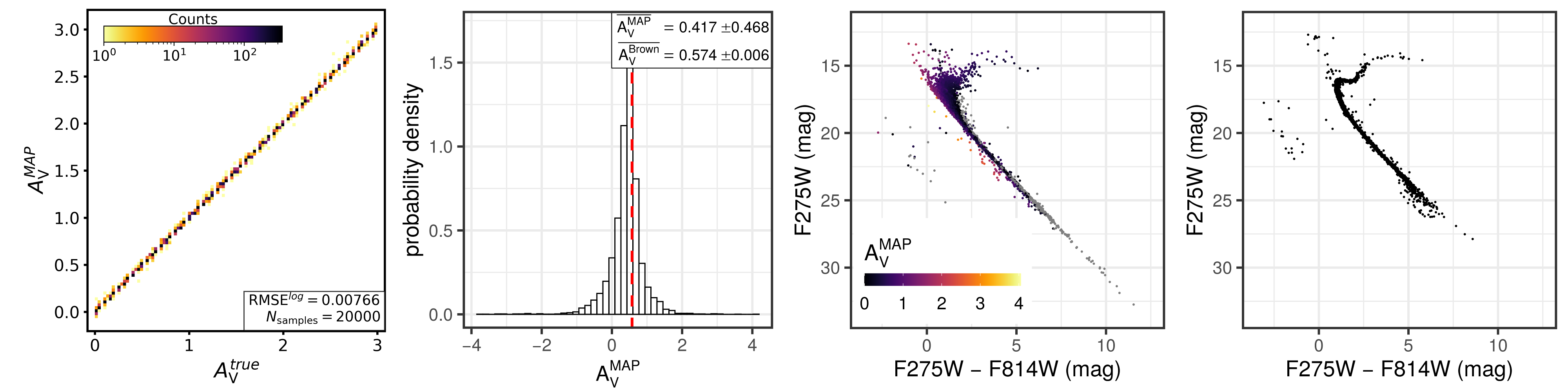}
            \caption{The \textbf{first} diagram shows the MAP estimates of extinction against the true values for 20,000 test observations as predicted by a cINN trained on 'NGC6397\_I'. The \textbf{second} panel shows a histogram of the predicted $A_\mathrm{V}$ MAP estimates of the NGC\,6397 HST data in comparison to the known mean extinction \citep[red dashed line;][]{Brown2018}. The \textbf{third} and \textbf{fourth} panel show extinction corrected UV-I CMDs of the NGC\,6397 HST data, once corrected via the MAP extinction estimates and once by sampling from $A_\mathrm{V} = 0.5735 \pm 0.0062$ \citep[as suggested by][]{Brown2018}. In the third panel the stars are additionally colour coded according to $A_\mathrm{V}^\mathrm{MAP}$ and the grey points indicate stars for which the MAP estimate is an unphysical value below 0.}
            \label{fig:NGC6397_I_Av_prediction}
        \end{figure*}
        
        As mentioned in the introduction we keep the regression problem as simple as possible for this study. Therefore, we adopt a single value for the metallicity of each cluster and assume that individual stellar extinctions are known. Nevertheless, given the way our training sets are constructed we can easily move the extinction from the observable to the physical parameter space. Because of that, and also in view of future development of this method, we perform one 'feasibility' test for both 'Wd2\_I' and 'NGC6397\_I' where the cINN trains to predict extinction instead of taking it as an input. 
        
        Without further modifications to our approach, the prediction of extinction does not work very well for 'Wd2\_I'. Not only does the predicted extinction on the synthetic test set exhibit a large RMSE ($\sim 1.7\,\mathrm{mag}$) for the point estimates, the prediction of the remaining parameters also suffers greatly (e.g. age RMSE increases to 1.3 dex). 
        
        In the 'NGC6397\_I' case, however, we find that the cINN can easily predict the extinction. Here the RMSE of the point estimates is only 0.008 mag (first panel in Figure \ref{fig:NGC6397_I_Av_prediction}), and there is no significant degradation of the predictive capabilities for the other parameters. Part of the failure of the Wd2 model can likely be attributed to the much larger extinction range, 12\,mag, adopted for 'Wd2\_I' training set in comparison to the only 3\, mag range for the 'NGC6397\_I' case. Another possibility, however, could be that the 'Wd2\_I' training set only uses two photometric filters, too few to properly constrain the extinction. A more optimised architecture of the cINN may be required in these cases. 
        
        Given this promising outcome for the extinction prediction of the 'NGC6397\_I' cINN model on the synthetic data, we further evaluate its performance by predicting the {\sl known} extinction of the real NGC\,6397 data. The results are shown in panels 2-4 of Figure \ref{fig:NGC6397_I_Av_prediction}. The histogram of the MAP estimates in panel 2 shows that the prediction for $A_V$ is fairly accurate, the mean being even within the narrow three sigma range determined by \cite{Brown2018}. There are a few cases where the cINN predicts an un-physical negative extinction value. As we have not enforced the extinction value to be positive during training, by e.g. taking its logarithm, this could easily be remedied in further optimization. The third and fourth panel show the predicted extinction corrected CMD of NGC\,6397 (colour-coded according to $A_\mathrm{V}^\mathrm{MAP}$) and the extinction corrected CMD retrieved by randomly sampling the extinction values of \cite{Brown2018}. The overall shape matches fairly well, especially the turn-off points are in good agreement. The cases where the cINN prediction fails (grey points in the third panel) are mostly related to the issue with the PARSEC isochrones, that we have discussed in Section \ref{sec:NGC6397_prediction}.
        
        Nevertheless, this test shows that extinction prediction is very well within the capability of the cINN method, at least if enough photometric filters are available as features, the considered extinction range is small enough, or a combination of both.
    
    \subsection{Other}
        Beside the prediction of extinction there are several physical effects, such as variability or photometric uncertainties, that provide room for extensions to our approach. While the latter can be taken into account using a weighted sampling strategy within the uncertainties of the observational data, developing an intrinsic uncertainty propagation mechanism would be a powerful extension to the approach. 
        
        In this paper we have also presented one approach to incorporating prior knowledge (specifically metallicity and age) into our method by curating training sets accordingly. Taking the young cluster Westerlund2 as an example, i.e. comparing the results of 'Wd2\_I' and 'Wd2\_II', this procedure does not necessarily change the cINN outcome. The alternative of treating prior knowledge intrinsically rather than through training set modification is another possible extension that could benefit the approach. A possible way to do so could be through modification of the target distribution of one or more latent variables. Within the approach we have presented, using Gaussian distributions as targets is the simplest choice, but an arbitrary one. In principle any distribution (for which a log-likelihood can be defined) can serve as the target during training of the cINN. Therefore, we plan to investigate if setting prior knowledge, e.g. a distribution of plausible ages for a given cluster, as the target distribution of one of the latent variables will have the desired effect of incorporating additional prior information more effectively. 
        
\section{Summary and Conclusions}
    \label{sec:conclusion}
    In this introductory paper we present the first application of a novel invertible neural network approach to the task of predicting physical parameters for individual stars based on photometric observations. In many such inverse regression problems the mapping from the physical parameters of interest $\mathbf{x}$ to the associated observables $\mathbf{y}$ is subject to an inherent information loss that induces degeneracies as $\mathbf{y}$ no longer captures all variance of $\mathbf{x}$. To retain this information otherwise lost the conditional invertible neural network (cINN) encodes all variance of $\mathbf{x}$ that is not covered by $\mathbf{y}$ in latent (not observable) variables $\mathbf{z}$ by learning a mapping from $\mathbf{x}$ to $\mathbf{z}$ conditioned on $\mathbf{y}$. Due to the invertible architecture of this network, after learning this forward mapping it automatically provides a solution for the inverse mapping $\mathbf{x} = g(\mathbf{z};\mathbf{y})$, and by sampling the latent variables $\mathbf{z}$ one obtains estimates for the full posterior distributions $p(\mathbf{x}|\mathbf{y})$ of interest. 
    
    We introduce cINNs to the analysis of photometric data in this pilot study by training and testing on synthetic data from the PARSEC stellar evolutionary models \citep{Bressan2012} and performing a benchmark analysis on real observational data obtained by the Hubble Space Telescope for the young cluster Westerlund 2 and the old globular cluster NGC\,6397. These clusters are chosen to cover the extremes of the cluster range, i.e. very young and very old, in order to gain first insights into the systematics of our approach, but not to conduct an exhaustive analysis of the whole spectrum of possible cluster parameters. We construct the synthetic training sets by adopting isochrone model tables of the correct metallicity for Westerlund 2 and NGC\,6397, respectively, with the aim to predict age, initial and current mass, luminosity, effective temperature and surface gravity of each cluster star. To overcome sampling issues in the $M_\mathrm{ini}$ and age spaces of the isochrone tables, we first oversample each individual isochrone using a spline interpolation and then extract sample points to fill up underpopulated areas within the parameter space. To simplify this regression problem we use extinction as an observable parameter, in addition to the available photometry. To account for extinction within the synthetic training set, assuming every isochrone model point to be a synthetic star, we add multiple examples of the same star with different amounts of extinction to the training sets.
    
    In order to evaluate how the cINN prediction on the real data behaves when we include prior knowledge about the age of the respective clusters, we construct two training sets for each cluster, one encompassing the entire age range of our theoretical models, from $\log(\mathrm{age/yr}) = 5$ to $10.13$, the other with a reduced age range close to the actual cluster age ($\log(\mathrm{age/yr}) = 5$ to 8 for Wd2 and 9 to 10.13 for NGC\,6397). To derive point estimates from the predicted posterior distributions we use kernel density estimation with bandwidths determined according to Silverman's rule of thumb. In this way we find the most likely values of physical parameters (maximum a posteriori estimates) in the marginalised distributions. We ascertain the training performance of our four models on a test set of 20.000 random synthetic observations that are excluded for the training process. Using this sample we determine for each parameter the median calibration error, the uncertainty at 68\% confidence as well as the standard and normalized root mean squared error (RMSE/NRMSE) between the maximum a posteriori (MAP) parameter estimates and the known true parameter values. Using a simple nearest neighbour approach on the training data we also approximate a re-simulation error for the predicted posteriors on the synthetic test set. Furthermore, to ascertain how well our models generalise to new populations, we test our models on synthetic data from two different stellar evolution models, namely isochrone tables from MIST \citep{Dotter2016_MIST, Choi2016_MIST} and Dartmouth \citep{Dotter2008_Dartmouth}. \\
    Our main results from the tests on \textit{synthetic data} are the following:
    \begin{itemize}
        \item[i)] Once trained ($\sim 2\,\mathrm{h}$) the cINN can rapidly predict a posterior distribution for a single star. Using GPU acceleration on a Nvidia GTX 1080 the cINN can predict about 35 posterior distributions with 4096 samples per second.
        \item[ii)] On the synthetic test data the prediction of initial/current mass, luminosity, effective temperature and surface gravity works extraordinarily well with posterior distributions that are narrowly constrained around the true values and low RMSEs of the derived MAP parameter estimates. 
        \item[iii)] Predicting the stellar age is a more difficult task. The predicted posteriors tend to be broader and often exhibit multi-modalities, revealing ample degeneracies in the age prediction. While we can confirm that the true value is part of the predicted distribution in more than 99\% of the cases, there are several instances where the true solution does not coincide with the most likely outcome of the posterior, falling into a second peak instead. In itself this is not problematic as a true posterior describes all possible parameters that could explain a given observation, such that the most likely prediction does not have to be the one that generated the given observation. However, as a consequence we find significantly more cases in the age prediction where our point estimates deviate from the true value. The intermediate age range from $\log(\mathrm{age/yr}) = 6.5$ to 8.5 is the least affected, whereas the predictions for very old stars ($>1\,\mathrm{Gyr}$) show a notable amount of instances where the MAP estimate is off by $\sim 0.4\,$dex on average. When fewer photometric filters (2 instead of 5) are available, as in the case of our Wd2 training set, we also observe more deviations for the very young stars ($<10\,\mathrm{Myr}$).
        Nevertheless, overall these cases where the MAP estimate deviates strongly from the true value are still a \textit{minority}.
        \item[iv)] Our nearest neighbour re-simulation approximation returns small errors, confirming the validity of the predicted posterior distributions and identified degeneracies.
        \item[v)] The predictive performance of the cINN improves (especially for the age) when more photometric filters are included in the observables. However, even with perfect information (17 photometric filters here) the prediction of age for old stars (1 Gyr and older) still remains highly challenging.
        \item[vi)] Our models generalise overall very well to the synthetic data of other stellar evolution models and perform particularly well in recovering luminosity, surface temperature and gravity. Ages and masses are also predicted fairly accurately for most samples, but tend to exhibit larger errors. Specifically, our models manage to also recover ages of main-sequence objects on these different synthetic test sets, providing confirmation that our models do not simply overfit their training data for these hard to predict cases. However, if there are significant discrepancies between the investigated models and the PARSEC isochrones, such as in our low metallicity Dartmouth test, the age prediction can fail severely. We conclude that our cINN model manages to recover ages for synthetic main-sequence stars through a combination of its latent variable approach and the perfect synthetic photometry. For real main-sequence observations, however, we suggest to treat the predicted ages with caution.
        
    \end{itemize}
    Applied to \textit{observed data} of Westerlund 2 and NGC\,6397 we find:
    \begin{itemize}
        \item[vii)] The cINN predictions based on the HST data of Westerlund 2 return excellent results. With a median of $1.27_{-0.94}^{+3.62}\,\mathrm{Myr} $ of the age MAP estimates and a most likely value of $1.04_{-0.90}^{+8.48}\,\mathrm{Myr} $ from the sum of all age posteriors, the cINN results are in good accordance with the previously determined age of Westerlund 2 of $1.04 \pm 0.72\,\mathrm{Myr}$ \citep{Zeidler2016}. Furthermore, the cINN correctly recognises that stars located on the turn-on could potentially also be RGB stars and thus returns multi-modal age posterior distributions, highlighting this degeneracy. 
        \item[viii)] Based on the cINN mass estimates, we are able to construct the IMF of Wd2 and fit its high-mass slope. We find a value of $\alpha = 2.39\pm 0.2$, which corresponds to the Salpeter slope within one sigma and is in accordance with the previously determined slope of the present day mass function of $\alpha = 2.53 \pm 0.05$ \citep{Zeidler2017}. We also find evidence for mass segregation based on the mass segregation ratios $\Lambda_\mathrm{MSR}$ and $\Gamma_\mathrm{MSR}$ of individual stars, again confirming previous results by \cite{Zeidler2017}.
        \item[ix)] For NGC\,6397 the cINN predictions are not as good as for Westerlund 2. While certain properties are recovered, such as the predicted Hertzsprung-Russel diagram which traces the isochrone corresponding to the known clusters age relatively well, there are glaring issues with the prediction of the cluster age. The majority of stars is predicted to be much younger than the actual cluster age, while stars that are correctly identified as old tend to have an overestimated age. We identify the culprit for these unsatisfying results in the PARSEC evolutionary models, as they do not fit the observations of this globular cluster well enough. 
        This example highlights that the careful selection of the underlying physical model is of utmost importance for our approach.
        \item[x)] When enough photometric filters are available (e.g. 5 in the case of NGC\,6397) the cINN can also predict extinction very well, instead of using it as an input observable, without losing accuracy in the prediction of the other physical parameters.
    \end{itemize}
        Overall the results presented in this paper demonstrate that the cINN is a very powerful approach that can solve the problem of predicting physical parameters from photometry data if the underlying physical models are selected carefully to match the observations. In other words, the possibility of solving the inverse problems (from observations to the physical parameters of each star) relies heavily on the quality of the forward modeling (from physical parameters to synthetic observations). In the case of Westerlund 2 we correctly recover the main cluster properties using \textit{only 2} photometric filters and an estimate of stellar extinction as an input. The cINN method can successfully learn and highlight degeneracies that appear within the given problem, making it an excellent tool for tasks that are subject to degenerate mappings from physical to observable parameter space. 
        
        Given its excellent prediction efficiency we believe that the cINN approach could become a very valuable tool in the big data epoch of astronomy. In particular current and future all-sky/very-wide-field surveys like Pan-Starrs as well as upcoming observational facilities such as the Vera Rubin Observatory (LSST) or the Roman Space Telescope (formerly WFIRST) will provide enormous amounts of data, for which efficient and robust deep learning approaches, such as the cINN, will truly be able to show their strength. With this in mind we plan to employ the cINN approach on data from large HST surveys, such as the Hubble Tarantula Treasury Project or the Measuring Young Stars in Space and Time survey, to characterise more complex stellar populations in a subsequent study.
        
        In this paper we purposefully keep the regression problem as simple as possible, in particular we only consider the single metallicity case, using extinction as an input parameter and ignore photometric errors. We plan to address these effects, together with variability and binarity, in future studies. As demonstrated for the example of the old globular cluster NGC~6397, predicting extinction is already well within the capacity of the cINN. However, it might require some architecture optimization to support observations with a low number of filters or regions with a large range of differential extinction. Photometric errors can be taken into account to some degree at this stage already, at least at the prediction stage, by simply re-sampling the observations according to their errors and performing a weighted addition of the resulting posterior distributions. In the future we also plan to investigate if some intrinsic treatment is possible at the training stage of the network, e.g. by incorporating uncertainties in the training set. Another avenue that we aim to pursue is the possibility of considering prior knowledge as part of the training strategy rather than incorporating it in the training set. Ultimately our goal is to provide observers with a robust, efficient and general tool to analyse observations and retrieve the key physical parameters of their targets.

\section*{Acknowledgements}

VFK was funded by the Heidelberg Graduate School of Mathematical and Computational Methods for the Sciences (HGS MathComp), founded by DFG grant GSC 220 in the German Universities Excellence Initiative. VFK also acknowledges support from the International Max Planck Research School for Astronomy and Cosmic Physics at the University of Heidelberg (IMPRS-HD).\\
RSK acknowledges financial support from the German Research Foundation (DFG) via the collaborative research centre (SFB 881, Project-ID 138713538) 'The Milky Way System' (subprojects A1, B1, B2, and B8). He also thanks for funding from the Heidelberg Cluster of Excellence STRUCTURES in the framework of Germany's Excellence Strategy (grant EXC-2181/1 - 390900948) and for funding from the European Research Council via the ERC Synergy Grant ECOGAL (grant 855130) and the ERC Advanced Grant STARLIGHT (grant 339177).\\
P.Z. acknowledges support by the Forschungsstipendium (ZE 1159/1-1) of the German Research Foundation, particularly via the project 398719443.

\section*{Data Availability}
The HUGS data of N6397 used in this article are available in the Mikulski Archive for Space Telescopes, at http://dx.doi.org/10.17909/T9810F.\\
The HST data of Westerlund 2 underlying this article were provided by Elena Sabbi by permission. Data will be shared on request to the corresponding author with permission of Elena Sabbi.\\
The data outcomes underlying this article will be shared on reasonable request to the corresponding author.



\bibliographystyle{mnras}
\bibliography{Ksoll_StellarParamINN_references}{} 

\begin{thebibliography}{}
\makeatletter
\relax
\def\mn@urlcharsother{\let\do\@makeother \do\$\do\&\do\#\do\^\do\_\do\%\do\~}
\def\mn@doi{\begingroup\mn@urlcharsother \@ifnextchar [ {\mn@doi@}
  {\mn@doi@[]}}
\def\mn@doi@[#1]#2{\def\@tempa{#1}\ifx\@tempa\@empty \href
  {http://dx.doi.org/#2} {doi:#2}\else \href {http://dx.doi.org/#2} {#1}\fi
  \endgroup}
\def\mn@eprint#1#2{\mn@eprint@#1:#2::\@nil}
\def\mn@eprint@arXiv#1{\href {http://arxiv.org/abs/#1} {{\tt arXiv:#1}}}
\def\mn@eprint@dblp#1{\href {http://dblp.uni-trier.de/rec/bibtex/#1.xml}
  {dblp:#1}}
\def\mn@eprint@#1:#2:#3:#4\@nil{\def\@tempa {#1}\def\@tempb {#2}\def\@tempc
  {#3}\ifx \@tempc \@empty \let \@tempc \@tempb \let \@tempb \@tempa \fi \ifx
  \@tempb \@empty \def\@tempb {arXiv}\fi \@ifundefined
  {mn@eprint@\@tempb}{\@tempb:\@tempc}{\expandafter \expandafter \csname
  mn@eprint@\@tempb\endcsname \expandafter{\@tempc}}}

\bibitem[\protect\citeauthoryear{{Allison}, {Goodwin}, {Parker}, {Portegies
  Zwart}, {de Grijs}  \& {Kouwenhoven}}{{Allison} et~al.}{2009}]{Allison2009}
{Allison} R.~J.,  {Goodwin} S.~P.,  {Parker} R.~J.,  {Portegies Zwart} S.~F.,
  {de Grijs} R.,   {Kouwenhoven} M.~B.~N.,  2009, \mn@doi [\mnras]
  {10.1111/j.1365-2966.2009.14508.x}, \href
  {https://ui.adsabs.harvard.edu/abs/2009MNRAS.395.1449A} {395, 1449}

\bibitem[\protect\citeauthoryear{{Anthony-Twarog}, {Twarog}  \&
  {Suntzeff}}{{Anthony-Twarog} et~al.}{1992}]{AnthonyTwarog1992}
{Anthony-Twarog} B.~J.,  {Twarog} B.~A.,   {Suntzeff} N.~B.,  1992, \mn@doi
  [\aj] {10.1086/116140}, \href
  {https://ui.adsabs.harvard.edu/abs/1992AJ....103.1264A} {103, 1264}

\bibitem[\protect\citeauthoryear{Ardizzone, Kruse, Rother  \& Köthe}{Ardizzone
  et~al.}{2019a}]{Ardizzone2019a}
Ardizzone L.,  Kruse J.,  Rother C.,   Köthe U.,  2019a, in International
  Conference on Learning Representations. \url
  {https://openreview.net/forum?id=rJed6j0cKX}

\bibitem[\protect\citeauthoryear{Ardizzone, L{\"{u}}th, Kruse, Rother  \&
  K{\"{o}}the}{Ardizzone et~al.}{2019b}]{Ardizzone2019b}
Ardizzone L.,  L{\"{u}}th C.,  Kruse J.,  Rother C.,   K{\"{o}}the U.,  2019b,
  CoRR, abs/1907.02392

\bibitem[\protect\citeauthoryear{{Ascenso}, {Alves}, {Beletsky}  \&
  {Lago}}{{Ascenso} et~al.}{2007}]{Ascenso2007}
{Ascenso} J.,  {Alves} J.,  {Beletsky} Y.,   {Lago} M.~T.~V.~T.,  2007, \mn@doi
  [\aap] {10.1051/0004-6361:20066433}, \href
  {https://ui.adsabs.harvard.edu/abs/2007A&A...466..137A} {466, 137}

\bibitem[\protect\citeauthoryear{Bellinger, Angelou, Hekker, Basu, Ball  \&
  Guggenberger}{Bellinger et~al.}{2016}]{Bellinger2016}
Bellinger E.~P.,  Angelou G.~C.,  Hekker S.,  Basu S.,  Ball W.~H.,
  Guggenberger E.,  2016, \mn@doi [The Astrophysical Journal]
  {10.3847/0004-637x/830/1/31}, 830, 31

\bibitem[\protect\citeauthoryear{{Bressan}, {Marigo}, {Girardi}, {Salasnich},
  {Dal Cero}, {Rubele}  \& {Nanni}}{{Bressan} et~al.}{2012}]{Bressan2012}
{Bressan} A.,  {Marigo} P.,  {Girardi} L.,  {Salasnich} B.,  {Dal Cero} C.,
  {Rubele} S.,   {Nanni} A.,  2012, \mn@doi [\mnras]
  {10.1111/j.1365-2966.2012.21948.x}, \href
  {https://ui.adsabs.harvard.edu/abs/2012MNRAS.427..127B} {427, 127}

\bibitem[\protect\citeauthoryear{{Brown}, {Casertano}, {Strader}, {Riess},
  {VandenBerg}, {Soderblom}, {Kalirai}  \& {Salinas}}{{Brown}
  et~al.}{2018}]{Brown2018}
{Brown} T.~M.,  {Casertano} S.,  {Strader} J.,  {Riess} A.,  {VandenBerg}
  D.~A.,  {Soderblom} D.~R.,  {Kalirai} J.,   {Salinas} R.,  2018, \mn@doi
  [\apjl] {10.3847/2041-8213/aab55a}, \href
  {https://ui.adsabs.harvard.edu/abs/2018ApJ...856L...6B} {856, L6}

\bibitem[\protect\citeauthoryear{{Cantat-Gaudin} et~al.,}{{Cantat-Gaudin}
  et~al.}{2020}]{2020A&A...640A...1C}
{Cantat-Gaudin} T.,  et~al., 2020, \mn@doi [\aap]
  {10.1051/0004-6361/202038192}, \href
  {https://ui.adsabs.harvard.edu/abs/2020A&A...640A...1C} {640, A1}

\bibitem[\protect\citeauthoryear{{Cardelli}, {Clayton}  \& {Mathis}}{{Cardelli}
  et~al.}{1989}]{Cardelli1989}
{Cardelli} J.~A.,  {Clayton} G.~C.,   {Mathis} J.~S.,  1989, \mn@doi [\apj]
  {10.1086/167900}, \href
  {https://ui.adsabs.harvard.edu/abs/1989ApJ...345..245C} {345, 245}

\bibitem[\protect\citeauthoryear{{Carraro}, {Turner}, {Majaess}  \&
  {Baume}}{{Carraro} et~al.}{2013}]{Carraro2013}
{Carraro} G.,  {Turner} D.,  {Majaess} D.,   {Baume} G.,  2013, \mn@doi [\aap]
  {10.1051/0004-6361/201321421}, \href
  {https://ui.adsabs.harvard.edu/abs/2013A&A...555A..50C} {555, A50}

\bibitem[\protect\citeauthoryear{{Chabrier}}{{Chabrier}}{2002}]{Chabrier2002}
{Chabrier} G.,  2002, \mn@doi [\apj] {10.1086/324716}, \href
  {https://ui.adsabs.harvard.edu/abs/2002ApJ...567..304C} {567, 304}

\bibitem[\protect\citeauthoryear{{Chen}, {Girardi}, {Bressan}, {Marigo},
  {Barbieri}  \& {Kong}}{{Chen} et~al.}{2014}]{Chen2014}
{Chen} Y.,  {Girardi} L.,  {Bressan} A.,  {Marigo} P.,  {Barbieri} M.,   {Kong}
  X.,  2014, \mn@doi [\mnras] {10.1093/mnras/stu1605}, \href
  {https://ui.adsabs.harvard.edu/abs/2014MNRAS.444.2525C} {444, 2525}

\bibitem[\protect\citeauthoryear{{Chen}, {Bressan}, {Girardi}, {Marigo}, {Kong}
   \& {Lanza}}{{Chen} et~al.}{2015}]{Chen2015}
{Chen} Y.,  {Bressan} A.,  {Girardi} L.,  {Marigo} P.,  {Kong} X.,   {Lanza}
  A.,  2015, \mn@doi [\mnras] {10.1093/mnras/stv1281}, \href
  {https://ui.adsabs.harvard.edu/abs/2015MNRAS.452.1068C} {452, 1068}

\bibitem[\protect\citeauthoryear{{Choi}, {Dotter}, {Conroy}, {Cantiello},
  {Paxton}  \& {Johnson}}{{Choi} et~al.}{2016}]{Choi2016_MIST}
{Choi} J.,  {Dotter} A.,  {Conroy} C.,  {Cantiello} M.,  {Paxton} B.,
  {Johnson} B.~D.,  2016, \mn@doi [\apj] {10.3847/0004-637X/823/2/102}, \href
  {https://ui.adsabs.harvard.edu/abs/2016ApJ...823..102C} {823, 102}

\bibitem[\protect\citeauthoryear{{Da Rio}, Gouliermis  \& Gennaro}{{Da Rio}
  et~al.}{2010}]{DaRio2010}
{Da Rio} N.,  Gouliermis D.~A.,   Gennaro M.,  2010, \mn@doi [The Astrophysical
  Journal] {10.1088/0004-637x/723/1/166}, 723, 166

\bibitem[\protect\citeauthoryear{{Dib}, {Schmeja}  \& {Parker}}{{Dib}
  et~al.}{2018}]{Dib2018}
{Dib} S.,  {Schmeja} S.,   {Parker} R.~J.,  2018, \mn@doi [\mnras]
  {10.1093/mnras/stx2413}, \href
  {https://ui.adsabs.harvard.edu/abs/2018MNRAS.473..849D} {473, 849}

\bibitem[\protect\citeauthoryear{{Dinh}, {Sohl-Dickstein}  \& {Bengio}}{{Dinh}
  et~al.}{2016}]{Dinh2016}
{Dinh} L.,  {Sohl-Dickstein} J.,   {Bengio} S.,  2016, arXiv e-prints, \href
  {https://ui.adsabs.harvard.edu/abs/2016arXiv160508803D} {p. arXiv:1605.08803}

\bibitem[\protect\citeauthoryear{{Dotter}}{{Dotter}}{2016}]{Dotter2016_MIST}
{Dotter} A.,  2016, \mn@doi [\apjs] {10.3847/0067-0049/222/1/8}, \href
  {https://ui.adsabs.harvard.edu/abs/2016ApJS..222....8D} {222, 8}

\bibitem[\protect\citeauthoryear{{Dotter}, {Chaboyer}, {Jevremovi{\'c}},
  {Baron}, {Ferguson}, {Sarajedini}  \& {Anderson}}{{Dotter}
  et~al.}{2007}]{Dotter2007_Dartmouth}
{Dotter} A.,  {Chaboyer} B.,  {Jevremovi{\'c}} D.,  {Baron} E.,  {Ferguson}
  J.~W.,  {Sarajedini} A.,   {Anderson} J.,  2007, \mn@doi [\aj]
  {10.1086/517915}, \href
  {https://ui.adsabs.harvard.edu/abs/2007AJ....134..376D} {134, 376}

\bibitem[\protect\citeauthoryear{{Dotter}, {Chaboyer}, {Jevremovi{\'c}},
  {Kostov}, {Baron}  \& {Ferguson}}{{Dotter}
  et~al.}{2008}]{Dotter2008_Dartmouth}
{Dotter} A.,  {Chaboyer} B.,  {Jevremovi{\'c}} D.,  {Kostov} V.,  {Baron} E.,
  {Ferguson} J.~W.,  2008, \mn@doi [\apjs] {10.1086/589654}, \href
  {https://ui.adsabs.harvard.edu/abs/2008ApJS..178...89D} {178, 89}

\bibitem[\protect\citeauthoryear{Feigelson \& Babu}{Feigelson \&
  Babu}{2012}]{FeigelsonBabu2012}
Feigelson E.~D.,  Babu G.~J.,  2012, Modern Statistical Methods for Astronomy:
  With R Applications.
Cambridge University Press, \mn@doi{10.1017/CBO9781139015653}

\bibitem[\protect\citeauthoryear{Goodfellow, Bengio  \& Courville}{Goodfellow
  et~al.}{2016}]{GoodBengCour16}
Goodfellow I.~J.,  Bengio Y.,   Courville A.,  2016, Deep Learning.
MIT Press, Cambridge, MA, USA

\bibitem[\protect\citeauthoryear{{Gratton}, {Bragaglia}, {Carretta},
  {Clementini}, {Desidera}, {Grundahl}  \& {Lucatello}}{{Gratton}
  et~al.}{2003}]{Gratton2003}
{Gratton} R.~G.,  {Bragaglia} A.,  {Carretta} E.,  {Clementini} G.,  {Desidera}
  S.,  {Grundahl} F.,   {Lucatello} S.,  2003, \mn@doi [\aap]
  {10.1051/0004-6361:20031003}, \href
  {https://ui.adsabs.harvard.edu/abs/2003A&A...408..529G} {408, 529}

\bibitem[\protect\citeauthoryear{Hastie, Tibshirani  \& Friedman}{Hastie
  et~al.}{2009}]{Hastie2009}
Hastie T.,  Tibshirani R.,   Friedman J.,  2009, The elements of statistical
  learning: data mining, inference and prediction, 2 edn.
Springer, \url {http://www-stat.stanford.edu/~tibs/ElemStatLearn/}

\bibitem[\protect\citeauthoryear{Hyvärinen \& Oja}{Hyvärinen \&
  Oja}{2000}]{Hyvarinen2000}
Hyvärinen A.,  Oja E.,  2000, \mn@doi [Neural Networks]
  {https://doi.org/10.1016/S0893-6080(00)00026-5}, 13, 411

\bibitem[\protect\citeauthoryear{Ivezic, Connolly, VanderPlas  \& Gray}{Ivezic
  et~al.}{2014}]{Ivezic2014}
Ivezic Z.,  Connolly A.~J.,  VanderPlas J.~T.,   Gray A.,  2014, Statistics,
  Data Mining, and Machine Learning in Astronomy: A Practical Python Guide for
  the Analysis of Survey Data.
Princeton University Press

\bibitem[\protect\citeauthoryear{{Jackson}, {Deliyannis}  \&
  {Jeffries}}{{Jackson} et~al.}{2018}]{2018MNRAS.476.3245J}
{Jackson} R.~J.,  {Deliyannis} C.~P.,   {Jeffries} R.~D.,  2018, \mn@doi
  [\mnras] {10.1093/mnras/sty374}, \href
  {https://ui.adsabs.harvard.edu/abs/2018MNRAS.476.3245J} {476, 3245}

\bibitem[\protect\citeauthoryear{{J\o{}rgensen, B. R.} \& {Lindegren,
  L.}}{{J\o{}rgensen, B. R.} \& {Lindegren, L.}}{2005}]{JorgensenLindegren2005}
{J\o{}rgensen, B. R.} {Lindegren, L.} 2005, \mn@doi [A\&A]
  {10.1051/0004-6361:20042185}, 436, 127

\bibitem[\protect\citeauthoryear{{Kingma} \& {Dhariwal}}{{Kingma} \&
  {Dhariwal}}{2018}]{Kingma2018}
{Kingma} D.~P.,  {Dhariwal} P.,  2018, arXiv e-prints, \href
  {https://ui.adsabs.harvard.edu/abs/2018arXiv180703039K} {p. arXiv:1807.03039}

\bibitem[\protect\citeauthoryear{{Kounkel}, {Covey}  \& {Stassun}}{{Kounkel}
  et~al.}{2020}]{2020arXiv200407261K}
{Kounkel} M.,  {Covey} K.,   {Stassun} K.~G.,  2020, arXiv e-prints, \href
  {https://ui.adsabs.harvard.edu/abs/2020arXiv200407261K} {p. arXiv:2004.07261}

\bibitem[\protect\citeauthoryear{{Kraft} \& {Ivans}}{{Kraft} \&
  {Ivans}}{2003}]{Kraft2003}
{Kraft} R.~P.,  {Ivans} I.~I.,  2003, \mn@doi [\pasp] {10.1086/345914}, \href
  {https://ui.adsabs.harvard.edu/abs/2003PASP..115..143K} {115, 143}

\bibitem[\protect\citeauthoryear{Miller et~al.,}{Miller
  et~al.}{2015}]{Miller2015}
Miller A.~A.,  et~al., 2015, \mn@doi [The Astrophysical Journal]
  {10.1088/0004-637x/798/2/122}, 798, 122

\bibitem[\protect\citeauthoryear{{Nardiello} et~al.,}{{Nardiello}
  et~al.}{2018}]{Nardiello2018_HUGS}
{Nardiello} D.,  et~al., 2018, \mn@doi [\mnras] {10.1093/mnras/sty2515}, \href
  {https://ui.adsabs.harvard.edu/abs/2018MNRAS.481.3382N} {481, 3382}

\bibitem[\protect\citeauthoryear{{Olczak}, {Spurzem}  \& {Henning}}{{Olczak}
  et~al.}{2011}]{Olczak2011}
{Olczak} C.,  {Spurzem} R.,   {Henning} T.,  2011, \mn@doi [\aap]
  {10.1051/0004-6361/201116902}, \href
  {https://ui.adsabs.harvard.edu/abs/2011A&A...532A.119O} {532, A119}

\bibitem[\protect\citeauthoryear{{Olney} et~al.,}{{Olney}
  et~al.}{2020}]{2020AJ....159..182O}
{Olney} R.,  et~al., 2020, \mn@doi [\aj] {10.3847/1538-3881/ab7a97}, \href
  {https://ui.adsabs.harvard.edu/abs/2020AJ....159..182O} {159, 182}

\bibitem[\protect\citeauthoryear{Paszke et~al.,}{Paszke
  et~al.}{2017}]{Paszke2017_pytorch}
Paszke A.,  et~al., 2017, in NIPS Autodiff Workshop.

\bibitem[\protect\citeauthoryear{{Paxton}, {Bildsten}, {Dotter}, {Herwig},
  {Lesaffre}  \& {Timmes}}{{Paxton} et~al.}{2011}]{Paxton2011_MESA}
{Paxton} B.,  {Bildsten} L.,  {Dotter} A.,  {Herwig} F.,  {Lesaffre} P.,
  {Timmes} F.,  2011, \mn@doi [\apjs] {10.1088/0067-0049/192/1/3}, \href
  {https://ui.adsabs.harvard.edu/abs/2011ApJS..192....3P} {192, 3}

\bibitem[\protect\citeauthoryear{{Paxton} et~al.,}{{Paxton}
  et~al.}{2013}]{Paxton2013_MESA}
{Paxton} B.,  et~al., 2013, \mn@doi [\apjs] {10.1088/0067-0049/208/1/4}, \href
  {https://ui.adsabs.harvard.edu/abs/2013ApJS..208....4P} {208, 4}

\bibitem[\protect\citeauthoryear{{Paxton} et~al.,}{{Paxton}
  et~al.}{2015}]{Paxton2015_MESA}
{Paxton} B.,  et~al., 2015, \mn@doi [\apjs] {10.1088/0067-0049/220/1/15}, \href
  {https://ui.adsabs.harvard.edu/abs/2015ApJS..220...15P} {220, 15}

\bibitem[\protect\citeauthoryear{{Piotto} et~al.,}{{Piotto}
  et~al.}{2015}]{Piotto2015_HUGS}
{Piotto} G.,  et~al., 2015, \mn@doi [\aj] {10.1088/0004-6256/149/3/91}, \href
  {https://ui.adsabs.harvard.edu/abs/2015AJ....149...91P} {149, 91}

\bibitem[\protect\citeauthoryear{{Sabbi} et~al.,}{{Sabbi}
  et~al.}{2020}]{Sabbi2020}
{Sabbi} E.,  et~al., 2020, \mn@doi [\apj] {10.3847/1538-4357/ab7372}, \href
  {https://ui.adsabs.harvard.edu/abs/2020ApJ...891..182S} {891, 182}

\bibitem[\protect\citeauthoryear{{Schlegel}, {Finkbeiner}  \&
  {Davis}}{{Schlegel} et~al.}{1998}]{Schlegel1998}
{Schlegel} D.~J.,  {Finkbeiner} D.~P.,   {Davis} M.,  1998, \mn@doi [\apj]
  {10.1086/305772}, \href
  {https://ui.adsabs.harvard.edu/abs/1998ApJ...500..525S} {500, 525}

\bibitem[\protect\citeauthoryear{{Sharma}, {Kembhavi}, {Kembhavi}, {Sivarani},
  {Abraham}  \& {Vaghmare}}{{Sharma} et~al.}{2020}]{2020MNRAS.491.2280S}
{Sharma} K.,  {Kembhavi} A.,  {Kembhavi} A.,  {Sivarani} T.,  {Abraham} S.,
  {Vaghmare} K.,  2020, \mn@doi [\mnras] {10.1093/mnras/stz3100}, \href
  {https://ui.adsabs.harvard.edu/abs/2020MNRAS.491.2280S} {491, 2280}

\bibitem[\protect\citeauthoryear{{Silverman}}{{Silverman}}{1986}]{Silverman1986}
{Silverman} B.~W.,  1986, {Density estimation for statistics and data
  analysis}.
Chapman and Hall

\bibitem[\protect\citeauthoryear{{Tang}, {Bressan}, {Rosenfield}, {Slemer},
  {Marigo}, {Girardi}  \& {Bianchi}}{{Tang} et~al.}{2014}]{Tang2014}
{Tang} J.,  {Bressan} A.,  {Rosenfield} P.,  {Slemer} A.,  {Marigo} P.,
  {Girardi} L.,   {Bianchi} L.,  2014, \mn@doi [\mnras]
  {10.1093/mnras/stu2029}, \href
  {https://ui.adsabs.harvard.edu/abs/2014MNRAS.445.4287T} {445, 4287}

\bibitem[\protect\citeauthoryear{Valls-Gabaud}{Valls-Gabaud}{2014}]{VallsGabaud2014}
Valls-Gabaud D.,  2014, \mn@doi [European Astronomical Society Publications
  Series] {10.1051/eas/1465006}, 65, 225–265

\bibitem[\protect\citeauthoryear{{Vargas {\'A}lvarez}, {Kobulnicky}, {Bradley},
  {Kannappan}, {Norris}, {Cool}  \& {Miller}}{{Vargas {\'A}lvarez}
  et~al.}{2013}]{VargasAlvarez2013}
{Vargas {\'A}lvarez} C.~A.,  {Kobulnicky} H.~A.,  {Bradley} D.~R.,  {Kannappan}
  S.~J.,  {Norris} M.~A.,  {Cool} R.~J.,   {Miller} B.~P.,  2013, \mn@doi [\aj]
  {10.1088/0004-6256/145/5/125}, \href
  {https://ui.adsabs.harvard.edu/abs/2013AJ....145..125V} {145, 125}

\bibitem[\protect\citeauthoryear{{Vulic}, {Barmby}  \& {Gallagher}}{{Vulic}
  et~al.}{2018}]{Vulic2018}
{Vulic} N.,  {Barmby} P.,   {Gallagher} S.~C.,  2018, \mn@doi [\mnras]
  {10.1093/mnras/stx2626}, \href
  {https://ui.adsabs.harvard.edu/abs/2018MNRAS.473.4900V} {473, 4900}

\bibitem[\protect\citeauthoryear{{Zeidler} et~al.,}{{Zeidler}
  et~al.}{2015}]{Zeidler2015}
{Zeidler} P.,  et~al., 2015, \mn@doi [\aj] {10.1088/0004-6256/150/3/78}, \href
  {https://ui.adsabs.harvard.edu/abs/2015AJ....150...78Z} {150, 78}

\bibitem[\protect\citeauthoryear{{Zeidler}, {Grebel}, {Nota}, {Sabbi},
  {Pasquali}, {Tosi}, {Bonanos}  \& {Christian}}{{Zeidler}
  et~al.}{2016}]{Zeidler2016}
{Zeidler} P.,  {Grebel} E.~K.,  {Nota} A.,  {Sabbi} E.,  {Pasquali} A.,  {Tosi}
  M.,  {Bonanos} A.~Z.,   {Christian} C.,  2016, \mn@doi [\aj]
  {10.3847/0004-6256/152/4/84}, \href
  {https://ui.adsabs.harvard.edu/abs/2016AJ....152...84Z} {152, 84}

\bibitem[\protect\citeauthoryear{{Zeidler} et~al.,}{{Zeidler}
  et~al.}{2017}]{Zeidler2017}
{Zeidler} P.,  et~al., 2017, in {Charbonnel} C.,  {Nota} A.,  eds,  IAU
  Symposium Vol. 316, Formation, Evolution, and Survival of Massive Star
  Clusters. pp 55--60, \mn@doi{10.1017/S1743921315008972}

\makeatother
\end{thebibliography}



\newpage
\appendix

\section{General}
    \label{app:general}
    \begin{table}
        \centering
        \caption{Overview of the $A_\lambda/A_\mathrm{V}$ values derived from the \citep{Cardelli1989} extinction curve according to Eq. \eqref{eq:alambda} for the HST filters used in the Wd2 and NGC\,6397 observations.} 
        \begin{tabular}[width = \linewidth]{clr}
             \hline
             \hline
             & Filter & $\frac{A_\lambda}{A_\mathrm{V}}$ \\
             \hline
             & $\mathrm{F275W}_\mathrm{WFC3}$ &  1.94436 \\
             & $\mathrm{F336W}_\mathrm{WFC3}$ &  1.65798 \\
             $R_\mathrm{V} = 3.1$ & $\mathrm{F438W}_\mathrm{WFC3}$ &  1.33088 \\
             & $\mathrm{F606W}_\mathrm{ACS}$  &  0.92246 \\
             & $\mathrm{F814W}_\mathrm{ACS}$  &  0.60593 \\
             \hline
             & $\mathrm{F218W}_\mathrm{WFC3}$ & 2.53769 \\ 
             & $\mathrm{F225W}_\mathrm{WFC3}$ & 2.21539 \\
             & $\mathrm{F275W}_\mathrm{WFC3}$ & 1.75064 \\
             & $\mathrm{F336W}_\mathrm{WFC3}$ & 1.49531 \\
             & $\mathrm{F390W}_\mathrm{WFC3}$ & 1.39453 \\
             & $\mathrm{F438W}_\mathrm{WFC3}$ & 1.28651 \\
             & $\mathrm{F475W}_\mathrm{WFC3}$ & 1.15971 \\
             & $\mathrm{F555W}_\mathrm{WFC3}$ & 1.03555 \\
             $R_\mathrm{V} = 3.8$ & $\mathrm{F606W}_\mathrm{WFC3}$ & 0.93420 \\
             & $\mathrm{F625W}_\mathrm{WFC3}$ & 0.88084 \\
             & $\mathrm{F775W}_\mathrm{WFC3}$ & 0.67977 \\
             & $\mathrm{F814W}_\mathrm{WFC3}$ & 0.62821 \\
             & $\mathrm{F105W}_\mathrm{WFC3}$ & 0.39924 \\
             & $\mathrm{F110W}_\mathrm{WFC3}$ & 0.34595 \\ 
             & $\mathrm{F125W}_\mathrm{WFC3}$ & 0.30448 \\
             & $\mathrm{F140W}_\mathrm{WFC3}$ & 0.25550 \\
             & $\mathrm{F160W}_\mathrm{WFC3}$ & 0.21792 \\
             \hline
             \hline
        \end{tabular}
        \label{tab:ex_law}
    \end{table}

    \begin{figure*}
        \centering
        \includegraphics[width = 0.975\linewidth]{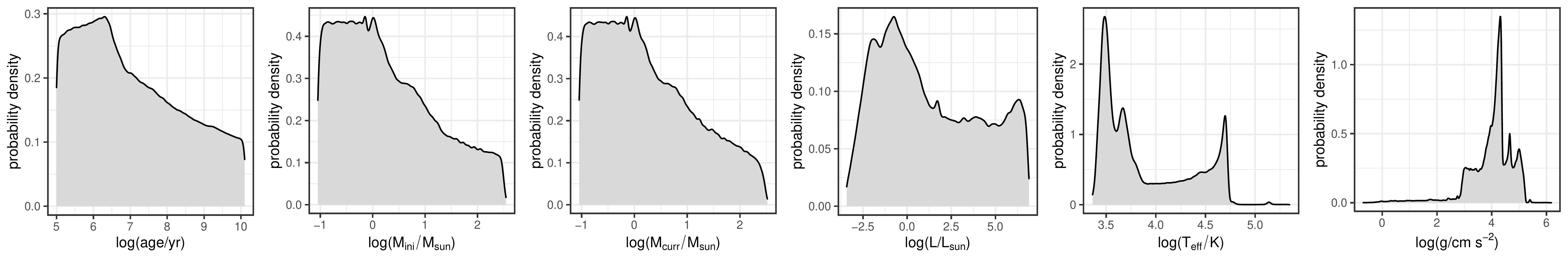}
        \includegraphics[width = 0.975\linewidth]{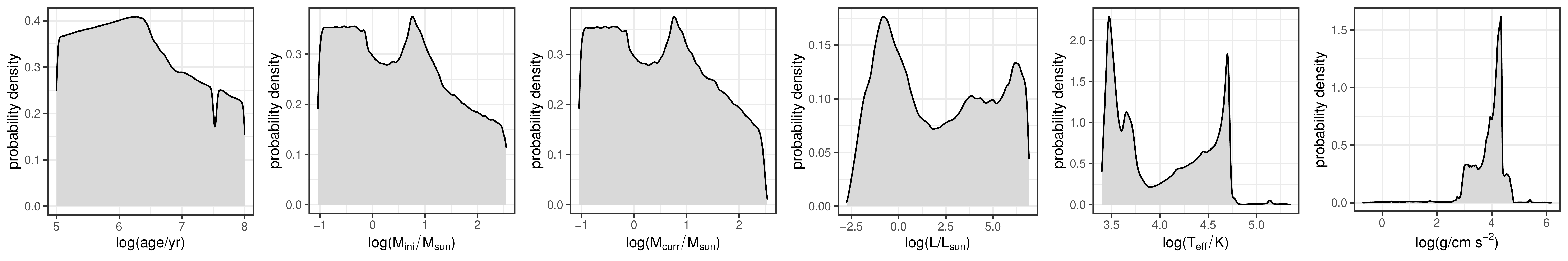}
        \includegraphics[width = 0.975\linewidth]{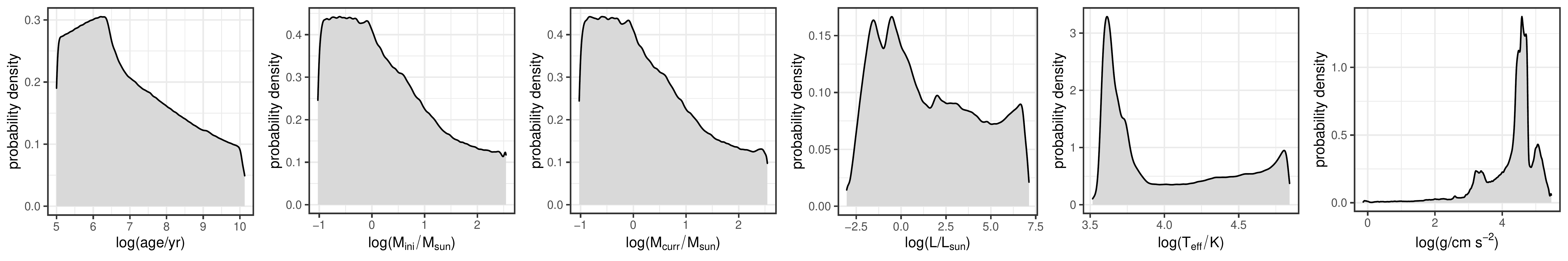}
        \includegraphics[width = 0.975\linewidth]{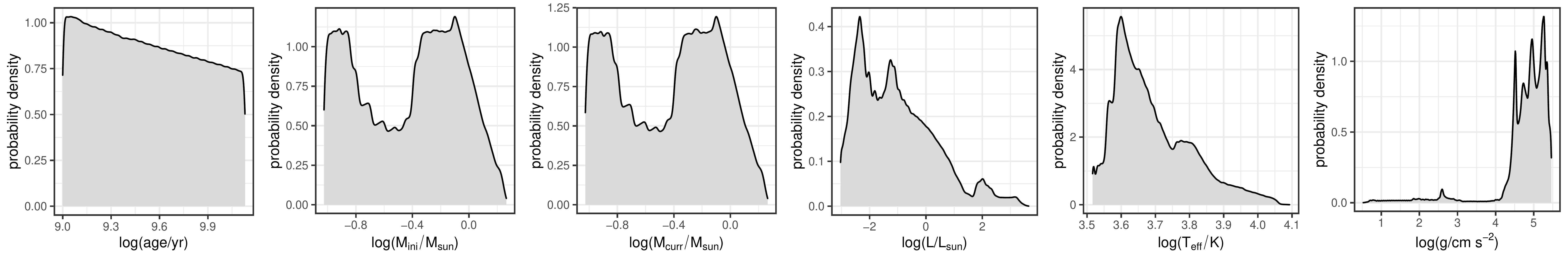}
        \caption{Prior distributions of all physical parameters in our training sets. From top to bottom the training sets are 'Wd2\_I', 'Wd2\_II', 'NGC6397\_I' and 'NGC6397\_II', respectively.}
        \label{fig:TS_Priors}
    \end{figure*}
    Table \ref{tab:ex_law} summarises the $A_\lambda/A_\mathrm{V}$ values we have derived according to Equation \eqref{eq:alambda} for all of the HST filters following the \cite{Cardelli1989} Milky Way extinction curve. The wavelength dependent coefficients $a_\lambda(x)$ and $b_\lambda(x)$, where $x = \lambda^{-1}(\mu\mathrm{m}^{-1})$, are given by Equations 2-4 in \cite{Cardelli1989} and are defined as follows.
    For the infrared regime $0.3\,\mu\mathrm{m}^{-1} \leq x \leq 1.1 \mu\mathrm{m}^{-1}$ they are given by
    \begin{equation}
        \begin{split}
            a(x) &= 0.574 x^{1.61}, \\
            b(x) &= -0.527 x^{1.61}. \\
        \end{split}
    \end{equation}
    In the optical and NIR regime, $1.1\,\mu\mathrm{m}^{-1} \leq x \leq 3.3 \mu\mathrm{m}^{-1}$, follows
    \begin{equation}
        \begin{split}
            a(x) &= 1 + 0.17699y - 0.50447y^2 - 0.02427y^3 + 0.72085y^4 \\ 
                 &+ 0.01979y^5 - 0.77530y^6 + 0.32999y^7, \\
            b(x) &= 1.41338y + 2.28305y^2 + 1.07233y^3 - 5.38434y^4 \\
                 &- 0.62251y^5 + 5.30260y^6 - 2.09002y^7,
        \end{split}
    \end{equation}
    where $y = x - 1.82$. \\
    Finally, for the UV regime, $3.3\,\mu\mathrm{m}^{-1} \leq x \leq 8 \mu\mathrm{m}^{-1}$, they are defined as
    \begin{equation}
        \begin{split}
            a(x) &= 1.752 - 0.316x - \frac{0.104}{(x-4.67)^2+0.341} + F_a(x), \\
            b(x) &= -3.090 + 1.825x + \frac{1.206}{(x-4.62)^2+0.263} + F_b(x),  \\
        \end{split}
    \end{equation}
    where $F_a(x) = F_b(x) = 0$ for $x < 5.9$ and 
    \begin{equation}
        \begin{split}
            F_a(x) &= -0.04473 (x-5.9)^2 - 0.009779 (x-5.9)^3, \\
            F_b(x) &= 0.2130 (x-5.9)^2 + 0.1207 (x-5.9)^3 \\
        \end{split}
    \end{equation}
    for $5.9 \leq x \leq 8$.
    
    Figure \ref{fig:TS_Priors} shows the prior distributions of all six phyiscal parameters, i.e. age, initial $M_\mathrm{ini}$ and current mass $M_\mathrm{curr}$, luminosity $L$, surface temperature $T_\mathrm{eff}$, surface gravity $g$, for the four training sets 'Wd2\_I', 'Wd2\_II', 'NGC6397\_I' and 'NGC6397\_II' that we employ in this study. 

\section{'Wd2\_I'}
    \label{app:Wd2_I}
    \begin{figure*}
        \centering
        \includegraphics[width = 0.75 \linewidth]{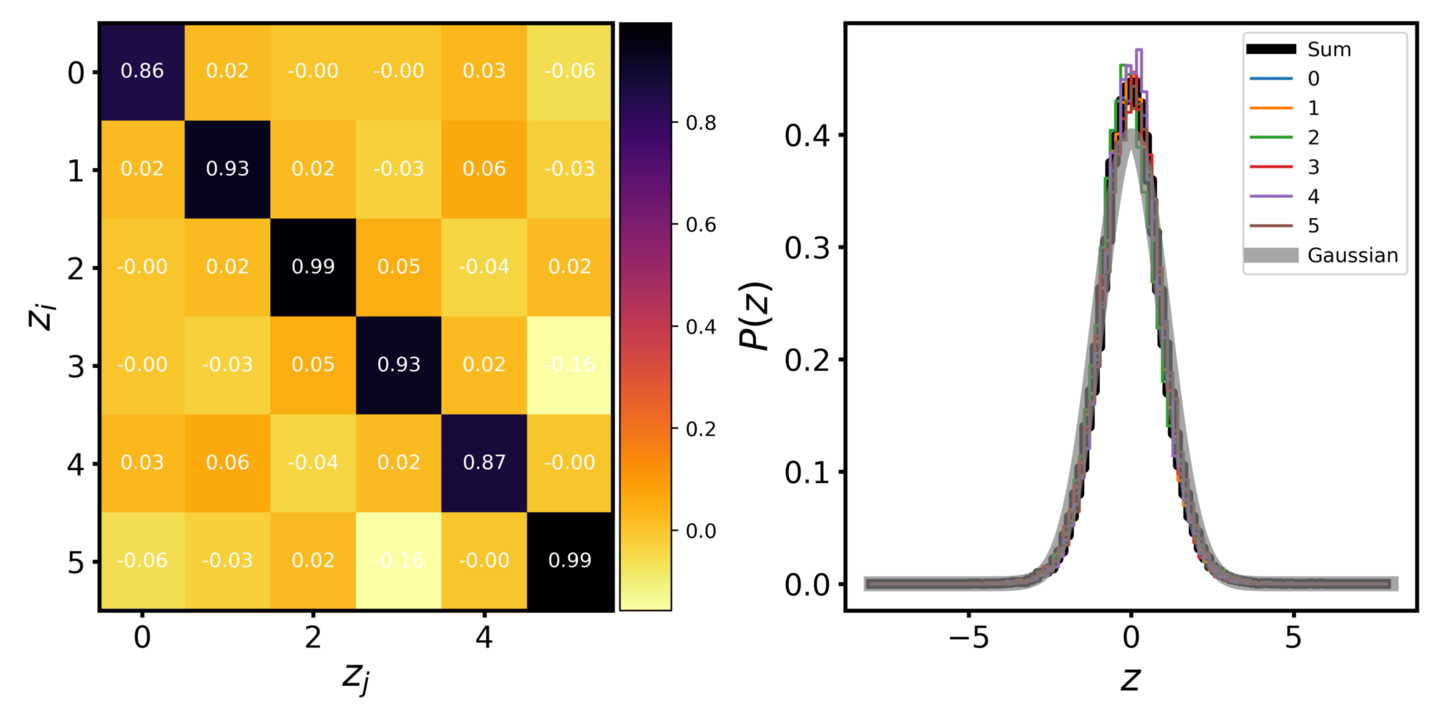}
        \caption{\textbf{Left:} The covariance matrix of the latent variables evaluated on 20,000 test observations provided by the cINN model trained on 'Wd2\_I'. \textbf{Right:} Histograms of the individual latent variable distributions. The black line indicates the distribution of the sum of all latent variables, while the grey line shows the target Normal distribution for reference.}
        \label{fig:Wd2_I_Z_covariance}
    \end{figure*}
    
    \begin{figure*}
        \centering
        \includegraphics[width = \linewidth]{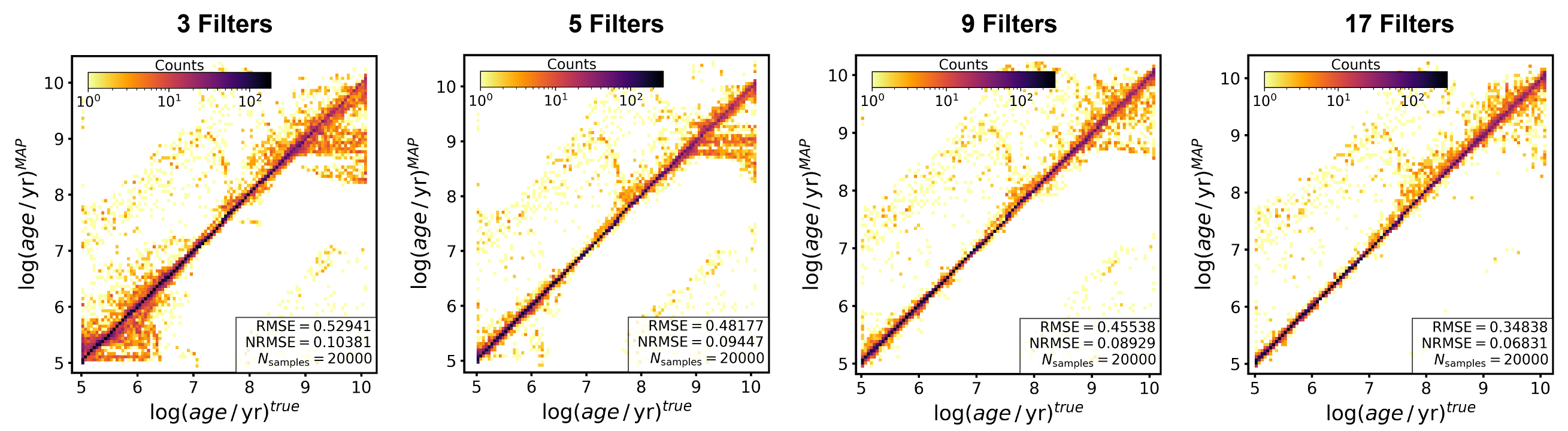}
        \caption{2D histograms of the MAP estimates of the age against the true value for 20,000 observations from the test set as predicted by the 'Wd2\_I' cINN model with increased amounts of features compared to our standard 'Wd2\_I' setup. On top of the extinction the columns indicate increasing numbers of photometric filters used as features. The 3 filter case entails the HST filters F555W, F814W, F160W, the 5 filter one adds F275W and F336W to that, and the 9 filter further includes F438W, F606W, F775W and F110W on top of the previous. The final 17 filter case entails all previous filters in addition to F218W, F225W, F390W, F475W, F625W, F105W, F125W and F140W. This sequence shows that the point estimate accuracy of the cINN improves with increasing number of available features as the RMSE decreases as well as the number of predictions that fall off the perfect 1-to-1 correlation.}
        \label{fig:Wd2_I_MAPvsTrue_Age_multifilter}
    \end{figure*}
    
    \begin{figure*}
        \centering
        \includegraphics[width = \linewidth]{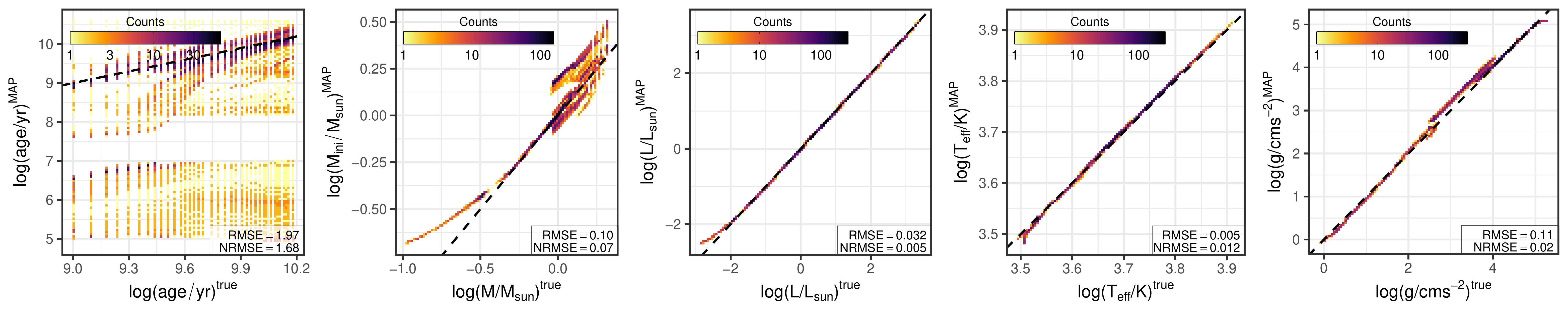}
        \caption{2D histograms of the MAP predictions for the physical parameters on $\sim$10,000 samples from the Dartmouth isochrone tables as predicted by the 'Wd2\_I' cINN model. Note that the NRMSEs are normalised to the parameter ranges of the Dartmouh ground truth here instead of the ranges of the 'Wd2\_I' PARSEC training data.}
        \label{fig:Wd2_I_DM_MAP_vs_True}
    \end{figure*}
    
    \begin{figure*}
        \centering
        \includegraphics[width = 0.8\linewidth]{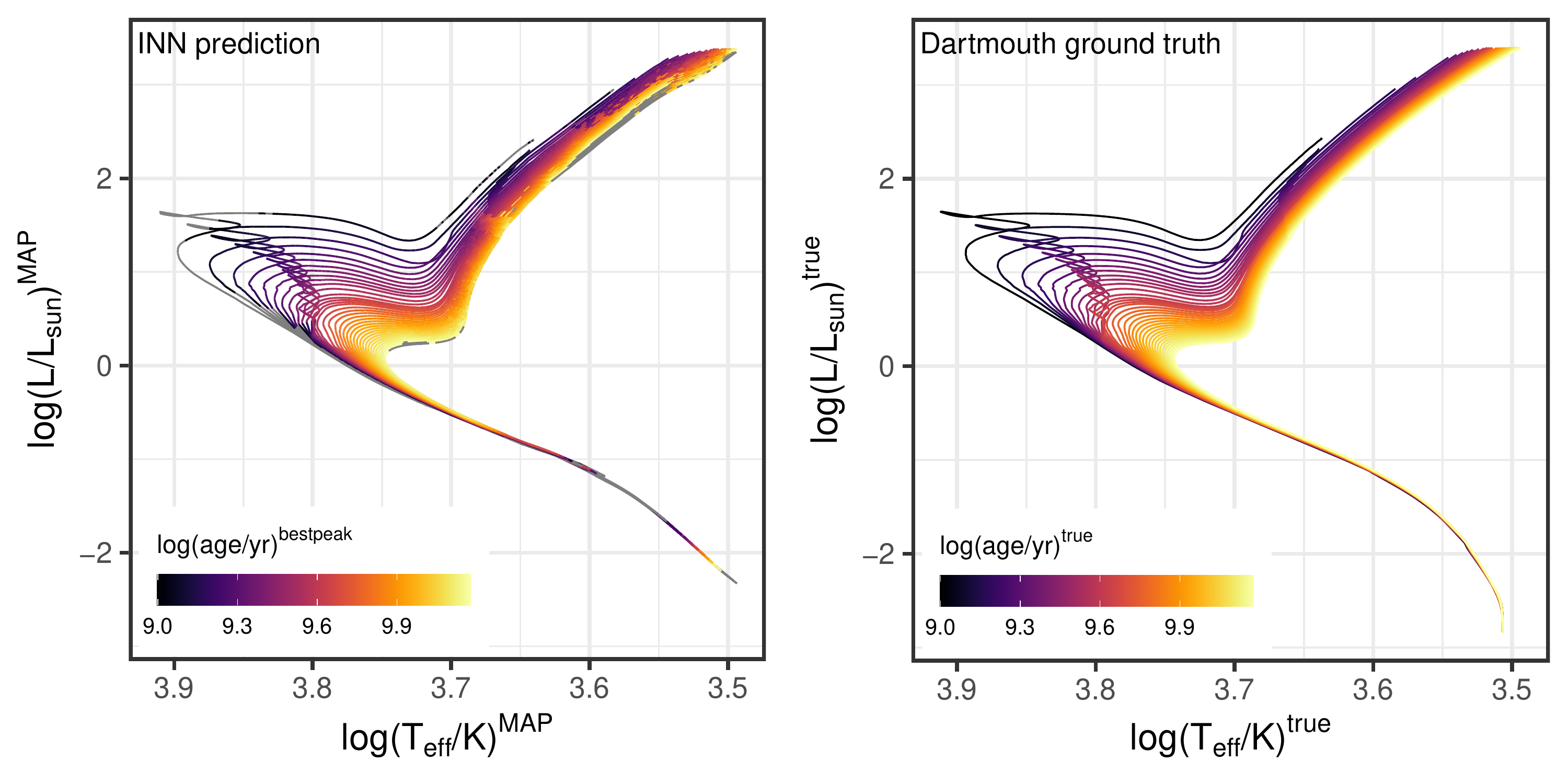}
        \caption{\textbf{Left:} cINN prediction for the HRD of the Dartmouth isochrones using model 'Wd2\_I'. Note that MAP estimates are used for luminosity and effective temperature here, but the colour code that indicates the predicted age does not correspond to the MAP age prediction but rather the best fitting peak of the predicted age posterior. The latter is done to take multi-modal age posteriors into account. \textbf{Right:} Ground truth HRD of the Dartmouth isochrones, colour coded according to their age.}
        \label{fig:Wd2_I_DM_HRD}
    \end{figure*}
    
    This appendix provides additional result diagrams for cINN models trained on 'Wd2\_I' and variations thereof.
    
    Figure \ref{fig:Wd2_I_Z_covariance} shows the covariance matrix of the latent variables in the left panel and their corresponding histograms in comparison to the target normal distribution in the right panel as evaluated on the 20,000 test observations using the trained 'Wd2\_I' model. These two diagrams serve as an example for convergence of the cINN model. 
    
    Corresponding to the series of posterior against true value diagrams presented in Figure \ref{fig:Wd2_I_PostVsTrue_multifilter} the respective MAP vs. true diagrams are shown in Figure \ref{fig:Wd2_I_MAPvsTrue_Age_multifilter}. Both these figures demonstrate how the cINN predictive performance improves with an increase of the number of photometric filters used as input.
    
    Figure \ref{fig:Wd2_I_DM_MAP_vs_True} shows the 'Wd2\_I' cINN prediction results on 10,000 samples from synthetic data of the Dartmouth isochrone tables. This Figure highlights that the cINN manages to recover most physical parameters quite accurately on this synthetic data set derived from a stellar evolution model that treats the underlying physics different than the PARSEC models, on which our cINN training sets are based. There are, however, some larger discrepancies in the low mass regime, where the PARSEC and Dartmouth models deviate most strongly from each other. While the age predictions appear significantly worse than on the MIST synthetic data at first glance, the median absolute error between prediction and ground truth is only 0.2 dex. Additionally, taking multi-modalities in the predicted age posteriors into account, most ages can actually be recovered quite accurately. The latter is highlighted in Figure \ref{fig:Wd2_I_DM_HRD} comparing the predicted to the ground truth HRD for the MIST isochrones.

\section{'Wd2\_II'}
    \label{app:Wd2_II}
    \begin{figure*}
        \centering
        \includegraphics[width = 0.85\linewidth]{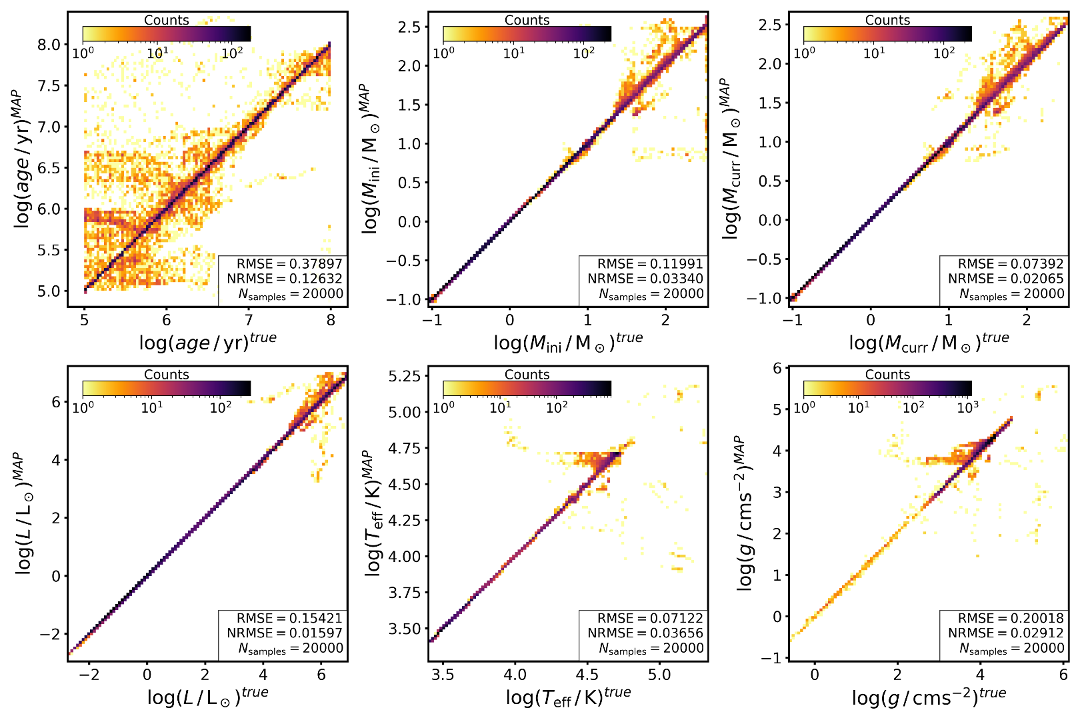}
        \caption{2D histograms of the MAP estimates plotted against the true values for the six physical parameters we predict with the cINN trained on 'Wd2\_II' for 20,000 cases from our test set. From top left to bottom right we show age, $M_\mathrm{ini}$, $M_\mathrm{curr}$, $L$, $T_\mathrm{eff}$ and $g$. }
        \label{fig:Wd2_II_MAP_vs_true}
    \end{figure*}
    
    \begin{figure*}
        \centering
        \includegraphics[width = 0.85\linewidth]{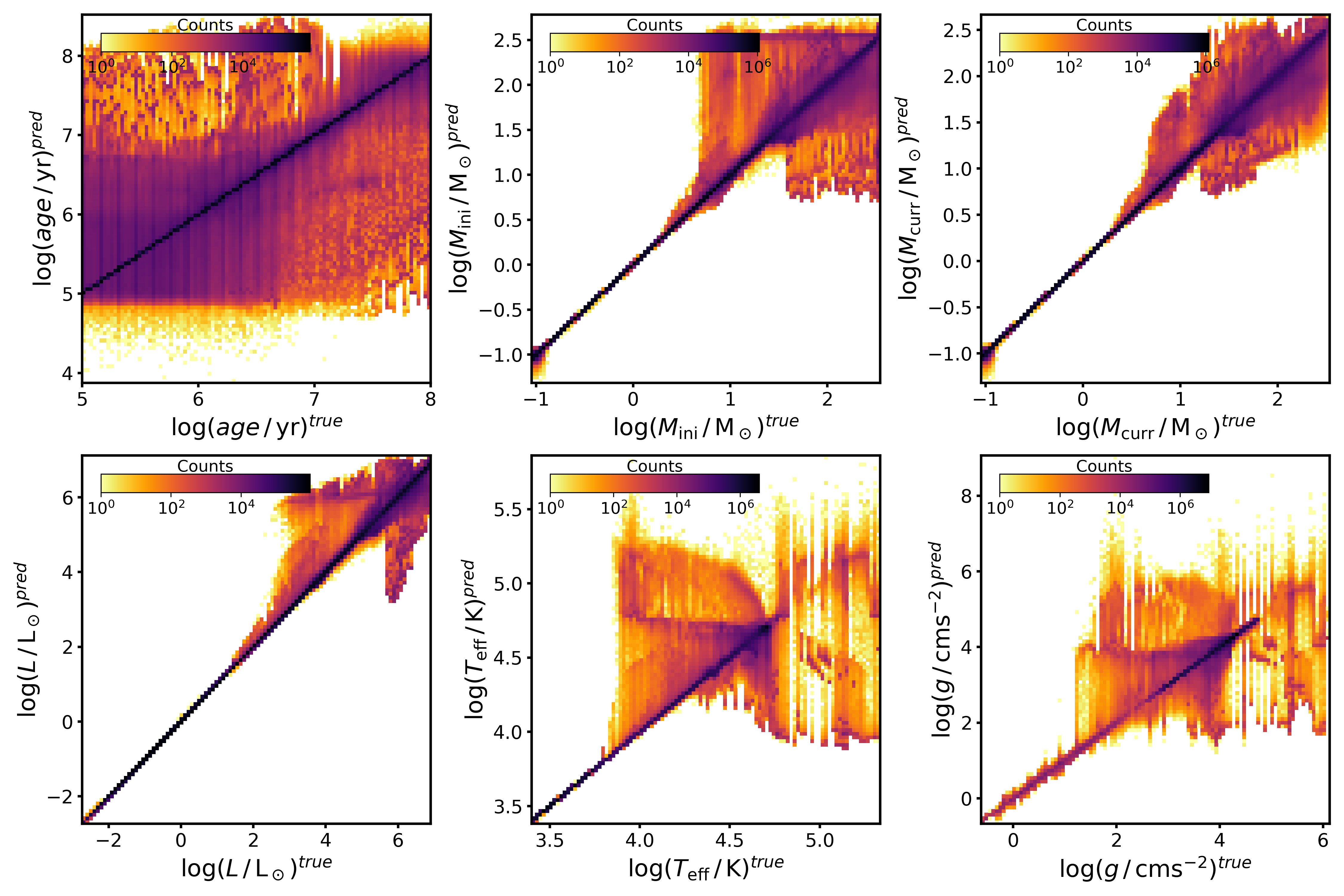}
        \caption{2D histograms of the entire predicted posteriors plotted against the true value of the six physical parameters provided by the cINN trained on 'Wd2\_II' for 20,000 cases from the respective test set. From top left to bottom right age, $M_\mathrm{ini}$, $M_\mathrm{curr}$, $L$, $T_\mathrm{eff}$ and $g$ are shown.}
        \label{fig:Wd2_II_Posteriors_vs_true}
    \end{figure*}
    
    \begin{figure*}
        \centering
        \includegraphics[width = 0.75\linewidth]{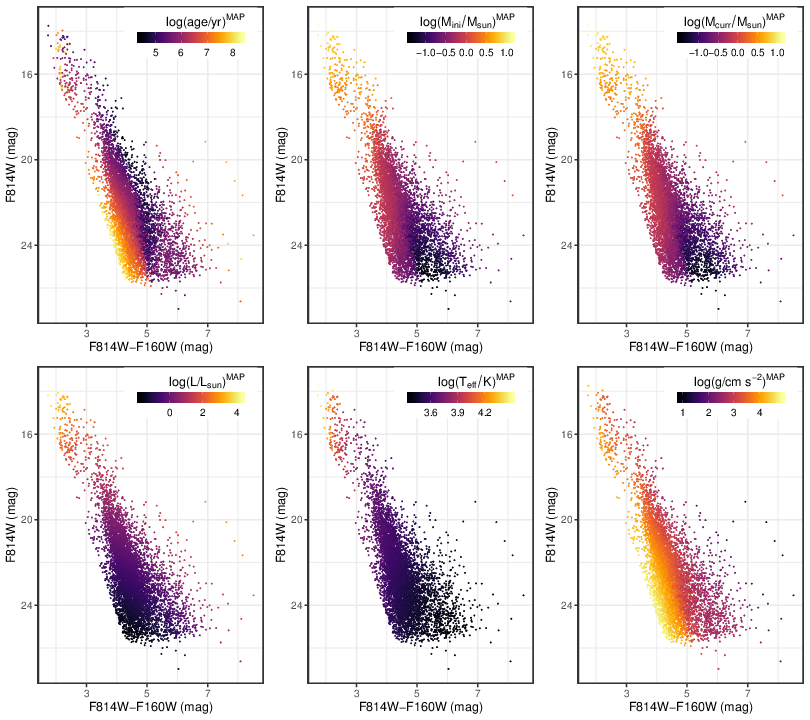}
        \caption{Optical colour magnitude diagrams of the Westerlund 2 HST data, colour coded according to the MAP estimates for the six physical parameters predicted with the cINN trained on 'Wd2\_II'.}
        \label{fig:Wd2_II_MAP_CMD}
    \end{figure*}
    
        \begin{figure}
        \centering
        \includegraphics[width = \linewidth]{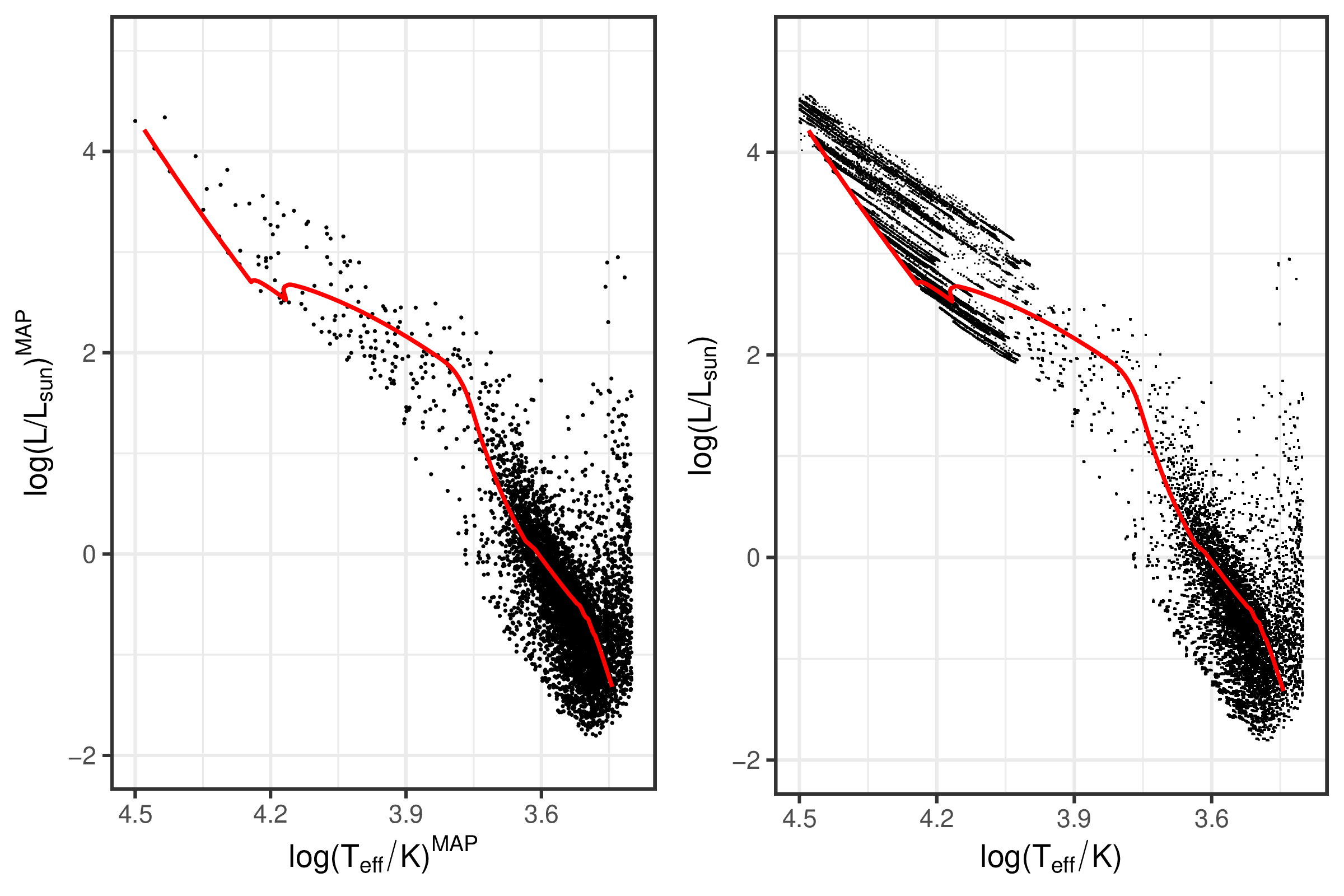}
        \caption{Predicted Hertzsprung-Russel-Diagram for the Westerlund 2 cluster constituents provided by the cINN model trained on 'Wd2\_II'. The \textbf{left} panel shows the HRD based on the maximum a posteriori predictions for $\log(L)$ and $\log(T_\mathrm{eff})$, while in the \textbf{right} panel the entire posterior distributions of these two parameters are plotted for every star. The red line in both diagrams indicates a 1 Myr isochrone for comparison.}
        \label{fig:Wd2_II_pred_HRD}
    \end{figure}
    
    \begin{figure*}
        \centering
        \includegraphics[width = 0.85\linewidth]{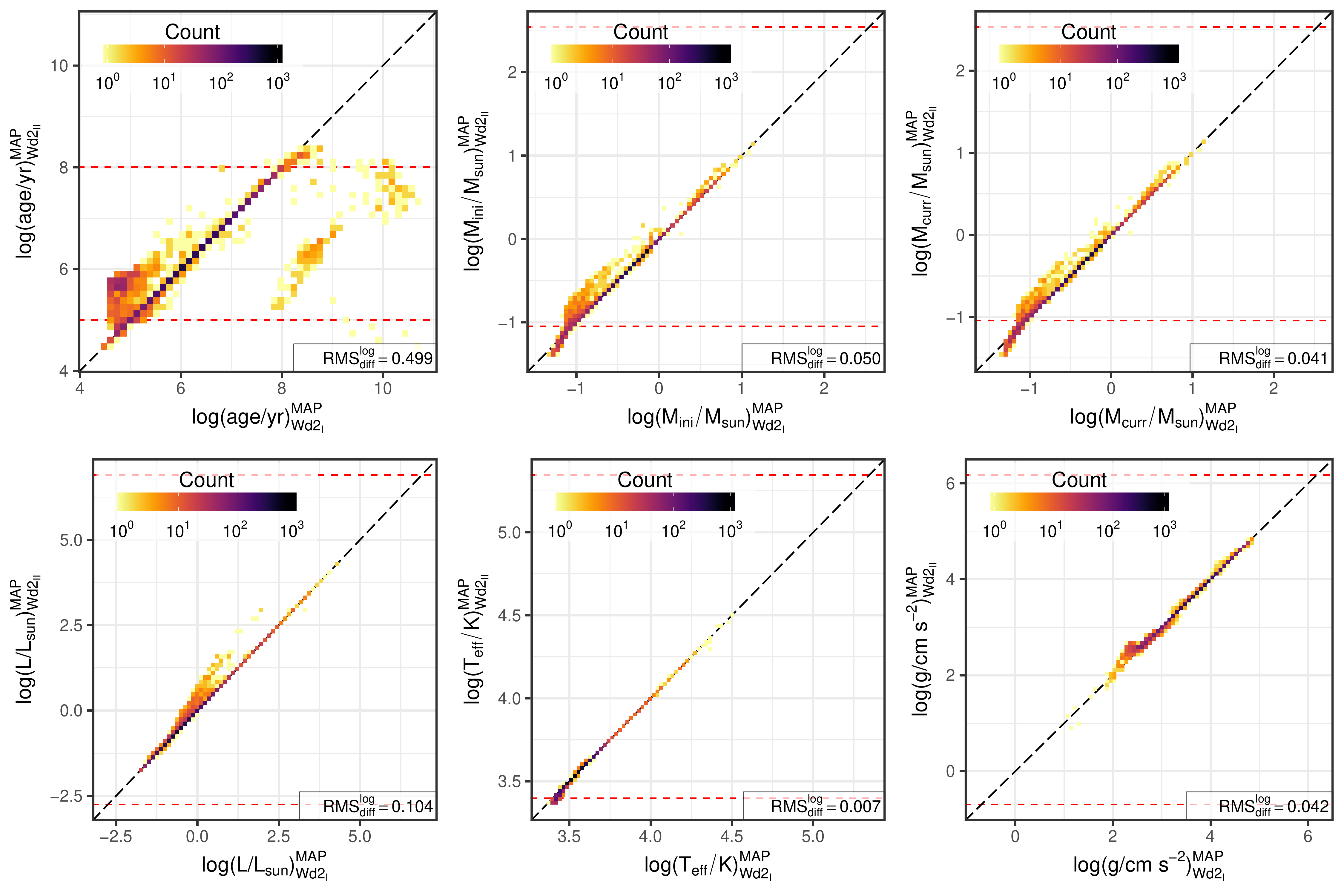}
        \caption{2D histogram of the comparison between the MAP estimates for the physical parameters of the Westerlund 2 stars between the two cINN models trained on 'Wd2\_I' and 'Wd2\_II', respectively. Note that the black dashed line indicates a perfect 1-to-1 correlation, while the red dashed lines indicate the limits of the 'Wd2\_II' training set for each parameter.}
        \label{fig:Wd2_I_vs_Wd2_II}
    \end{figure*}
    
    \begin{figure*}
        \centering
        \includegraphics[width = 0.4\linewidth]{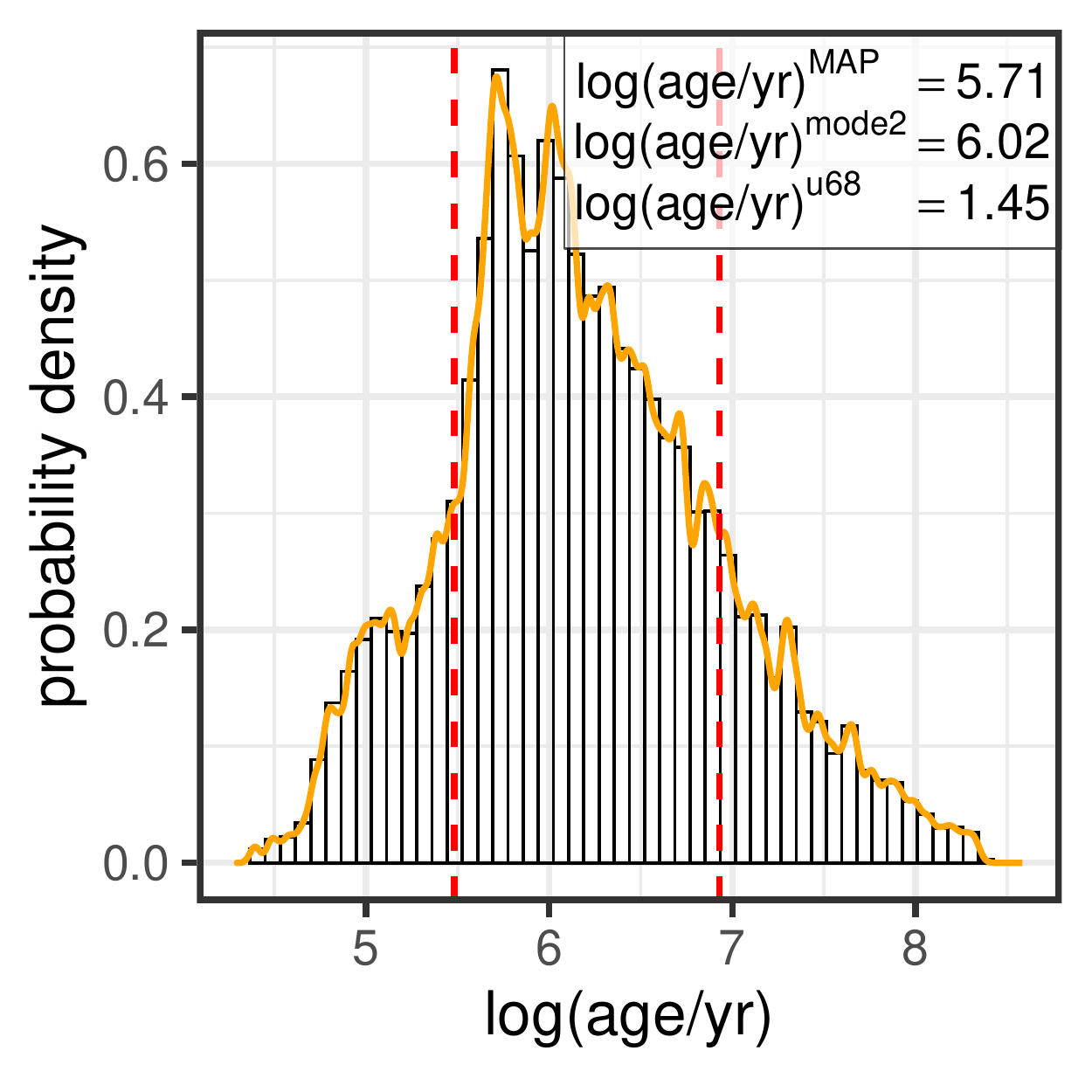}
        \caption{Histogram of the sum of the age posterior distributions of all Westerlund 2 cluster stars as predicted by the 'Wd2\_II' cINN model. The orange line indicates a kernel density fit to this cumulative posterior distribution to determine the most likely cluster age. The red dashed lines mark the width of the 68\% confidence interval.}
        \label{fig:Wd2_II_logAgeKDE}
    \end{figure*}
    
    In this appendix we provide additional plots and further discussion for the cINN trained on 'Wd2\_II'.
    
    Figures \ref{fig:Wd2_II_MAP_vs_true} and Figures \ref{fig:Wd2_II_Posteriors_vs_true} present the MAP and posterior against true diagrams for 'Wd2\_II' corresponding to Figures \ref{fig:Wd2_I_MAP_vs_true} and \ref{fig:Wd2_I_Posteriors_vs_true} presented in the main paper for 'Wd2\_I'. These two diagrams show that the final cINN model on 'Wd2\_II' does not differ significantly from the 'Wd2\_I' solution. Aside from a few more outlier cases that likely cause the increased RMSEs of the physical parameters beside age, the point estimate performance shows similar successes and flaws, especially the same increased number of outliers in the age prediction for the young stars. As already indicated by comparable median uncertainties (except age) the posterior distributions are equally unaffected by the age cut in the 'Wd2\_II' training set. The predicted age posteriors also do not show significant changes except for the obvious limitation due to the smaller age range of 'Wd2\_II' and a rare tendency to extrapolate down to $\log(\mathrm{age}/\mathrm{yr}) = 4$, which we do not observe for 'Wd2\_I'. The more limited age range of course eliminates some of the degeneracies that may have caused some of the false MAP estimates in the 'Wd2\_I' prediction, like for example the few 10 to 100 Myr old stars in Figure \ref{fig:Wd2_I_MAP_vs_true} that have a MAP prediction somewhere between 100 Myr to 1 Gyr. As these are rare cases in the previous test, however, it seems safe to say that the decrease in RMSE and median uncertainty is primarily caused by the fact that the posterior distributions should span a smaller range of maximum 3 dex instead of the 5 dex in 'Wd2\_I'. The almost equal NRMSEs confirm this.
    
    Corresponding to Figure \ref{fig:Wd2_I_MAP_CMD} we show the 'Wd2\_II' prediction results for all physical parameters in terms of the MAP estimates in the CMDs of Figure \ref{fig:Wd2_II_MAP_CMD}. 
    
    Analogous to Figure \ref{fig:Wd2_I_HRD} we show the predicted HRD for the observational Wd2 data as given by the 'Wd2\_II' model in Figure \ref{fig:Wd2_II_pred_HRD}.
    
    Figure \ref{fig:Wd2_I_vs_Wd2_II} shows the comparison of the MAP estimates between the two models. In general the deviations in the MAP for all physical parameters are fairly insignificant except for age, with RMS deviations on the order of a few 0.01 dex. In the case of the age prediction the deviations appear more severe, on average about 0.499 dex, likely caused by the cut-off around $\log(\mathrm{age/yr}) = 8$ for 'Wd2\_II'. Overall most predictions (note the logarithmic colour coding) fall onto the identity mapping, except for a set of about 500 stars which are placed at $\log(\mathrm{age}/\mathrm{yr}) \sim 5.7$ by the  'Wd2\_II' model, while the 'Wd2\_I' extrapolated an age below 0.1\,Myr here. However, as the median relative deviation between the two MAP estimates is only 0.3\% it is safe to say that the two models do not differ significantly.
        
    We conclude that the cINN does not get confused in any significant way if there are more potentially degenerate mappings in the training set, as the full model predicts the same physical parameters as the model that incorporates prior knowledge to decrease the amount of degeneracies. 
    
    Figure \ref{fig:Wd2_II_logAgeKDE} presents the derivation of a cluster age of Wd2 based on the sums of the age posteriors for the predictions by the 'Wd2\_II' model. Here we find a similar result as for the 'Wd2\_I' model shown in Figure \ref{fig:Wd2_I_logAgeKDE}, except for a split peak with maxima at 0.5\,Myr and 1.04\,Myr in the sum of posterior distributions. As these two solutions are part of one major peak they likely belong to the same mode of the distribution, located at the Wd2 cluster age. This demonstrates again that 'Wd2\_II' agrees well with 'Wd2\_I'.

\section{'NGC6397\_I'}
    \label{app:NGC6397_I}
    \begin{figure*}
        \centering
        \includegraphics[width = 0.85\linewidth]{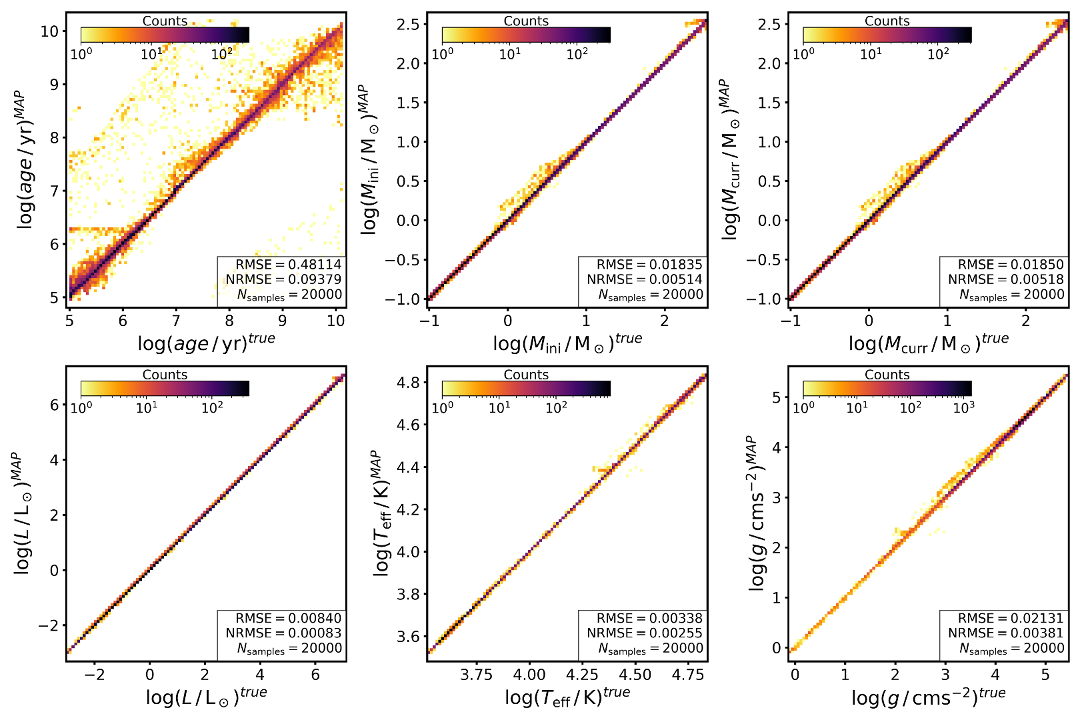}
        \caption{2D histograms of the MAP estimates plotted against the true values for all six predicted physical parameters as given by the cINN trained on 'NGC6397\_I' for 20,000 test set observations. }
        \label{fig:NGC6397_I_MAP_vs_true}
    \end{figure*}
    
    \begin{figure*}
        \centering
        \includegraphics[width = 0.85\linewidth]{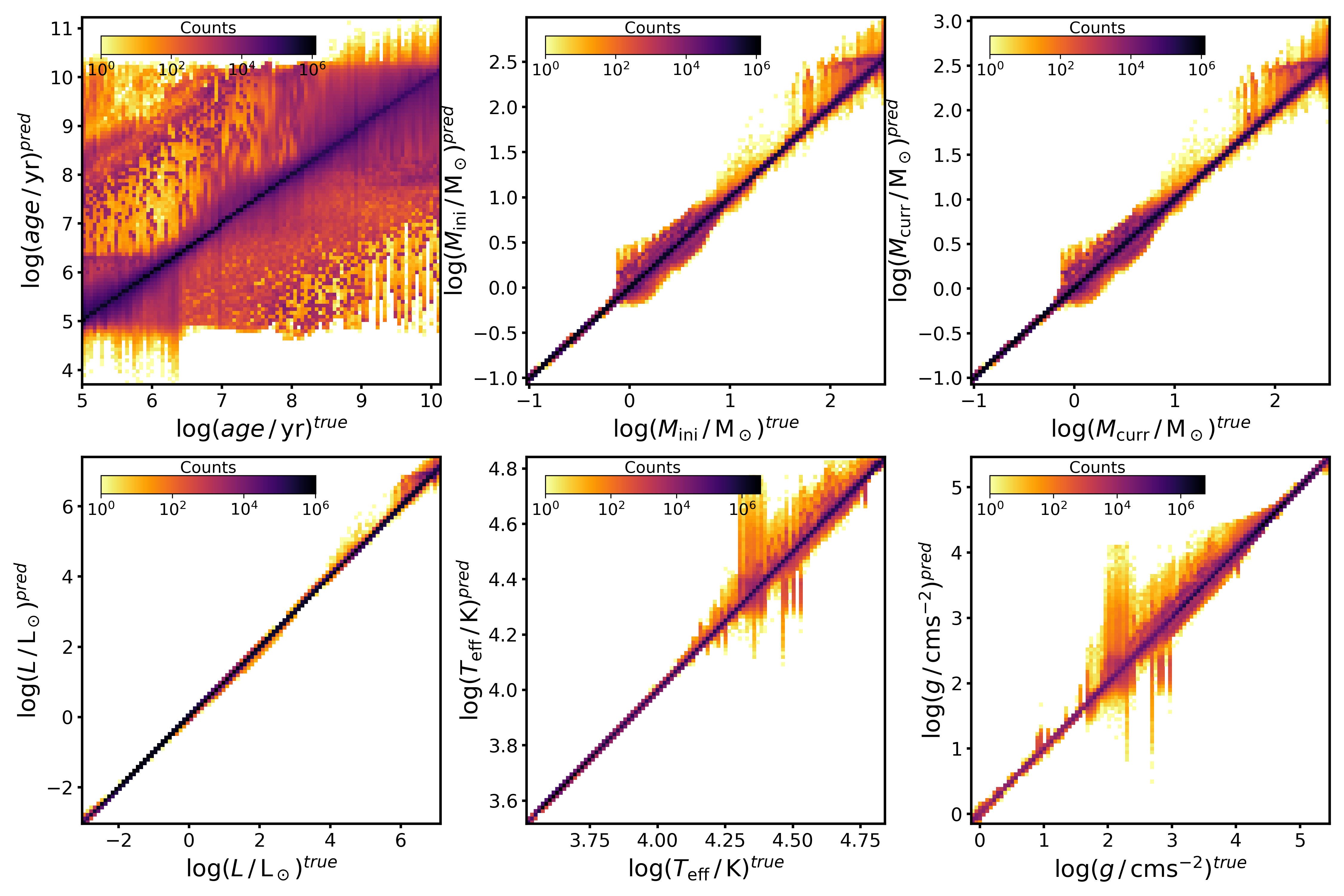}
        \caption{2D histograms of the entire predicted posteriors plotted against the true value of the six physical parameters given by the cINN trained on 'NGC6397\_I' for 20,000 cases from the respective test set. From top left to bottom right age, $M_\mathrm{ini}$, $M_\mathrm{curr}$, $L$, $T_\mathrm{eff}$ and $g$ are shown.}
        \label{fig:NGC6397_I_Posterior_Vs_true}
    \end{figure*}
    
    \begin{figure*}
        \centering
        \includegraphics[width = 0.85\linewidth]{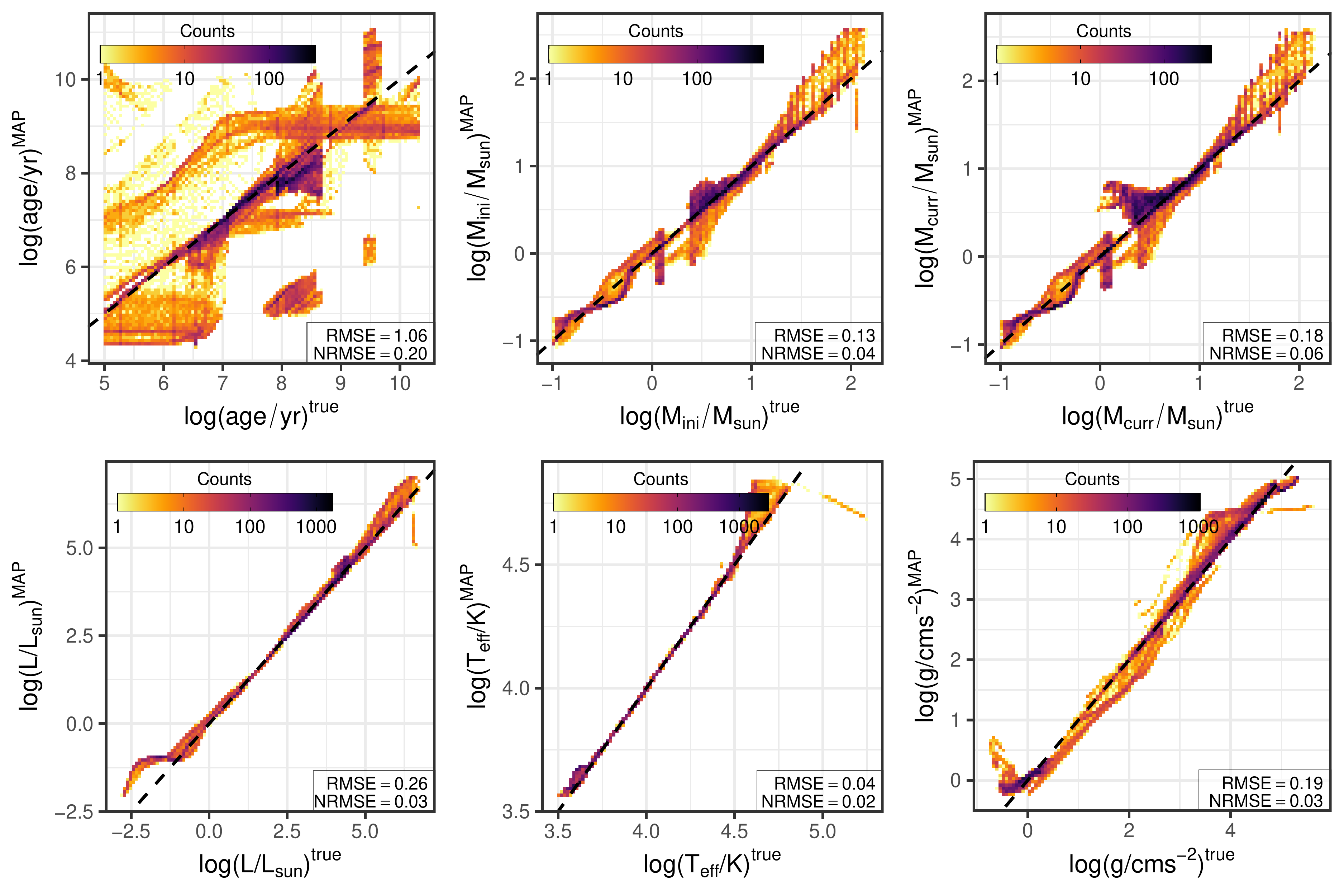}
        \caption{2D histograms of the MAP estimates for the physical parameters of 40,000 samples of the synthetic MIST isochrone tables as predicted by the 'NGC6397\_I' cINN model. Note that the NRMSEs are normalised to the physical parameter ranges of the MIST ground truth here instead of the 'NGC6397\_I' PARSEC training data.}
        \label{fig:NGC6397_I_MIST_MAP_vs_True}
    \end{figure*}
    
    \begin{figure*}
        \centering
        \includegraphics[width = 0.8\linewidth]{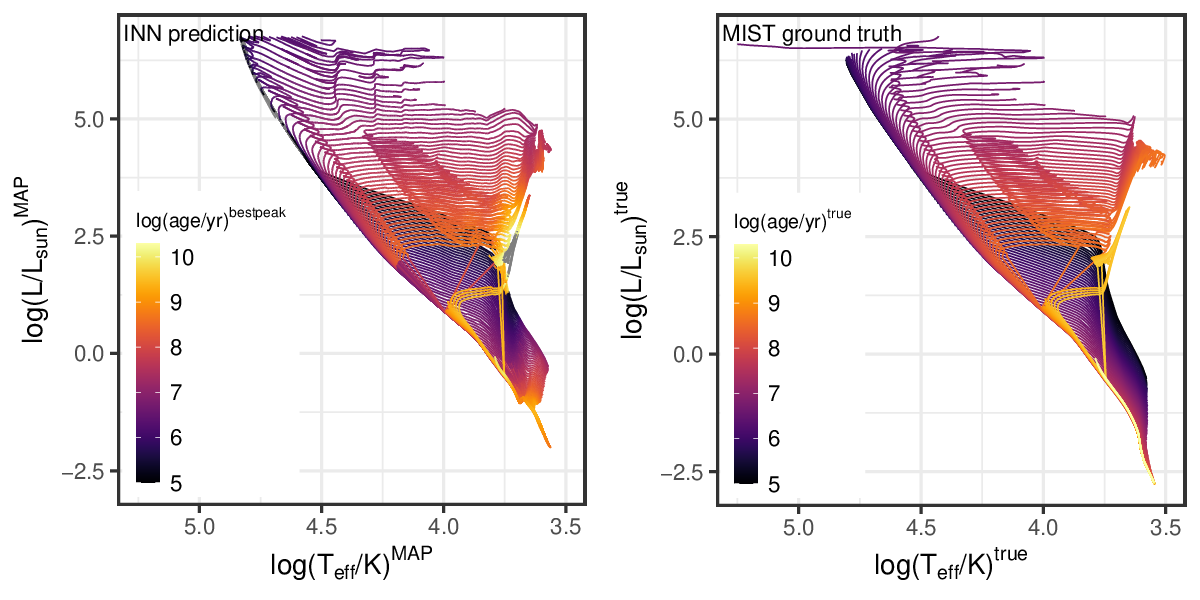}
        \caption{\textbf{Left:} cINN prediction for the HRD of the MIST isochrones using model 'NGC6397\_I'. Note that MAP estimates are used for luminosity and effective temperature here, but the colour code that indicates the predicted age does not correspond to the MAP age prediction but rather the best fitting peak of the predicted age posterior. The latter is done to take multi-modal age posteriors into account. \textbf{Right:} Ground truth HRD of the MIST isochrones, colour coded according to their age.}
        \label{fig:NGC6397_I_MIST_HRD}
    \end{figure*}
    
    \begin{figure*}
        \centering
        \includegraphics[width = \linewidth]{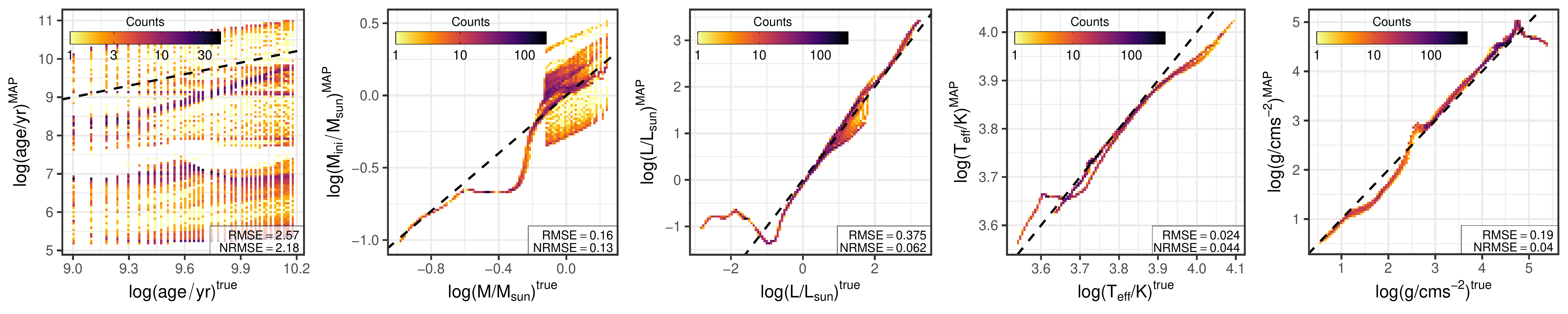}
        \caption{2D histograms of the MAP estimates for the physical parameters of 10,000 samples of the synthetic Dartmouth isochrones as predicted by the 'NGC6397\_I' cINN model. Note that the NRMSEs are normalised to the physical parameter ranges of the MIST ground truth here instead of the 'NGC6397\_I' PARSEC training data.}
        \label{fig:NGC6397_I_DM_MAP_vs_True}
    \end{figure*}
    
    \begin{figure*}
        \centering
        \includegraphics[width = 0.8\linewidth]{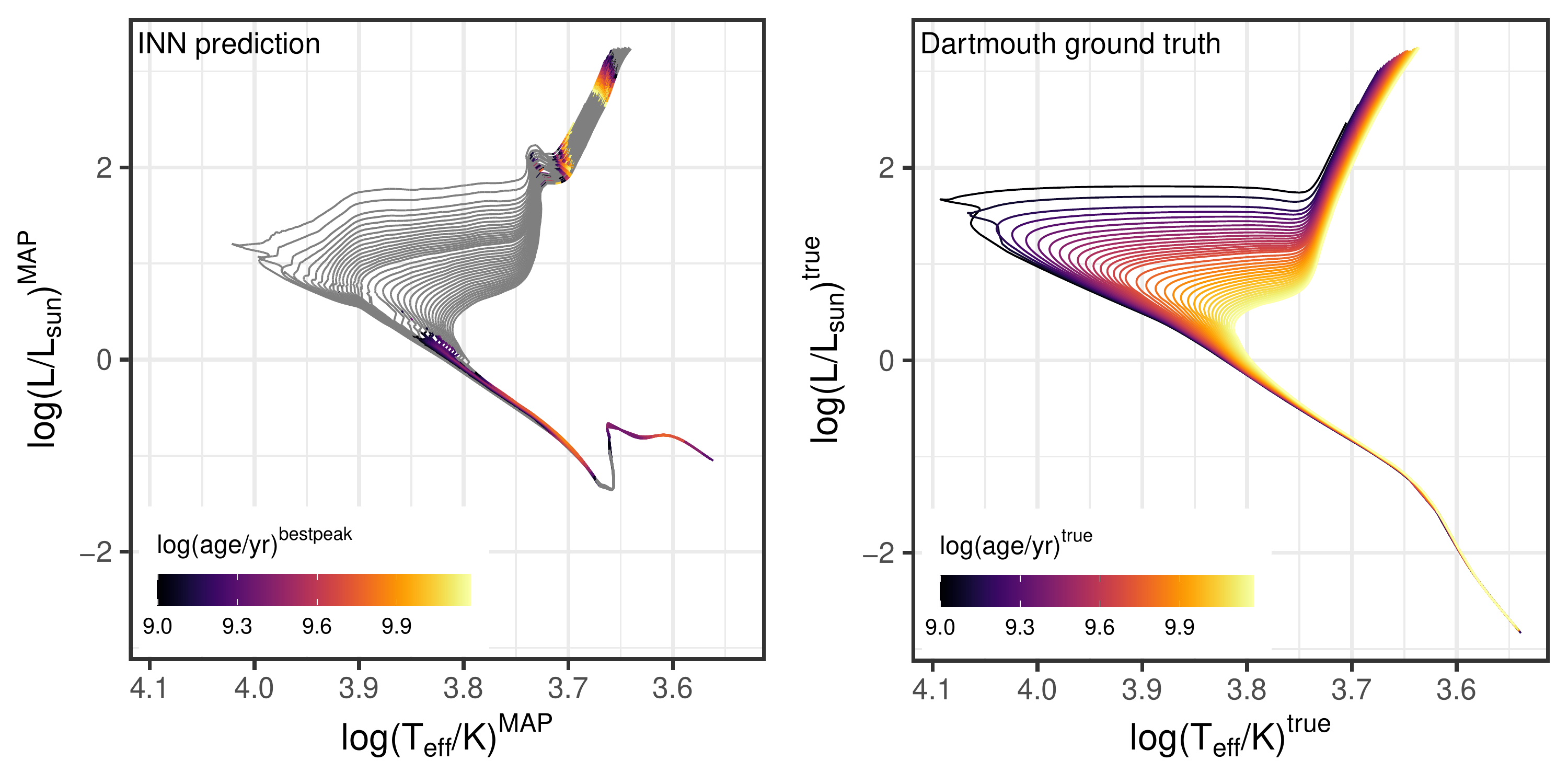}
        \caption{\textbf{Left:} cINN prediction for the HRD of the Dartmouth isochrones using model 'NGC6397\_I'. Note that MAP estimates are used for luminosity and effective temperature here, but the colour code that indicates the predicted age does not correspond to the MAP age prediction but rather the best fitting peak of the predicted age posterior. The latter is done to take multi-modal age posteriors into account. \textbf{Right:} Ground truth HRD of the MIST isochrones, colour coded according to their age.}
        \label{fig:NGC6397_I_DM_HRD}
    \end{figure*}
    
    Appendix \ref{app:NGC6397_I} provides complementary diagrams and discussion for the 'NGC6397\_I' cINN results.
    
    Figure \ref{fig:NGC6397_I_MAP_vs_true} provides the MAP vs. true diagram for 'NGC6397\_I' analogous to Figures \ref{fig:Wd2_I_MAP_vs_true} and \ref{fig:Wd2_II_MAP_vs_true}. Figure \ref{fig:NGC6397_I_Posterior_Vs_true} shows the corresponding posterior vs. true diagram (c.f. Figures \ref{fig:Wd2_I_Posteriors_vs_true} and \ref{fig:Wd2_II_Posteriors_vs_true}). These two figures show that the 'NGC6397\_I' model delivers overall similar results to 'Wd2\_I', except for slight improvements in prediction accuracy. These can likely be accredited to the larger coverage of 5 filters in 'NGC6397\_I' over the 2 filters in 'Wd2\_I' as a comparison of Figure \ref{fig:NGC6397_I_MAP_vs_true} to the 5 filter experiment for 'Wd2\_I' in Figure \ref{fig:Wd2_I_MAPvsTrue_Age_multifilter} confirms. The same comparison holds for Figure \ref{fig:NGC6397_I_Posterior_Vs_true} vs. the 5 filter results for 'Wd2\_I' displayed in Figure \ref{fig:Wd2_I_PostVsTrue_multifilter}. 
    
    Figures \ref{fig:NGC6397_I_MIST_MAP_vs_True} to \ref{fig:NGC6397_I_DM_HRD} present the 'NGC6397\_I' cINN model prediction results for the MIST and Dartmouth synthetic isochrones, respectively. They correspond to Figures \ref{fig:Wd2_I_MIST_MAP_vs_True} and \ref{fig:Wd2_I_HRD}, shown in the main paper. As already discussed in Section \ref{sec:synth_ms_age_pred} these diagrams here show that the 'NGC6397\_I' is similarly successful on the MIST data set as 'Wd2\_I', although suffering from overall larger prediction errors, but fails quite severely on the Dartmouth data due to the significant model discrepancies between Dartmouth and PARSEC.
    
\section{'NGC6397\_II'}
    \label{app:NGC6397_II}
    \begin{figure*}
        \centering
        \includegraphics[width = 0.85\linewidth]{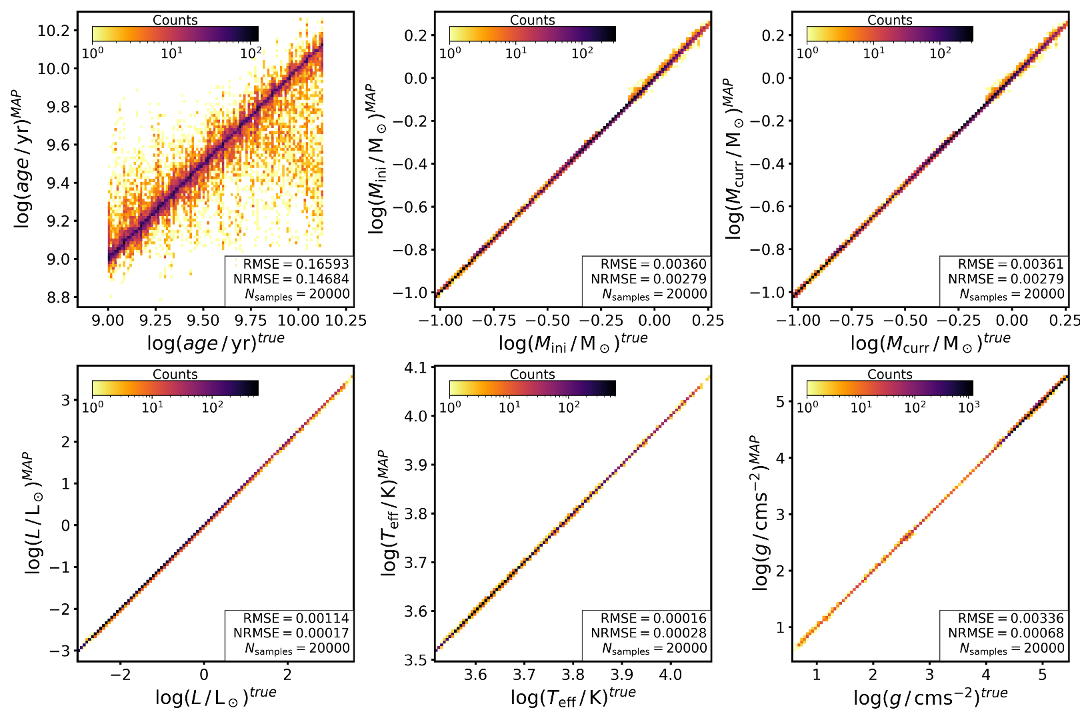}
        \caption{2D histograms of the MAP estimates plotted against the true values for the six physical parameters we predict with the cINN trained on 'NGC6397\_II' for 20,000 cases from our test set. From top left to bottom right we show age, $M_\mathrm{ini}$, $M_\mathrm{curr}$, $L$, $T_\mathrm{eff}$ and $g$.}
        \label{fig:NGC6397_II_MAP_vs_true}
    \end{figure*}
    
    \begin{figure*}
        \centering
        \includegraphics[width = 0.85\linewidth]{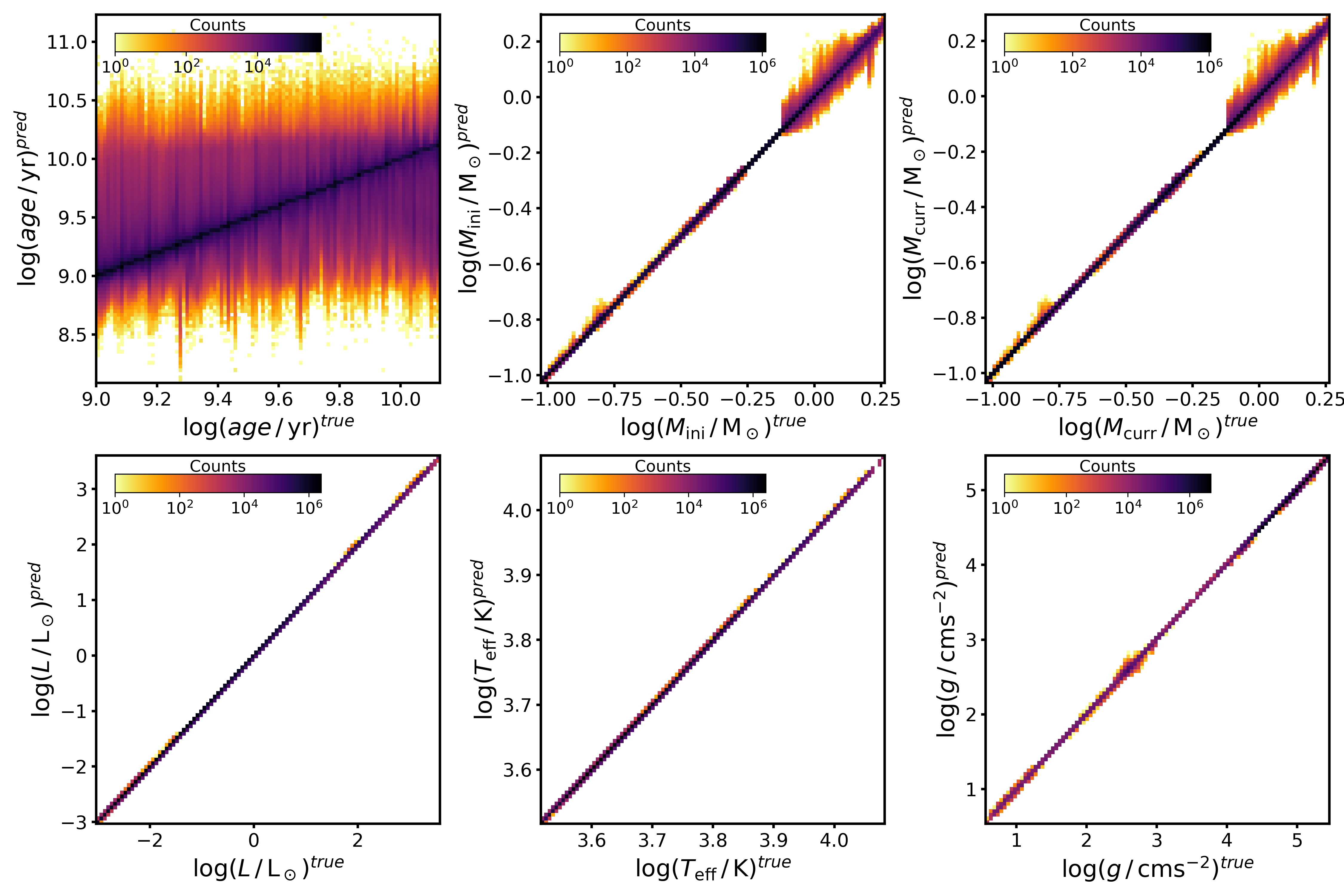}
        \caption{2D histograms of the entire predicted posteriors plotted against the true value of the six physical parameters provided by the cINN trained on 'NGC6397\_II' for 20,000 cases from the respective test set. From top left to bottom right age, $M_\mathrm{ini}$, $M_\mathrm{curr}$, $L$, $T_\mathrm{eff}$ and $g$ are shown.}
        \label{fig:NGC6397_II_Posteriors_vs_true}
    \end{figure*}
    
    \begin{figure*}
        \centering
        \includegraphics[width = 0.75\linewidth]{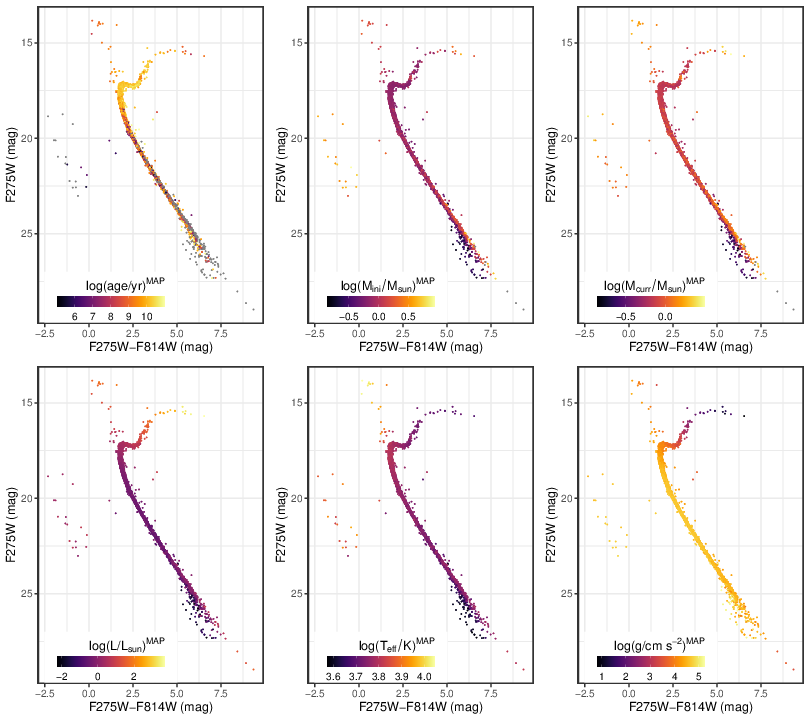}
        \caption{UV-I colour magnitude diagrams of NGC\,6397, colour coded according to the MAP estimates for the six physical parameters predicted with the cINN trained on 'NGC6397\_II'. Note that the grey points are those stars for which the prediction falls outside of the range provided by the respective colour bar of each panel.}
        \label{fig:NGC6397_II_MAP_CMD}
    \end{figure*}
   
    \begin{figure*}
    	\centering
    	\includegraphics[width = 0.75\linewidth]{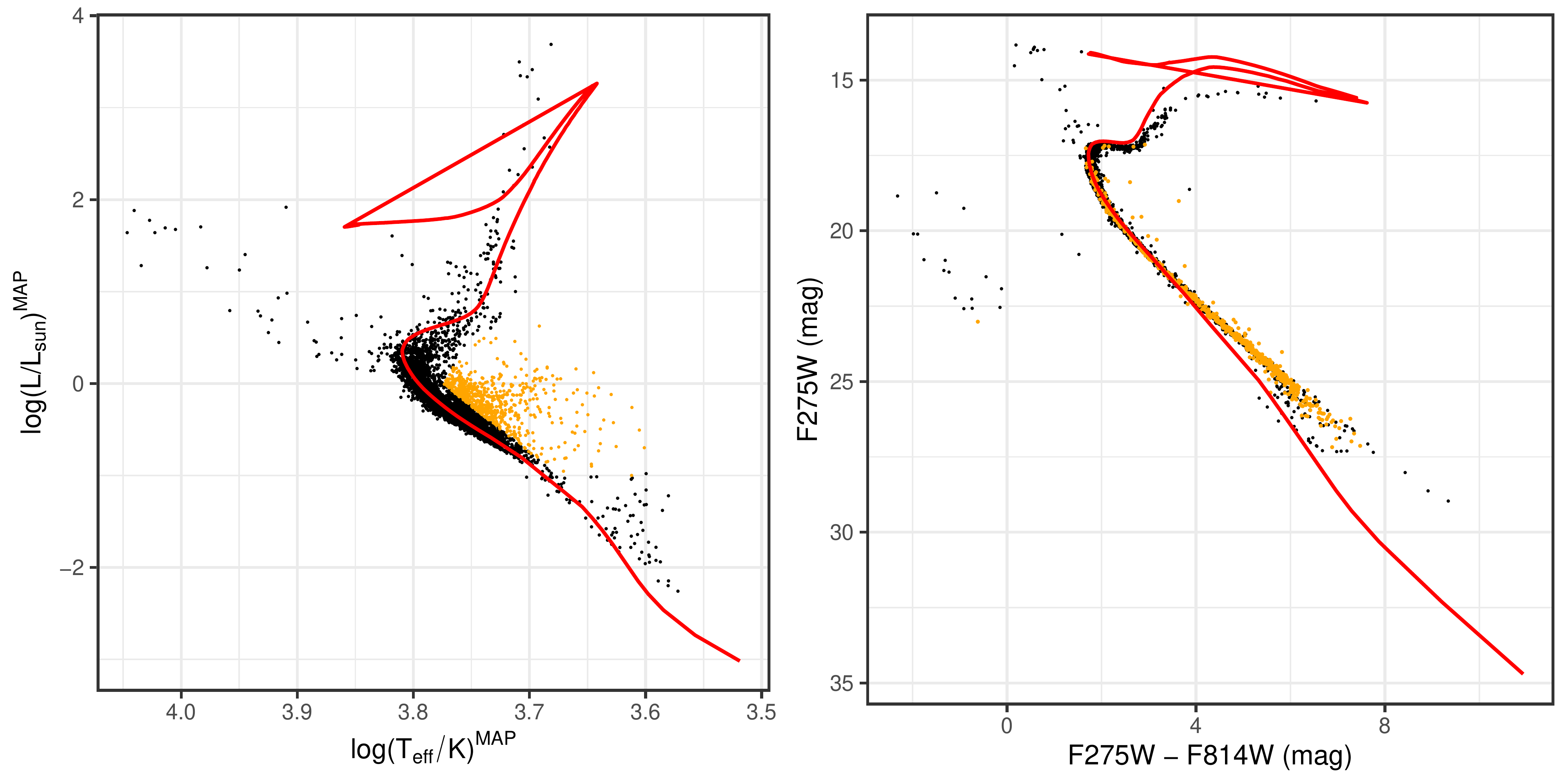}
    	\caption{The \textbf{left} panel shows the predicted HRD of NGC\,6397 based on the MAP estimates of $\log(L)$ and $\log(T_\mathrm{eff})$ by the cINN trained on 'NGC6397\_II'. Highlighted in orange are stars which we deem outliers. The \textbf{right} panel shows the corresponding UV-I CMD of NGC\,6397 indicating the position of these outliers in relation to the remaining observational data. The red line in both diagrams marks a 13\,Gyr isochrone, the supposed age of NGC\,6397, for comparison. These two diagrams show that the outlier predictions in the HRD are predominantly located in the CMD where the model and observational data deviate the most from each other.}
    	\label{fig:NGC6397_II_HRD_outliers}
    \end{figure*}
    
    \begin{figure*}
        \centering
        \includegraphics[width = 0.375\linewidth]{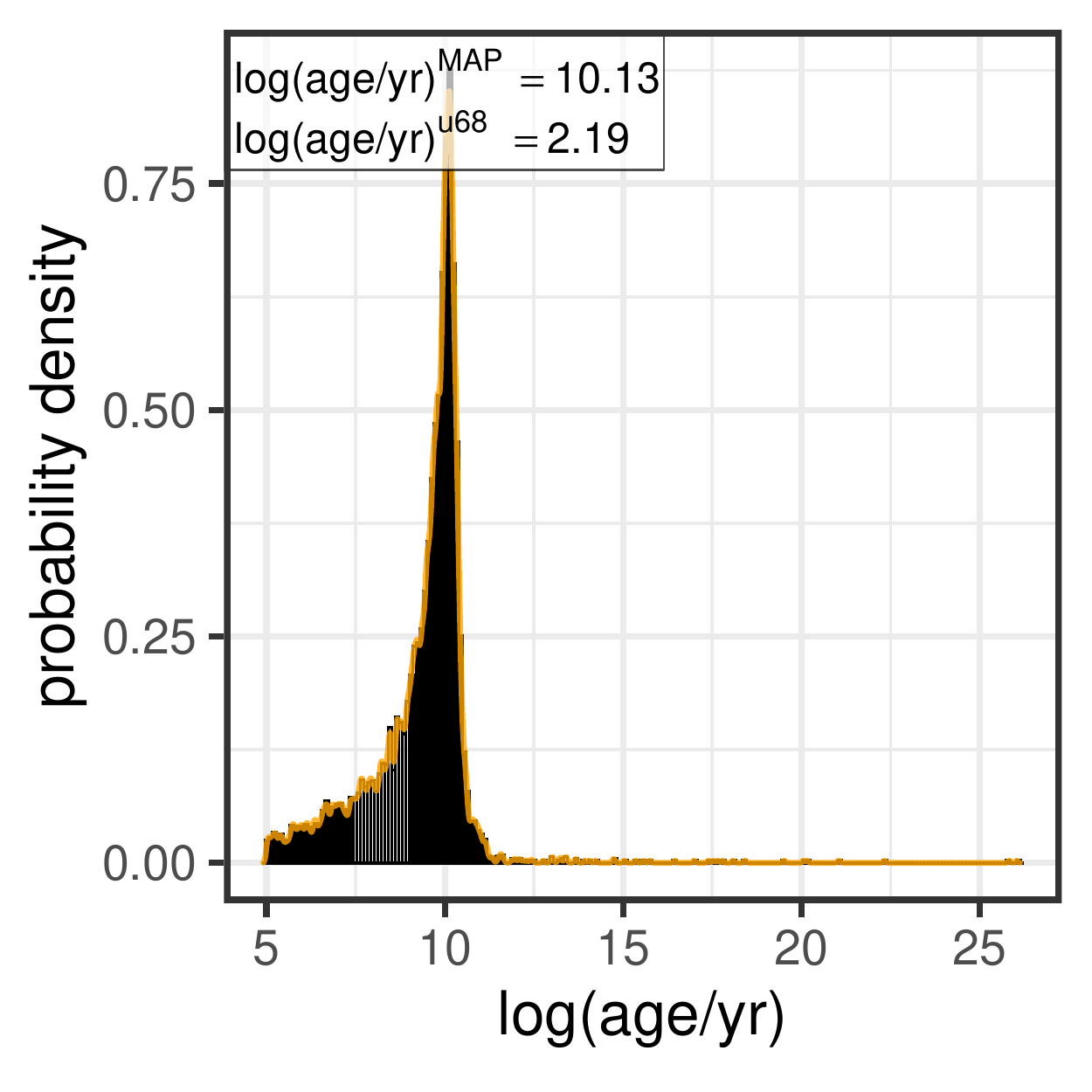}
        \caption{Histogram of the sum of the age posterior distributions of all NGC\,6397 stars as predicted by the 'NGC6397\_II' cINN model. The orange line indicates a kernel density fit to this cumulative posterior distribution to determine the most likely cluster age.}
        \label{fig:NGC6397_II_logAgeKDE}
    \end{figure*}

\begin{figure*}
	\centering
	\includegraphics[width = 0.85\linewidth]{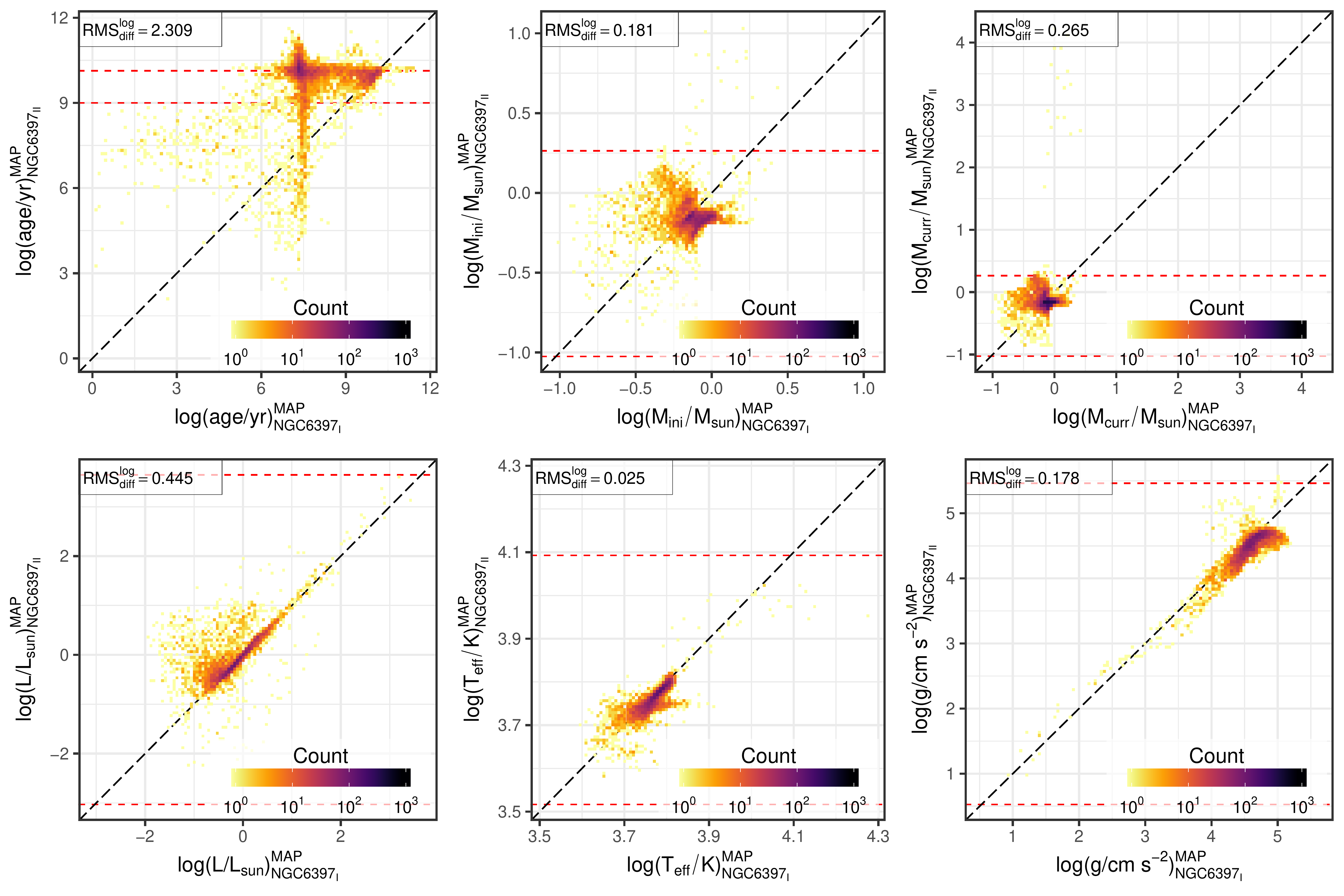}
	\caption{2D histogram of the comparison between the MAP estimates of the physical parameters as predicted by the cINN models trained on 'NGC6397\_I' and 'NGC6397\_II' respectively. The black dashed line indicates a perfect 1-to-1 correlation, while the red dashed lines mark the limits in parameter space of the training set 'NGC6397\_II'.}
	\label{fig:NGC6397_I_vs_NGC6397_II}
\end{figure*}

    This appendix presents additional diagrams and discussion concerning the results of the 'NGC6397\_II' cINN model.
    
    Figures \ref{fig:NGC6397_II_MAP_vs_true} and \ref{fig:NGC6397_II_Posteriors_vs_true} show the MAP and posterior vs. true diagrams for 'NGC6397', respectively, corresponding to Figures \ref{fig:Wd2_I_MAP_vs_true}, \ref{fig:Wd2_I_Posteriors_vs_true} for 'Wd2\_I', Figures \ref{fig:Wd2_II_MAP_vs_true}, \ref{fig:Wd2_II_Posteriors_vs_true} for 'Wd2\_II' and Figures \ref{fig:NGC6397_I_MAP_vs_true}, \ref{fig:NGC6397_I_Posterior_Vs_true} for 'NGC6397\_I'.
    
    Figures \ref{fig:NGC6397_II_MAP_CMD} shows the CMD coloured according to the MAP estimates analogous to Figure \ref{fig:NGC6397_I_MAP_pred_CMD} for 'NGC6397\_I'. In correspondence to the top left panel of Figure \ref{fig:NGC6397_II_MAP_CMD} Figure \ref{fig:NGC6397_II_Age_outliers} in the main paper provides a breakdown of the age prediction results in the CMD, distinguishing under- and overestimates from the reasonable outcomes. Figure \ref{fig:NGC6397_II_Age_outliers} also indicates that the elimination of all pre-main sequence examples in 'NGC6397\_II' helps the cINN to recognise the turn-off and RGB stars as old objects (c.f. Figure \ref{fig:NGC6397_I_Age_MAP_CMD_below_above_GYR_hists} for 'NGC6397\_I'). As the right panel demonstrates, discrepancies between the observed and modelled RGBs are again a likely cause for the cINN age overestimates for a number of RGB constituents. While the age predictions have arguably somehow improved, the prediction of effective temperature and luminosity appear to suffer slightly with the 'NGC6397\_II' cINN. Figure \ref{fig:NGC6397_II_HRD_outliers} shows the corresponding predicted HRD in the left panel, indicating about 650 outliers (orange points) to the right of the 13\,Gyr isochrone, which do not appear in the corresponding 'NGC6397\_I' diagram. The right panel in the same figure suggests that these outliers are, again, mainly LMS stars located where model and observations disagree the most.
    
    Figure \ref{fig:NGC6397_II_logAgeKDE} presents the cluster age derivation, analogous to Figure \ref{fig:NGC6397_I_logAgeKDE} shown in the main paper for 'NGC6397\_I'.
    Here it is worth mentioning that a cluster age of $13.4_{-13.2}^{+3.4}\,$Gyr, determined from the most likely value of the sum of all individual age posteriors (excluding the posteriors of the total failure cases with $\log(\mathrm{age/yr})^\mathrm{MAP} < 5$), is actually fairly plausible despite the described issues with the age prediction. Even neglecting these problems, however, the large uncertainties of this estimate make this outcome unsatisfactory.
    
    The last Figure \ref{fig:NGC6397_I_vs_NGC6397_II} exhibits the 2D histogram comparing the MAP estimates for the physical parameters between the 'NGC6397\_I' and 'NGC6397\_II' predictions, analogous to the comparison between 'Wd2\_I' and 'Wd2\_II' presented in Figure \ref{fig:Wd2_I_vs_Wd2_II} in Appendix \ref{app:Wd2_II}. Overall the differences are more significant than those resulting from the comparison of 'Wd2\_I' and 'Wd2\_II'. The predictions of $L$, $T_\mathrm{eff}$ and $g$ appear to be the least affected by the change in model, being quite close to a 1-to-1 correlation, although we find a median relative deviation of 20\% for $L$ and 18\% for $g$. At a first glance the predictions of $M_\mathrm{ini}$ and $M_\mathrm{curr}$ look more scattered around the 1-to-1 correlation, but with a median relative deviation of about 18.6\% the difference is of similar magnitude. 
    
\bsp	
\label{lastpage}
\end{document}